\documentclass[a4paper,fleqn]{cas-sc}

\usepackage[numbers]{natbib}
\usepackage{subcaption}
\usepackage{float}
\usepackage{placeins}
\usepackage{needspace}

\usepackage{graphicx} % Required for inserting images
\usepackage{multirow}
\usepackage{bm}
\usepackage{lineno}
\makeatletter
\newcommand*\patchAmsMathEnvironmentForLineno[1]{%
  \expandafter\let\csname old#1\expandafter\endcsname\csname #1\endcsname
  \expandafter\let\csname oldend#1\expandafter\endcsname\csname end#1\endcsname
  \renewenvironment{#1}%
    {\linenomath\csname old#1\endcsname}%
    {\csname oldend#1\endcsname\endlinenomath}%
}
\newcommand*\patchBothAmsMathEnvironmentsForLineno[1]{%
  \patchAmsMathEnvironmentForLineno{#1}%
  \patchAmsMathEnvironmentForLineno{#1*}%
}
\AtBeginDocument{%
  \patchBothAmsMathEnvironmentsForLineno{equation}%
  \patchBothAmsMathEnvironmentsForLineno{align}%
  \patchBothAmsMathEnvironmentsForLineno{flalign}%
  \patchBothAmsMathEnvironmentsForLineno{alignat}%
  \patchBothAmsMathEnvironmentsForLineno{gather}%
  \patchBothAmsMathEnvironmentsForLineno{multline}%
}
\makeatother
\usepackage[title]{appendix}

\begin{document}
\let\WriteBookmarks\relax
\def\floatpagepagefraction{1}
\def\textpagefraction{.001}
\shorttitle{}
\shortauthors{Guodan Dong et~al.}

\title [mode = title]{Multi-scale Dynamic Wake Modeling and Prediction of Floating Offshore Wind \texorpdfstring{{\color{black}Turbine Wakes}}{Turbine Wakes} via Physics-Informed Neural Networks and Fourier Neural Operators}                      

\author[1,2,3]{Guodan Dong}[style=chinese, orcid=0000-0001-8488-8294]
\ead{gd.dong@hhu.edu.cn}
\credit{Conceptualization of this study, Methodology, Software}

\affiliation[1]{College of Renewable Energy, Hohai University, Changzhou, 213200, China}
\affiliation[2]{College of Water Conservancy and Hydropower Engineering, Nanjing, 210098, China}
\affiliation[3]{National Technology Innovation Center for Wind Power, Hohai University, Changzhou, 213200, China}

\author[1]{Jianhua Qin}[style=chinese]
\cormark[1]
\credit{Data curation, Methodology, Software, Writing - Original draft preparation}
\ead{J.Qin@hhu.edu.cn}
\cortext[cor1]{Corresponding author}
% \author[1,2]{Jiawei Zhang}[style=chinese]
% \credit{Conceptualization of this study, Methodology, Software, Writing}

\author[1,2,3]{Chang Xu}[style=chinese]
% \cormark[2]
\ead{zhuifengxu@163.com}
\credit{Conceptualization of this study, Methodology, Writing}
% \cortext[cor2]{Corresponding author}

\begin{abstract}
Multi-scale dynamic wake modeling and prediction are essential for wake assessment and optimization of floating offshore wind turbines (FOWTs). In this study, wakes of FOWTs under coupled surge and pitch motions across a range of Strouhal numbers ($St_{\rm p}$) are modeled via physics-informed neural networks (PINNs) and Fourier neural operators (FNOs). Using a dataset generated by large-eddy simulations with the actuator line model (LES-AL), the capabilities of the two frameworks are compared. Results demonstrate that dominant dynamic wake structures, such as wake meandering, can be captured by both frameworks. However, FNOs outperform PINNs in terms of training efficiency and accuracy in mean-wake reconstruction and instantaneous multiscale prediction.
Over the prediction period, the FNO model achieves a normalized streamwise velocity mean absolute error (MAE) of 0.0222, 58.0\% below the PINN value of 0.0529. Furthermore, wakes reconstructed by the PINNs exhibit a smoothing effect, leading to underestimated fluctuations in wake-center position and wake half-width. For the reconstruction interval, power spectral density (PSD) and pre-multiplied PSD (FPSD) analysis are performed. PSD reveals that FNOs resolve the primary meandering frequency ($St_{\rm p}$), its higher-order harmonics ($2St_{\rm p}, 3St_{\rm p}$), and retain the resolved temporal spectral distribution. However, the PINN spectra attenuate more rapidly at higher resolved frequencies ($St > 1.0$). The FPSD indicates that the energy contained in the meandering and its harmonic frequencies, as modeled by PINNs, is relatively low compared to that in LES and FNOs. These findings suggest that FNOs are promising for the multi-scale dynamic modeling of FOWT wakes.
\end{abstract}

% % Research highlights
% \begin{highlights}
% \item 
% \item 
% \item 
% \end{highlights}


\maketitle
%\linenumbers

\section{Introduction} \label{sec:Intro}
With the accelerating development of offshore wind energy, the investigation of floating offshore wind turbines (FOWTs) has become a prominent focus, given that more than 70\% of global wind resources are located in deep-water regions~\citep{GWEC2025Report}. Previous studies have shown that wake effects in {\color{black}onshore wind farms} can induce power losses of 40\% to 80\%~\citep{barthelmie2007modelling,dong2026reinforcement} and typically elevate fatigue loads by more than 10\%, with extreme scenarios reaching up to 80\%~\citep{wang2023evolution}. {\color{black}For FOWTs, wake effects are more dynamic and complex under six-degree-of-freedom (6-DOF) motions (especially surge and pitch) of the wind turbine, significantly influencing downstream turbine performance and structural integrity.}
Therefore, the accurate and computationally efficient modeling and prediction of these multi-scale dynamic wakes of FOWTs is a critical prerequisite for wind farm layout optimization and control~\citep{yang2026wind,sant2026progress}.

The modeling of FOWTs has recently received significant attention and can be categorized into three groups~\citep{stevens2017flow}: (i) analytical wake models, (ii) computational fluid dynamics (CFD) methods, and (iii) data-driven approaches. 
Analytical FOWT wake models are in their early stages. To capture the non-stationary, time-varying wakes of FOWTs, various dynamic models incorporating 6-DOF motions have recently been proposed, including the 3$D$ dynamic model accounting for wind shear~\citep{huanqiang2024investigation}, the high-order Gaussian model considering 6-DOF motions~\citep{wenfeng2025investigation}, and a surge-specific periodic model~\citep{he2025novel}. However, analytical models fail to resolve the intricate, multi-scale turbulent structures present in FOWT wakes~\citep{luo2025innovative, liu2026phywakenet, yang2026wind}. 
CFD methods, on the other hand, can resolve these structures and are therefore frequently used as high-fidelity benchmarks~\citep{zhang2025advanced}. Among them, large-eddy simulations (LES) have been increasingly preferred for wind energy applications~\citep{dong2023characteristics, dong2026impacts}. To reduce computational costs, wind turbines are commonly parameterized using actuator methods, such as the actuator line (AL), actuator disk (AD), and actuator surface (AS) models~\citep{martinez2015large, dong2022predictive}. In this work, high-fidelity wake datasets of FOWTs are generated using the LES-AL approach~\citep{li2025effects}. 

In advanced applications, the analytic model and CFD methods often struggle to balance high-fidelity with computational efficiency~\citep{luo2025innovative}. Therefore, data-driven methods have emerged as a promising alternative and can be broadly categorized into~\citep{goccmen2025data}: (i) reduced-order modeling (ROM)~\citep{taira2017modal} and (ii) machine learning (ML) approaches. Traditionally, ROMs such as proper orthogonal decomposition (POD), dynamic mode decomposition (DMD), and spectral proper orthogonal decomposition (SPOD) were utilized to identify coherent structures and physical mechanisms \cite{taira2017modal, li2022onset}. While hybrid approaches combining ROM and ML (such as POD combined with long short-term memory (LSTM) networks \cite{zhou2023high} or fully connected neural networks (FCNN) \cite{luo2025innovative}) have improved predictive capabilities, they remain sensitive to data fidelity and expert-dependent mode selection.

With the rapid development of ML, wind turbine wake modeling has evolved along three main paradigms: (i) purely data-driven learning, (ii) physics-informed neural networks (PINNs), and (iii) operator learning, including Fourier neural operators (FNOs).
{\color{black}Within the purely data-driven paradigm, a three-dimensional U-Net surrogate combined with a wake-superposition model has been used to predict single-turbine and wind-farm wake fields~\citep{parekh2023uncertainty}, while transformer-enhanced conditional generative adversarial networks have been developed for three-dimensional yawed wind turbine wakes~\citep{li2024transformerwake}. In related bluff-body flows, CNN autoencoders coupled with LSTM networks have reproduced the unsteady wake of a circular cylinder~\citep{hasegawa2020cnn}. These studies demonstrate the breadth of data-driven wake modeling, although model performance remains dependent on the coverage and fidelity of the training data.}

To enhance physical consistency, PINNs have gained attention by embedding governing equations into the loss function \cite{raissi2019physics} and have increasingly been applied to wind-energy flow modeling. Early applications focused on sparse wind-field reconstruction, where LiDAR measurements were combined with physical constraints to recover spatio-temporal wind fields~\cite{zhang2021three}. Physics-informed learning has also been explored for wind-farm applications, {\color{black}including wind-field modeling~\cite{cobelli2023physics}, digital-twin development~\cite{zhang2023digital}, and flow-field prediction using physics-informed deep architectures~\cite{hasanpoor2025physics}.}
{\color{black}Methodological advances have strengthened the standard PINN formulation through Fourier-feature embeddings and periodic activation functions to reduce spectral bias~\citep{wang2021fourierfeatures,sitzmann2020siren}, adaptive loss weighting and residual connections to improve optimization~\citep{wang2021gradient,wang2024piratenets}, and space--time domain decomposition, including XPINNs~\citep{jagtap2020xpinn}.}
Subsequent studies extended PINN-based approaches toward wind turbine wake reconstruction~\cite{zhang2025advanced}. Wang et al.~\cite{wang2024dynamic}, Zhang and Zhao~\cite{zhang2024reconstruction}, and Luo et al.~\cite{luo2025innovative} {\color{black}demonstrated the ability of PINNs to reconstruct dynamic wakes from sparse observations.} Recent efforts have further improved PINN-based wake modeling, including turbulence modeling through $k$--$\epsilon$ Reynolds-averaged Navier--Stokes (RANS) equations~\cite{gafoor2025physics}, multi-scale wake reconstruction with transfer learning~\cite{wang2026multi}, {\color{black}adaptive physics-informed architectures~\cite{jin2026sparse}, and wind turbine wakes of complex terrain~\cite {zhang2026novel}.} {\color{black}Despite these advances, coordinate-based PINNs can remain costly to train for turbulent flows and may underresolve high-frequency structures~\citep{wang2021fourierfeatures,luo2025innovative,zhang2025novel}. Moreover, to the best of our knowledge, their capability for modeling multi-scale dynamic wakes of FOWTs under coupled surge--pitch motions has not been systematically explored.}

Operator learning offers a complementary strategy by approximating mappings between function spaces rather than pointwise input--output relations~\citep{kovachki2023neural,azizzadenesheli2024neural}. Within this framework, DeepONet represents operators through branch and trunk networks~\citep{lu2021learning}, whereas FNO parameterizes integral kernels in Fourier space and evaluates the associated convolutions efficiently using the fast Fourier transform (FFT)~\citep{li2020fourier}. The resulting mapping is well suited to spatially coherent, multi-scale dynamics on regular grids. With the increasing development, FNO-based models have been applied to meteorological forecasting~\citep{pathak2022fourcastnet}, airfoil dynamic stall~\citep{meng2023fast}, and {\color{black}turbomachinery flow-field prediction~\citep{li2025attention}}.
For turbulent flow evolution, an implicit U-Net-enhanced FNO has been evaluated in LES of forced and decaying homogeneous isotropic turbulence and temporally evolving turbulent mixing layers, with assessments of velocity spectra, flow statistics, and coherent structures~\citep{li2023iufno}. Related work extended this formulation to three-dimensional turbulent channel flows~\citep{wang2024channelFNO}. In wake flows, FNOs have forecast experimentally measured subcritical cylinder wakes~\citep{renn2023forecasting}. Wind turbine applications include a spatio-temporal FNO trained on dynamic wake meandering (DWM) model data across inflow and yaw conditions~\citep{zhang2025novel}, and a channel-assisted FNO for yawed far-wake prediction~\citep{lee2025channelFNO}.

{\color{black}To the best of our knowledge, existing PINN and FNO wake studies have examined wakes of wind turbines without coupled motion. Wang et al.~\citep{wang2024dynamic} reconstructed yaw-induced wake dynamics from sparse LiDAR observations using a PINN. Zhang et al.~\citep{zhang2025novel} developed an FNO for dynamic wake prediction using DWM data under varying inflow and control conditions, while Lee and Lee~\citep{lee2025channelFNO} predicted yawed far wakes using an FNO trained on LES--actuator-disk data. These studies establish the capabilities of the individual approaches, but their relative performance for FOWT wakes under coupled surge--pitch motion remains insufficiently understood. Such motion produces more complex, multiscale wake dynamics in which motion-induced coherent structures interact with wake meandering and smaller-scale fluctuations, placing additional demands on wake reconstruction and prediction.}
{\color{black}Therefore, the present FNO model is trained on high-fidelity LES-AL results~\citep{li2025effects} for wake fields induced by coupled surge-pitch motions.} These motions are examined across a range of Strouhal numbers ($St_{\rm p} = fD/U_{\infty} \in [0, 0.6]$, where $f$ is the motion frequency, $D$ is the rotor diameter, and $U_{\infty}$ is the inflow velocity). Such a range of $St_{\rm p}$ is chosen because previous studies~\cite{dong2023characteristics, li2025effects} demonstrated that wake meandering (a large-scale dynamic wake motion of the entire far-wake in the spanwise and vertical direction) can be triggered under such motions. The results provide a promising pathway for applying PINNs and FNOs to multi-scale dynamic wake modeling and prediction of FOWTs.
{\color{black}The present study addresses this gap through a systematic PINN--FNO comparison of FOWT wake reconstruction and prediction across six prescribed surge--pitch conditions. Full-field errors, wake-center and half-width statistics, and resolved temporal spectra connect model accuracy to the wake scales retained by each approach. Both models recover the dominant large-scale wake motion, while the FNO more accurately reproduces wake fluctuations, meandering harmonics, and higher resolved frequencies during reconstruction and yields lower velocity errors over the unseen prediction interval. This comparison establishes their relative capabilities for representing the multiscale dynamics of motion-driven FOWT wakes. Held-out-frequency tests further assess the transfer of the FNO to unseen motion conditions.}

The remainder of this paper is organized as follows: Section~\ref{sec:Methods} details the methodology and experimental setup, including the flow data generation via LES-AL, the physical problem configuration under coupled surge and pitch motions, and the mathematical architectures of the PINN and FNO frameworks. Section~\ref{sec:Results} compares the models through wake field reconstruction and prediction, error and wake statistics analyses, and PSD and FPSD analyses. 
% comparing the models across four dimensions: computational efficiency and convergence, instantaneous flow field reconstruction and long-term prediction, quantitative local velocity profiles, and the capture of physical wake parameters such as meandering and deficit. This section also includes a detailed spectral fidelity analysis to illustrate the models' differing capabilities in resolving multiscale turbulent signatures. 
Finally, Section~\ref{sec:Conclusions} summarizes the key findings.

\section{Methods and Setups} \label{sec:Methods}
This section describes the generation of reference high-fidelity wake fields using large-eddy simulation with an actuator-line model (LES--AL) and the setup of the coupled surge-pitch cases. The physics-informed neural network (PINN) and Fourier neural operator (FNO) formulations, data partitions, and training protocols used in the comparison are then specified.

    \subsection{Flow data generation}
% The reference dataset for the multi-scale dynamic wakes generated by the coupled surge--pitch FOWT motions. The numerical method and its verification are presented first, followed by the computational domain, verification, mesh-independent study, motion conditions, and data sampling and temporal partition for PINNs and FNOs.
\subsubsection{Computational methods}
The generation of high-fidelity wind turbine wake data is a prerequisite for training and evaluating the PINN and FNO models. {\color{black}The simulations are performed within the OpenFOAM framework, where the turbulent flow is resolved using LES, and the aerodynamic effects of the turbine blades, nacelle, and tower are represented by the AL model~\cite{li2025effects}. The transient \texttt{pimpleFoam} solver is used with PIMPLE pressure--velocity coupling and the momentum predictor. The temporal derivative is discretized using the Euler scheme. The velocity-convection and Laplacian terms use \texttt{Gauss limitedLinearV} and \texttt{Gauss linear corrected}, respectively. The generalized geometric--algebraic multigrid (GAMG) solver is used for pressure, while the velocity equations are solved using the preconditioned bi-conjugate gradient (PBiCG) method with a diagonal incomplete lower--upper (DILU) preconditioner and a tolerance of $10^{-8}$.} The filtered governing Navier-Stokes equations can be shown as:
\begin{align}
     \frac{\partial \widetilde{u}_i}{\partial x_i}&=0, \\
      \frac{\partial \widetilde{u}_i}{\partial t}
    +\frac{\partial \widetilde{u}_i \widetilde{u}_j}{\partial x_j}
    &=-\frac{1}{\rho}\frac{\partial \widetilde{p}}{\partial x_i}
    +\nu\frac{\partial^2 \widetilde{u}_i}{\partial x_j\partial x_j}
    -\frac{\partial \tau_{ij}}{\partial x_j}
    + \frac{ f_{{\rm WT},\,i}}{\rho} ,
    \label{eq:NS} 
\end{align}
where $\widetilde{u}_i$ is the filtered velocity, $\widetilde{p}$ is the filtered pressure, $\rho$ is the air density, $\nu$ is the kinematic viscosity, $f_{{\rm WT},\, i}$ is {\color{black}the body force} introduced by wind turbine, and {\color{black}$\tau_{ij}=\widetilde{u_i u_j}-\widetilde{u}_i \widetilde{u}_j$ {\color{black}is the subgrid-scale (SGS) stress tensor introduced by filtering. Its deviatoric component is modeled using the dynamic Smagorinsky formulation}~\cite{dong2026impacts} as $ \tau_{ij}-\frac{1}{3}\tau_{kk}\delta_{ij}
=-2\nu_{\mathrm{sgs}}\widetilde{S}_{ij}$, where $\widetilde{S}_{ij} =\frac{1}{2}\left(\frac{\partial \widetilde{u}_i}{\partial x_j}
+\frac{\partial \widetilde{u}_j}{\partial x_i}\right)$ is the strain-rate tensor and  $\nu_{\mathrm{sgs}}=(C_s\Delta)^2|\widetilde{S}|$ is the SGS eddy viscosity, $C_s$ is the Smagorinsky coefficient computed via the dynamic procedure~\cite{germano1991dynamic}, $\Delta=V_{\rm cell}^{1/3}=(\Delta x\Delta y\Delta z)^{1/3}$ is the filter size given by the cubic root of the cell volume ($V_{\rm cell}$), $\Delta x$, $\Delta y$, and $\Delta z$ are the size of the meshes in the $x, y$, and $z$ directions, respectively. $|\widetilde{S}|=\left(2\widetilde{S}_{ij}\widetilde{S}_{ij}\right)^{1/2}$ is the magnitude of the strain-rate tensor.}

In the present work, the AL model~\cite{sorensen2002numerical,li2025effects,frontera2022cfd} is used to model turbine blades, the nacelle, and the tower. The model discretizes the turbine blades into several rotating segments, replacing them with actuator points where body forces are applied. {\color{black}Each of the three blades is represented by 68 actuator elements. The tower and nacelle/hub are represented by 40 and 36 elements, respectively. The tower tapers from 6.0~m at its base to 3.87~m at hub height, and the nacelle/hub is represented by a 6.0~m long, 3.0~m diameter cylinder. Blade geometry and sectional airfoil data follow the NREL 5-MW specification~\citep{jonkman2009definition}, using cylindrical root sections, DU airfoils, and NACA64\_A17 outboard sections~\citep{dong2023characteristics}. Glauert tip and root corrections~\citep{glauert1935airplane} are also used. {\color{black}The present simulations use static airfoil polars without a dynamic-stall model, with lift and drag coefficients obtained by linear interpolation between tabulated angles of attack.}} Through this method, the boundary layer of the turbine blades is not resolved, thereby significantly reducing computational costs. The forces at the actuator points are derived from the drag ($C_D$) and lift ($C_L$) coefficients for the airfoil section, as shown $ \bm{F}_{\rm AL} = \frac{1}{2}\,\rho \,c \,|\bm{U}_{\mathrm{rel}}|^2\,(C_L \,\bm{e}_L+C_D \,\bm{e}_D)\,{\rm d}\bm{r}$, where $\bm{F}_{\rm AL}$ denotes the fluid force acting on the AL point, $c$ represents the airfoil chord length at the blade radius $r$, $\bm{U}_{\rm rel}$ {\color{black}indicates} the relative wind speed, $\bm{e}_L$ and $\bm{e}_D$ are the unit vectors in the directions of lift and drag, respectively, and $d\bm{r}$ is the length of blade element. For the prescribed surge--pitch motion considered here, the relative velocity $\bm{U}_{\mathrm{rel}}$ can be calculated as $ \bm{U}_{\mathrm{rel}} = \bm{U}_{\mathrm{in}}+\bm{U}_{\mathrm{motion}}+\bm{U}_{\mathrm{rotation}}
$, in which $\bm{U}_{\mathrm{in}}$ represents the incoming velocity at the AL point, $\bm{U}_{\mathrm{motion}}$ denotes the velocity of the AL point due to the turbine motion, $\bm{U}_{\mathrm{rotation}}$ is the velocity of the rotor.
{\color{black}
The Eulerian body force can be expressed as $\bm{f}_{\mathrm{WT}} = \sum_{i=1}^{N} \bm{F}_{\rm AL}\,\eta$. Here, $N$ is the total number of actuator points. The vector $\bm{F}_{AL}$ denotes the discrete force at actuator point $n$, $\eta=\frac{1}{\varepsilon^{3}\, \pi^{3/2}} \,{\rm \,e}^{-(d / \varepsilon)^{2}},$ is the three-dimensional isotropic Gaussian function~\citep{sorensen2002numerical} and is used to distribute the discrete body forces to regularize the gradient of the body force applied by adjacent AL points. Besides, $\varepsilon$ is the projection width, {\color{black}$d=\left|\bm{r}-\bm{r}_n\right|$ is the distance between the Eulerian cell center $\bm{r}$ and the instantaneous position $\bm{r}_n$ of the $n$-th AL point}, and the location where the body force is applied~\citep{martinez2015large, martinez2017optimal}.
For each actuator point, the projection width is prescribed as ${\color{black}\varepsilon=\max \left[ 0.25c,\, C_{D}c/2,\, m_f\Delta \right]}$, where $c$ is the local blade chord at actuator point $n$ and the Gaussian mesh factor is set to ${\color{black}m_f=2}$.
}

\begin{figure}
    \centering
    \includegraphics[width=0.7\linewidth]{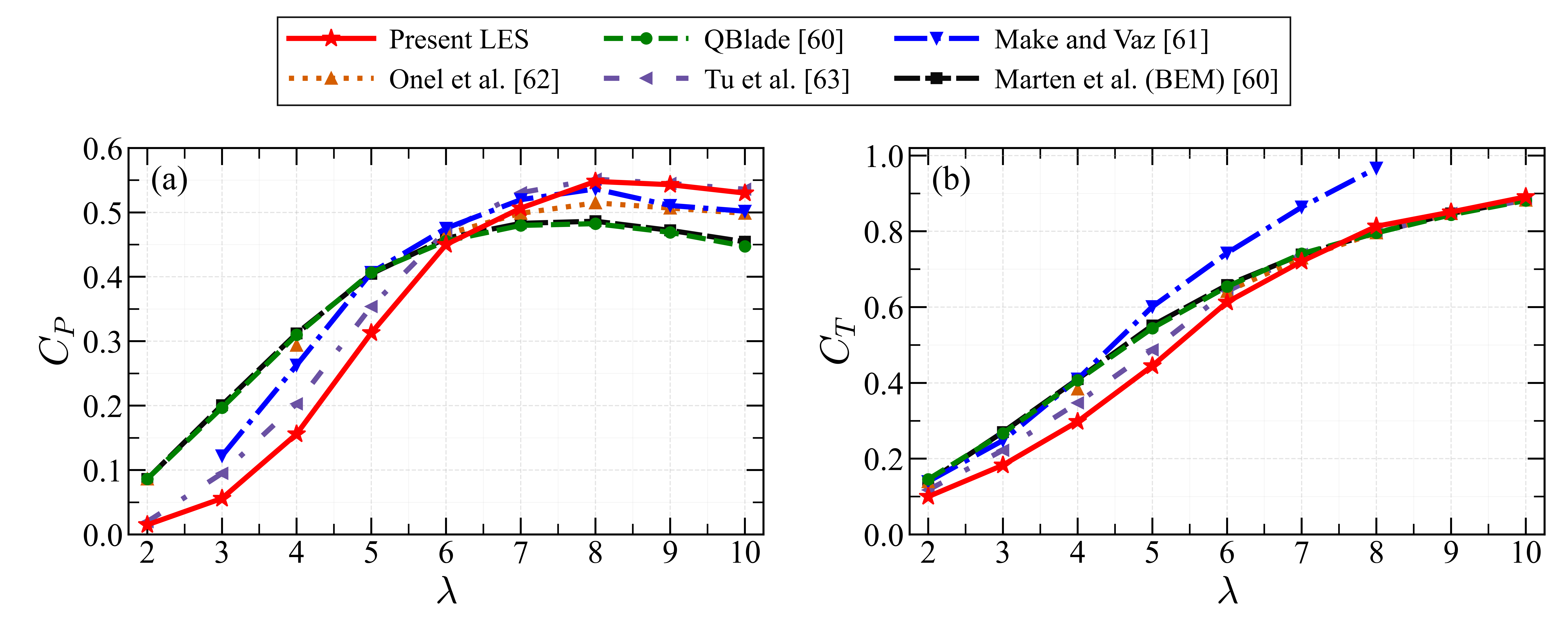}
    \caption{{\color{black}Comparison of the present (a) $C_P$ and (b) $C_T$ with published results over $\lambda=2$-10. }}
    \label{fig:cpct_validation}
\end{figure}
\begin{figure}
    \centering
    \includegraphics[width=0.6\linewidth]{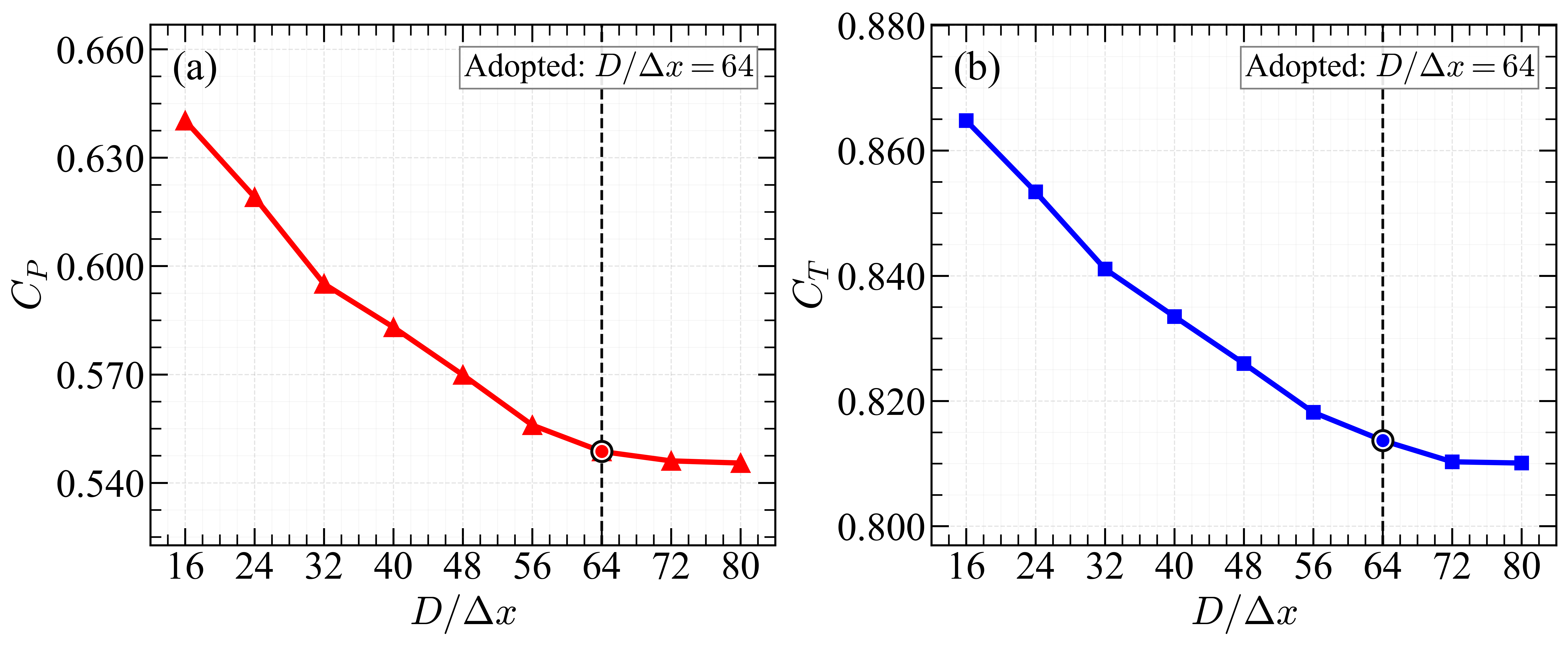}
    \caption{{\color{black}Grid convergence study of (a) $C_P$ and (b) $C_T$.}}
    \label{fig:grid_independence}
\end{figure}
\Needspace{5\baselineskip}
\subsubsection{{\color{black}Verification}}\label{sec:verification}
{\color{black} Figure~\ref{fig:cpct_validation} compares the power coefficient ($C_P=P/(0.5\rho A U_{\infty}^{3})$) and thrust coefficient ($C_T=F_T/(0.5\rho A U_{\infty}^{2})$) obtained in the present work with results of previous studies~\cite{marten2013qblade, make2015analyzing, onel2021investigation, tu2023aerodynamic}. The present solution reproduces the increase in $C_P$ toward a broad maximum near the selected operating tip speed ratio ($\lambda = \Omega R/U_{\infty}$, where $ \Omega$ and $R$ are {\color{black}the rotor angular velocity and rotor radius}, respectively) and the continued increase in $C_T$ with $\lambda$. At $\lambda=8$, the present values are $C_P=0.548$ and $C_T=0.813$. As seen, for all $ \lambda$ values, our results are in good agreement with the published results.
A grid convergence study has also been performed with results presented in Fig.~\ref{fig:grid_independence}. Both $C_P$ and $C_T$ approach stable values as the mesh is refined. Relative to the next finer $D/\Delta x=72$ mesh, the differences in $C_P$ and $C_T$ are 0.48\% and 0.42\%, respectively. Relative to the finest $D/\Delta x=80$ mesh, the corresponding differences are 0.59\% and 0.44\%. To balance computational efficiency and accuracy, a grid resolution of $D/\Delta x=64$ is adopted in the present study.
The effects of temporal resolution and Gaussian force projection width were also examined. At $\Delta t=0.05$, 0.025, and 0.0125~s, $C_P$ is 0.5601, 0.5487, and 0.5481, while $C_T$ is 0.8208, 0.8137, and 0.8133, respectively. Since the results at $\Delta t=0.025$ and 0.0125~s are nearly identical, $\Delta t=0.025$~s is adopted to balance accuracy and computational efficiency. At $m_f=1$, 2, and 4, $C_P$ is 0.5460, 0.5487, and 0.5883, while $C_T$ is 0.8128, 0.8137, and 0.8357, respectively. Since the results at $m_f=1$ and 2 are close, $m_f=2$ is adopted in the present simulations.}
{\color{black}Previous work~\citep{li2025effects} employed the same LES--AL approach and showed close agreement with the transverse wake-velocity profiles of Tu et al.~\citep{tu2023aerodynamic} at $x/D=1$, 3, and 5 for the NREL 5-MW turbine under uniform inflow of 11.4~m/s and $\lambda=8$. This agreement supports the capability of the numerical approach to reproduce the downstream velocity-deficit distribution and complements the present rotor-load comparison. {\color{black}Note that these wake-profile comparisons provide supporting validation evidence for the LES--AL methodology under fixed-turbine conditions.}}

\subsubsection{Problem setup}
\begin{figure}
    \centering
    \includegraphics[width=0.9\linewidth]{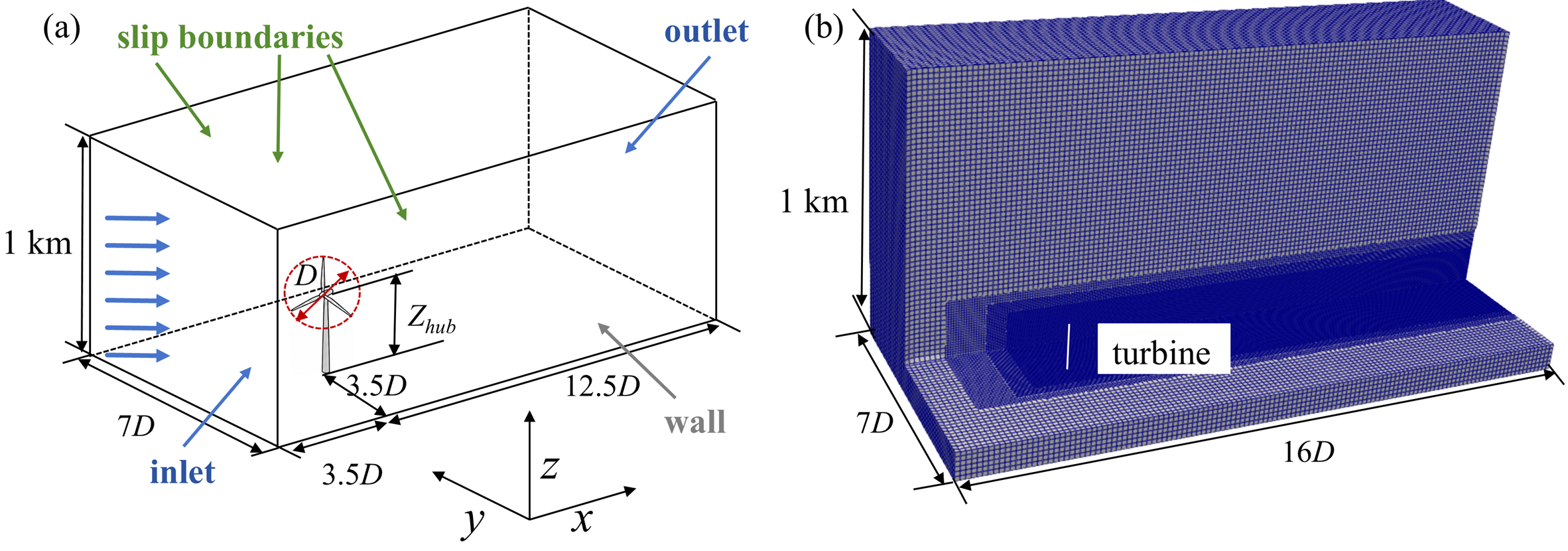}
    \caption{Computational setup: (a) domain and boundary conditions and (b) three-dimensional mesh details.}
    \label{fig:Schematic_main}
\end{figure}

Figure~\ref{fig:Schematic_main}(a) shows the computational domain, with dimensions $L_x=16D$, $L_y=7D$, and $L_z=1$~km in the streamwise, spanwise, and vertical directions, respectively, where $D$ is the rotor diameter. The turbine is located $3.5D$ downstream of the inlet and centered spanwise, with its hub at $(0,0,z_{\rm hub})$. {\color{black}As illustrated in this figure, a fixed uniform velocity and zero-normal-gradient pressure are imposed at the inlet. Zero-normal-gradient velocity and fixed reference pressure are applied at the outlet. The bottom is treated as an idealized hydraulically smooth no-slip wall with zero prescribed roughness, while the two lateral boundaries and the top use slip conditions. No precursor simulation, atmospheric-surface-layer parameterization, or wall function is applied, and the near-ground layer is not used as a validation target in the present hub-height wake analysis~\citep{li2025effects}.}
The base mesh has uniform spacing $\Delta x_0=\Delta y_0=\Delta z_0=D/8$. As shown in Fig.~\ref{fig:Schematic_main}(b), three nested refinement regions are used to resolve the tip vortices and wake dynamics. 
% Their opposing corners are $[-2.5D,-2D,0]$ and $[12.5D,2D,1.5D+z_{\rm hub}]$, $[-1.5D,-1.75D,0]$ and $[12.5D,1.75D,1.25D+z_{\rm hub}]$, and $[-D,-1.5D,0]$ and $[12.5D,1.5D,D+z_{\rm hub}]$, respectively. 
{\color{black}The first refinement region spans from $[-2.5D, -2D, 0]$ to $[12.5D, 2D, 1.5D + z_{\rm hub}]$, the second spans from $[-1.5D, -1.75D, 0]$ to $[12.5D, 1.75D, 1.25D + z_{\rm hub}]$, and the third spans from $[-D, -1.5D, 0]$ to $[12.5D, 1.5D, D + z_{\rm hub}]$.}
The finest spacing is $\Delta x=\Delta y=\Delta z=D/64$ throughout the wake refinement region, and the complete mesh contains approximately 20 million cells. The rotor-loading grid sensitivity of this resolution is documented in Section~\ref{sec:verification}, and {\color{black}a comparable wake-region grid scale was used in previous work}~\citep{li2025effects}{\color{black}, whose wake-profile comparisons provide complementary evidence for the downstream velocity-deficit representation}.
{\color{black}The rotor swept area is approximately 1.4\% of the inlet cross-sectional area. The learning domain is restricted to $|y/D|\leq1.5$, and the downstream wake diagnostics extend to $x/D=10$, leaving $2D$ between the learning domain and each lateral boundary and $2.5D$ between the farthest analysis station and the outlet. The same domain strategy has been applied in previous LES studies of fixed, floating, and tandem wind-turbine wakes~\citep{li2022onset,dong2023characteristics,dong2026impacts}. {\color{black}The present domain extends $3.5D$ to either side of the rotor centre, well beyond the $|y/D|\leq1.5$ region used to analyse wake meandering. This lateral extent provides space to capture the meandering wake and is supported by the wake scales examined in previous studies~\citep{dong2023characteristics,dong2026impacts}.}}

The NREL 5~MW baseline offshore wind turbine is used, with its principal parameters listed in Table~\ref{tab:paras}. Further details can be found in Ref.~\citep{dong2023characteristics}.
FOWTs can undergo 6-DOF motions induced by combined wind, wave, and current forcing, and surge and pitch can strongly modify wake evolution and are most prevalent~\citep{li2022onset,wang2023evolution}. The prescribed coupled surge--pitch motions are investigated in the present work, with the surge displacement $x(t)$ and pitch angle $\theta(t)$ as: $x(t) = A_s \sin(2\pi f t), \,
    \theta(t) = A_{p} \sin(2\pi f t),$
{\color{black}where $A_s=0.04D$ and $A_p=5^\circ=0.0873$~rad are the surge and pitch amplitudes adopted from previous studies~\citep{li2022onset,li2025effects}. Both motions are defined about the rotation center at the turbine hub, $(0,0,z_{\rm hub})$. The corresponding surge velocity and pitch angular velocity are:}
\begin{align}
    v(t) = 2\pi f A_s \cos(2\pi f t), \\
    \dot{\theta}(t)=2\pi f A_p\cos(2\pi f t).
\end{align}
{\color{black}The common motion frequency is represented by $St_{\rm p}=fD/U_{\infty}$, where $U_{\infty}$ is the inflow velocity. Table~\ref{tab:Sts} lists the fixed case and the five moving cases with $St_{\rm p}=0.2$--0.6. This range spans the regime in which motion-induced wake meandering was identified in the preceding studies~\citep{li2022onset}.}
\begin{table}[width=.9\linewidth, pos=h]
\caption{Parameters of the NREL 5~MW turbine.}
\label{tab:paras}
\begin{tabular}{l c} 
\hline \hline
\textbf{Parameter} & \textbf{Value} \\
\hline
Rated power [MW]        & 5.0   \\
Hub height [m]          & 90.0  \\
Rotor diameter [m]      & 126.0 \\
Hub diameter [m]        & 3.0   \\
Cut-in wind speed [m/s] & 3.0   \\
Cut-out wind speed [m/s]& 25.0  \\
Rated wind speed [m/s]  & 11.4  \\
Number of airfoil types & 8     \\
\hline
\end{tabular}
\end{table}
\begin{table}
\caption{{\color{black}Simulated cases for the NREL 5~MW turbine ($U_{\infty}=11.4$~m/s, $A_s=0.04D$, and $A_p=5^\circ=0.0873$~rad). }}
\label{tab:Sts}
\begin{tabular*}{0.8\textwidth}{@{\extracolsep{\fill}} l c c c c c c @{}} 
\hline \hline
\color{black}\textbf{$St_p$} & fixed & 0.2 & 0.3 & 0.4 & 0.5 & 0.6 \\
\hline
\color{black}\textbf{$f$ (Hz)} & \color{black}-- & \color{black}0.0181 & \color{black}0.0271 & \color{black}0.0362 & \color{black}0.0452 & \color{black}0.0543 \\
\textbf{Period (s)} & - & 55.26 & 36.84 & 27.63 & 22.11 & 18.42 \\
\color{black}\textbf{Training cycles} & \color{black}-- & \color{black}3.60 & \color{black}5.40 & \color{black}7.20 & \color{black}9.00 & \color{black}10.80 \\
\color{black}\textbf{Prediction cycles} & \color{black}-- & \color{black}0.90 & \color{black}1.36 & \color{black}1.81 & \color{black}2.26 & \color{black}2.71 \\
\hline
\end{tabular*}
\end{table}

{\color{black}A uniform inlet velocity of $U_{\infty}=11.4$~m/s is imposed.} Because the inflow is uniform, $U_{\rm hub}=U_{\infty}$. The rotor operates at the selected tip-speed ratio, $\lambda=8$. {\color{black}The LES time step is $\Delta t=0.025$~s}.
The initial 250~s are {\color{black}excluded as a startup interval}, exceeding one nominal inlet to outlet flow through time{\color{black}~($L_x/U_\infty\simeq177$~s). Previous wake studies~\citep{dong2023characteristics,dong2026impacts} likewise separate flow development from subsequent analysis}. {\color{black}The nominal flow-through time provides a reference for the initialization duration, with statistical repeatability of the periodically forced wake not separately assessed in this study.} {\color{black}The horizontal wake fields used by the PINN and FNO models are extracted on the hub-height plane, $z=z_{\rm hub}$, and sampled every 1~s, yielding 251 snapshots at $t=250,\ldots,500$~s. The LES wake region has a finest spacing of $D/64$, and the extracted fields are interpolated onto a uniform, cell-centered $700\times150$ grid over $x/D\in[-1.5,12.5]$ and $y/D\in[-1.5,1.5]$, with an in-plane spacing of $\Delta x_{\rm ML}=\Delta y_{\rm ML}=D/50$.} {\color{black}Preliminary model-selection training} uses 250--425~s, with FNO target fields beginning at 254~s because four preceding fields are required{\color{black}. The} 426--449~s interval is reserved for {\color{black}model-selection evaluation} and determines the retained training duration. Once that duration is fixed, each final model is initialized from scratch and trained using the available data through 449~s. {\color{black}Throughout this paper, ``training'' denotes this final 250--449~s stage, and prediction is evaluated over the unseen {\color{black}450--500~s interval (51 snapshots)}. No field from {\color{black}this prediction interval} enters training, checkpoint selection, or training-duration selection. %Because this interval was inspected during model development before the protocol was fixed, it is treated conservatively as a development benchmark; blind confirmation requires a new LES continuation or an independent trajectory.
}

\subsection{Physics-informed neural network (PINN) algorithm}
\begin{figure}
    \centering
    \includegraphics[width=0.75\linewidth]{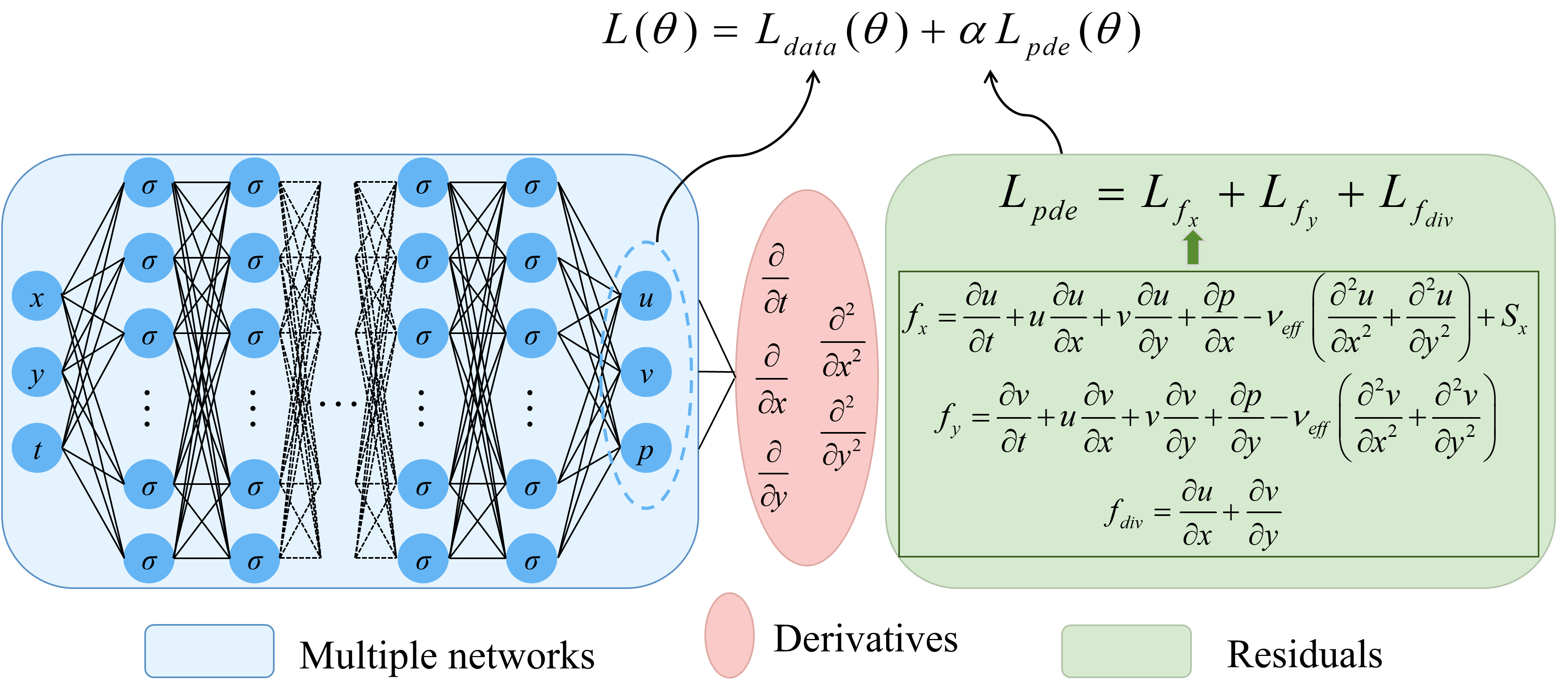}
    \caption{The PINN architecture.}
    \label{fig:pinn_architecture}
\end{figure}

{\color{black}The PINN model is applied to the horizontal hub-height plane, $z=z_{\rm hub}$, over $x/D\in[-1.5,12.5]$ and $y/D\in[-1.5,1.5]$, where $x$ and $y$ denote the streamwise and spanwise coordinates, respectively. Related plane-based physics-informed formulations have been adopted for wind-turbine wake reconstruction from sparse measurements~\citep{wang2024dynamic,luo2025innovative,zhang2025advanced,wang2026multi}. The hub-height plane passes through the rotor center and captures the velocity deficit, lateral wake meandering, and multiscale velocity fluctuations along the horizontal rotor diameter{\color{black}. These} features determine the time-dependent wake overlap with a downstream turbine and are therefore particularly relevant to its power capture and fatigue-load exposure. Because $z$ is fixed, it is not included as an input coordinate. The vertical velocity $w$ is excluded from the PINN outputs, supervised targets, and two-dimensional residuals. The model maps $\bm{x}=[x,y,t]^{\rm T}$ to $\hat{\bm{u}}=[u,v,p]^{\rm T}$. Only $u$ and $v$ are supervised at the labeled points. Pressure is an auxiliary output inferred through the momentum residuals and is determined through its gradients up to an arbitrary additive constant.} The network has five hidden layers, each with 128
neurons and a $\tanh$ activation, followed by a linear output layer.
Including the learned scalar thrust amplitude and logarithmic effective
viscosity, the model has 66,949 trainable real-valued parameters.
Figure~\ref{fig:pinn_architecture} shows the fully connected architecture of the PINN network. Its forward map is:
    \begin{equation}
        \begin{aligned}
            \bm{z}_0 &= \bm{x}, \\
            \bm{z}_k &= \sigma(\bm{W}_k \,\bm{z}_{k-1} + \bm{b}_k), \quad k = 1, 2, \dots, L-1, \\
            \hat{\bm{u}} &= \bm{W}_L \,\bm{z}_{L-1} + \mathbf{b}_L,
        \end{aligned}
    \end{equation}
where $\bm{W}_k \in \mathbb{R}^{n_k \times n_{k-1}}$ and $\bm{b}_k \in \mathbb{R}^{n_k}$ are the weight matrix and bias vector of layer $k$, respectively. In the present network, $\sigma=\tanh$ in every hidden layer and the output layer is linear. {\color{black}The $\tanh$ activation is fixed for all operating conditions{\color{black}. The} sensitivity analysis varies the physics-loss weight while keeping the network architecture and activation unchanged. The complete trainable parameter set is $\theta=\{\bm{W}_k,\bm{b}_k\}_{k=1}^{L}\cup\{C_T^{\rm PINN},\beta_\nu\}$, where $C_T^{\rm PINN}$ and $\beta_\nu$ parameterize the streamwise source and effective viscosity defined below.}
Before the residuals are formed, length, velocity, time, and pressure are scaled by $D$, $U_\infty$, $D/U_\infty$, and $\rho U_\infty^2$, respectively. The nondimensional variables are denoted by $x$, $y$, $t$, $u$, $v$, and $p$ below for conciseness. Although the corresponding rotor-scale Reynolds number is $Re_D=U_\infty D/\nu$, the molecular coefficient $1/Re_D$ is not imposed. Instead, the Laplacian coefficient is the learned positive dimensionless effective viscosity $\nu_{\rm eff}=\exp(\beta_\nu)$, which absorbs part of the discrepancy associated with the omitted transport terms for regularization purposes and is not interpreted as the molecular viscosity or as a physical turbine Reynolds number. Similar to previous wake-reconstruction studies~\cite{luo2025innovative,zhang2026novel}, the two-dimensional residuals are used to constrain in-plane continuity and momentum balance, rather than as a closed representation of the three-dimensional LES wake. The vertical velocity, out-of-plane transport, and SGS stress divergence are not included explicitly. Their influence is supplied indirectly through the supervised LES velocity targets. The scope of the PINN model is therefore restricted to reconstruction and prediction on the selected hub-height plane. The physics-informed residuals are:
\begin{align}
    f_x &= \frac{\partial u}{\partial t} + \left( u \frac{\partial u}{\partial x} + v \frac{\partial u}{\partial y} \right) + \frac{\partial p}{\partial x} - \nu_{\rm eff} \left( \frac{\partial^2 u}{\partial x^2} + \frac{\partial^2 u}{\partial y^2} \right) + S_{x} = 0, \\
    f_y &= \frac{\partial v}{\partial t} + \left( u \frac{\partial v}{\partial x} + v \frac{\partial v}{\partial y} \right) + \frac{\partial p}{\partial y} - \nu_{\rm eff} \left( \frac{\partial^2 v}{\partial x^2} + \frac{\partial^2 v}{\partial y^2} \right), \\
    f_{\rm div} &= \frac{\partial u}{\partial x} + \frac{\partial v}{\partial y} = 0,
    \label{eq:pinn_eq}
\end{align}
{\color{black}The source term can be written as:
\begin{equation}
 S_x=C_T^{\rm PINN}
 \exp\!\left[-\frac{x^2}{2(0.2)^2}\right]
 \left[1+\exp\{-20(0.5-|y|)\}\right]^{-1},
\end{equation}
where $C_T^{\rm PINN}$ is a learned scalar initialized to 0.6. Thus, $S_x$ is a prescribed, time-independent spatial mask with one trainable amplitude, rather than a scalar, an independently trainable field, or an additional network output. No $S_y$ is imposed because the prescribed surge translation and pitch rotation occur in the streamwise--vertical plane and do not prescribe a spanwise body force. Their unsteady effects on the in-plane wake are supplied by the time-resolved LES velocity labels. Introducing a free transverse forcing field would add an unidentifiable degree of freedom. The final $C_T^{\rm PINN}$ values for the fixed and
$St_{\rm p}=0.2$-0.6 cases are 0.8530, 0.8513, 0.8630, 0.8704, 0.8649, and
0.8666{\color{black}. The} corresponding effective viscosities are
$1.20\times10^{-4}$, $2.39\times10^{-4}$, $2.25\times10^{-4}$,
$2.92\times10^{-4}$, $3.70\times10^{-4}$, and $4.14\times10^{-4}$.
%These are reduced-order regularization parameters and are not interpreted as actuator-line thrust coefficients or molecular viscosities. The nondimensionalization uses $D=126$~m and $U_\infty=11.4$~m/s, but no fixed turbine Reynolds number is inserted into the reduced residual. Writing$1/\nu_{\rm eff}$ as an ``effective Reynolds number'' would only re-label a learned regularization parameter and is therefore avoided.
}
{\color{black}For optimization, the nondimensional input coordinates are further linearly mapped to $[-1,1]$.} The parameters are learned by minimizing $\mathcal{L}(\theta) = \mathcal{L}_{\text{data}}(\theta) + \alpha \mathcal{L}_{\text{PDE}}(\theta)$, with $\mathcal{L}_{\rm data}=\mathrm{SSE}(u)+\mathrm{SSE}(v)$, and
$\mathcal{L}_{\rm PDE}=\mathrm{SSE}(f_x)+\mathrm{SSE}(f_y)+
\mathrm{SSE}(f_{\rm div})$. Here, SSE denotes the sum of squared errors. The main comparison fixes $\alpha=1$. For each
case, 65,536 labeled $u$--$v$ space--time points are sampled without
replacement from the $700\times150$ hub-height fields over the 250--449~s candidate interval, while 524,288 additional
unlabeled collocation coordinates are generated by Latin-hypercube sampling. The data and PDE
streams each use batches of 65,536 points, giving eight updates per epoch.
Each operating condition is modeled by a trained
PINN with the same five-layer, 128-neuron $\tanh$ architecture. Adam uses a
learning rate of $10^{-3}$ and zero weight decay, without a scheduler. {\color{black}The
preliminary model-selection training uses 250-425~s, and 426-449~s is reserved for model-selection evaluation.
All candidates are trained for at most 160,000 updates. Among checkpoints
within 1\% of the minimum normalized $u$--$v$ root-mean-square error (RMSE) over the model-selection evaluation interval, the checkpoint
with the lowest streamwise-gradient relative $L_2$ error over the same interval is
selected to reduce oversmoothing. Once the update count is fixed, each
final model is trained from scratch on 250--449~s and evaluated once on the
{\color{black}450--500~s} fields. %The selected update counts for the fixed and $St_{\rm p}=0.2$--0.6 cases are 44,800, 51,200, 160,000, 92,000, 145,600, and 156,800, respectively. 
A sensitivity study with
$\alpha\in\{0.1,1,10\}$ for the fixed, $St_{\rm p}=0.4$, and $St_{\rm p}=0.6$ cases is
reported in Appendix~\ref{app:model-checks}. The primary PINN configuration is summarized in Table~\ref{tab:pinn_paras}.}

{\color{black}
\begin{table}
    \centering
    \caption{PINN hyperparameters.}
    \begin{tabular}{c c} \hline \hline 
    \textbf{Parameters} & \textbf{Values} \\
\hline
        Architecture/activation & $[3,128\times5,3]$/$\tanh$ \\
        Trainable parameters & 66,949 \\
        Analysis plane/grid & $z=z_{\rm hub}$/$700\times150$ ($D/50$) \\
        {\color{black}Model-selection training/evaluation} & 250--425/426--449~s \\
        {\color{black}Final training/prediction} & 250--449/{\color{black}450--500}~s \\
        Labeled/collocation design & 65,536 $u$--$v$/524,288 LHS PDE points \\
        %Pressure treatment & inferred through momentum residuals \\
        Data/PDE batch & 65,536/65,536 points \\
        Source/effective viscosity & fixed disk, learned amplitude/learned positive $\nu_{\rm eff}$ \\
        %Main physics weight $\alpha$ & 1 \\
        %Appendix sensitivity $\alpha$ & $0.1,1,10$ \\
        Maximum selection updates & 160,000 \\
        %Selected training updates & 44,800--160,000 \\
        %Checkpoint rule & within 1\% min. model-selection RMSE, then min. gradient error \\
        Initial learning rate/weight decay & $10^{-3}/0$ \\
        Optimizer/scheduler & Adam/none \\
        %Scaling & non-dimensional fields; inputs to $[-1,1]$ \\
        %Recorded seed & 1234 \\
        Final-training time per case & 16.58--61.14~min \\  \hline \hline
    \end{tabular}
    \label{tab:pinn_paras}
\end{table}
}

\subsection{Fourier neural operator (FNO) algorithm}
\begin{figure}
    \centering
    \includegraphics[width=0.7\linewidth]{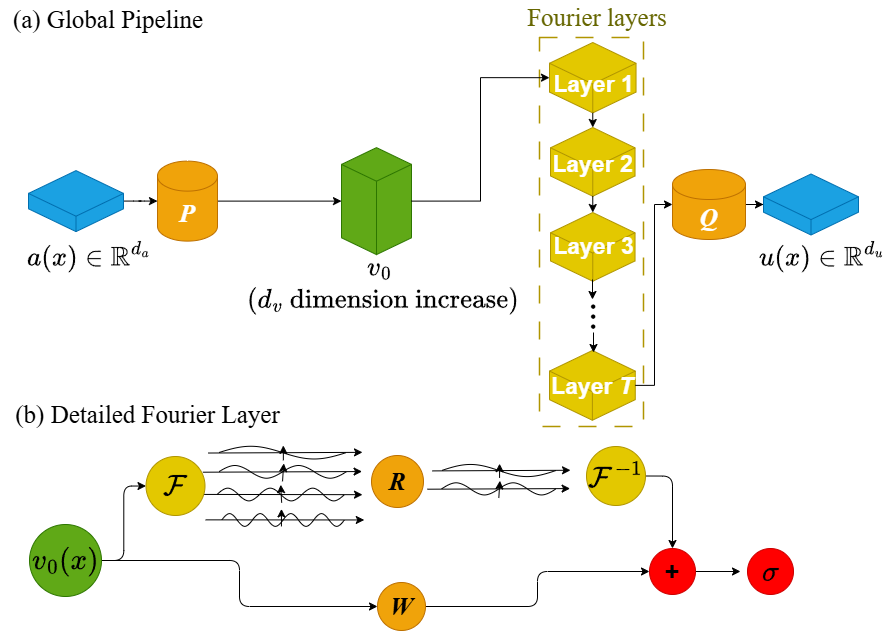}
    \caption{{\color{black}The FNO architecture.}}
    \label{fig:FNO_schematic}
\end{figure}

The FNO learns a mapping between function spaces by parameterizing an integral kernel in the Fourier domain. %Although the formulation is discretization compatible, all experiments in this study use the same spatial grid; mesh-resolution generalization is therefore not assessed. 
Figure~\ref{fig:FNO_schematic} shows the overall mapping and the operations within each Fourier layer.
A pointwise lifting map $P$ transforms the input field $a(x) \in \mathbb{R}^{d_{\rm in}}$ into the latent field $v_0(x) \in \mathbb{R}^{d_v}$. The latent field is then updated through $T$ Fourier layers:
\begin{equation}
    v_{t+1}(x) = \sigma \left( W_t v_t(x) + \mathcal{F}^{-1} \left( R_t \cdot \mathcal{F} v_t \right)(x) \right),
\end{equation}
where $W_t$ is a pointwise linear map, $\mathcal{F}$ and $\mathcal{F}^{-1}$ denote the Fourier transform and its inverse, and $R_t$ is a learnable complex valued tensor applied to the retained Fourier coefficients. Retaining $k_{x,\max}$ and $k_{y,\max}$ spatial modes provides nonlocal spectral coupling while preserving the $O(N\log N)$ complexity of the fast Fourier transform. A pointwise projection map $Q$ transforms $v_T$ to the output field. In this study, the resulting operator is trained to advance the wake state by one sampling interval~\citep{li2020fourier}.
For direct comparison with the separately trained PINNs, six independent FNOs are trained. 
{\color{black}Let $\bm q_t=[u_t,v_t,w_t]^{\rm T}$, and let $\bm q_t^{\rm std}$ denote its channel-wise standardized form based only on statistics from the model-selection training interval. At each time step, four consecutive standardized velocity fields are stacked into the input tensor, and the FNO model predicts the residual update:
\begin{equation}
    \widehat{\bm q}_{t+1}^{\rm std}
    =\bm q_t^{\rm std}
    +\mathcal{G}_{\Theta_{\rm FNO}}
    \left(\bm q_{t-3}^{\rm std},\ldots,\bm q_t^{\rm std}\right).
\end{equation}
Here, $\mathcal{G}_{\Theta_{\rm FNO}}$ is the learned FNO operator, parameterized by the trainable weights $\Theta_{\rm FNO}$, that maps four consecutive standardized wake states to a one-step state increment, and the hat denotes the predicted state. The sampling interval between successive wake fields is $\Delta t_{\rm FNO}=1~\mathrm{s}$. The four-frame history provides temporal-state information for the two-dimensional hub-height slice. Prediction over {\color{black}450--500~s} is initialized with the four LES fields at 446-449~s{\color{black}. Each} subsequent input contains the four most recent predicted fields. This autoregressive field-to-field formulation differs from the PINN mapping $[x,y,t]\mapsto[u,v,p]$, which predicts each field directly from its coordinates.}
{\color{black}The FNO model is optimized with a multi-step rollout loss{\color{black}. The} curriculum and checkpoint-selection protocol are documented in Appendix~\ref{app:model-checks}. %For a horizon $K$, the loss is the mean normalized squared field error over the $K$ sequentially predicted states. The curriculum uses $K=1,2,5,$ and 10 for 5, 5, 10, and 20 epochs, respectively, and the gradient norm is clipped at 1.0. During preliminary model selection, the epoch is chosen using the mean normalized RMSE over the contiguous 426--449~s model-selection evaluation interval. The final FNO is initialized from scratch and trained for the selected number of epochs using all fields through 449~s; prediction is then evaluated over 450--500~s.
}
The FNOs have four Fourier layers, 32 hidden channels, 64 retained modes in each spatial direction, GELU activation, grid positional embedding, and no domain padding. The Adam optimizer uses an initial learning rate of $10^{-3}$ and a weight decay of $10^{-4}$, and the batch size is 2. Deterministic seeds and training-only normalization statistics are stored with each checkpoint. {\color{black}The primary FNO configuration is summarized in Table~\ref{tab:fno_paras}.}
\begin{table}
    \centering
    \caption{FNO hyperparameters.}
    \begin{tabular}{c c} \hline \hline 
    \textbf{Parameters } & \textbf{Values} \\
\hline
        {\color{black}Grid and sampling} & {\color{black}$700\times150$ ($D/50$), $\Delta t=1$~s} \\
        %History/output channels & $4\times3/3$ ($u,v,w$) \\
        Fourier layers/hidden channels & $4/32$ \\
        Retained modes & $[64,64]$ \\
        Activation/padding & GELU/none \\
        {\color{black}Trainable parameters} & {\color{black}17,315,491} \\
        {\color{black}Model-selection training/evaluation targets} & 254--425/426--449~s \\
        {\color{black}Final training/prediction target fields} & 254--449/{\color{black}450--500}~s \\
        %{\color{black}Batch size} & {\color{black}2} \\
        %Multi-step curriculum $(K:\mathrm{epochs})$ & $1:5,2:5,5:10,10:20$ \\
        %Maximum/selected epochs & $40/25$--$40$ \\
        %Loss & mean $K$-step normalized MSE \\
        Initial learning rate & $10^{-3}$ \\
        Weight decay & $10^{-4}$ \\
        Optimizer/scheduler & Adam/cosine annealing \\
        %Additional normalization & train-only, per velocity channel \\
        %Primary seed & 20262023 \\
        %Checkpoint rule & minimum contiguous model-selection RMSE \\
        %Peak GPU memory per case & 7.39~GiB \\
        Final-training time per case & 2.54--5.55~min\\  \hline \hline
    \end{tabular}
    \label{tab:fno_paras}
\end{table}
{\color{black}Supplementary PINN and FNO sensitivity, ensemble, workload, inference, and persistence analyses are documented in Appendix~\ref{app:model-checks}, while the shared model and leave-one-case-out tests are reported in Appendix~\ref{app:transfer}.}
All experiments were performed on one workstation equipped with a 24-core Intel Core Ultra 9 285K CPU, 127.4~GiB of system memory, and an NVIDIA GeForce RTX 5090 D v2 GPU with 24~GiB of VRAM. The software environment comprised Windows 11, PyTorch 2.11, and CUDA 12.8.

%{\color{black}The fixed-case inference benchmark uses 30 warm-up calls followed by 200 timed calls. CUDA is synchronized immediately before and after each GPU call, and the input tensors remain resident on the measured device.}

\section{Results and Discussion} \label{sec:Results}
This section compares the PINN model and FNO model in terms of computational cost, wake-field reconstruction and prediction, time-averaged and instantaneous wake statistics, and spectral fidelity. The analysis proceeds from full-field errors and velocity deficit profiles to wake-center and half-width statistics, followed by power spectral density (PSD) and pre-multiplied PSD (FPSD) analyses of FOWT wake dynamics.

\subsection{Computational efficiency and wake-field reconstruction and prediction}
\begin{figure}
    \centering
    \includegraphics[width=0.85\linewidth]{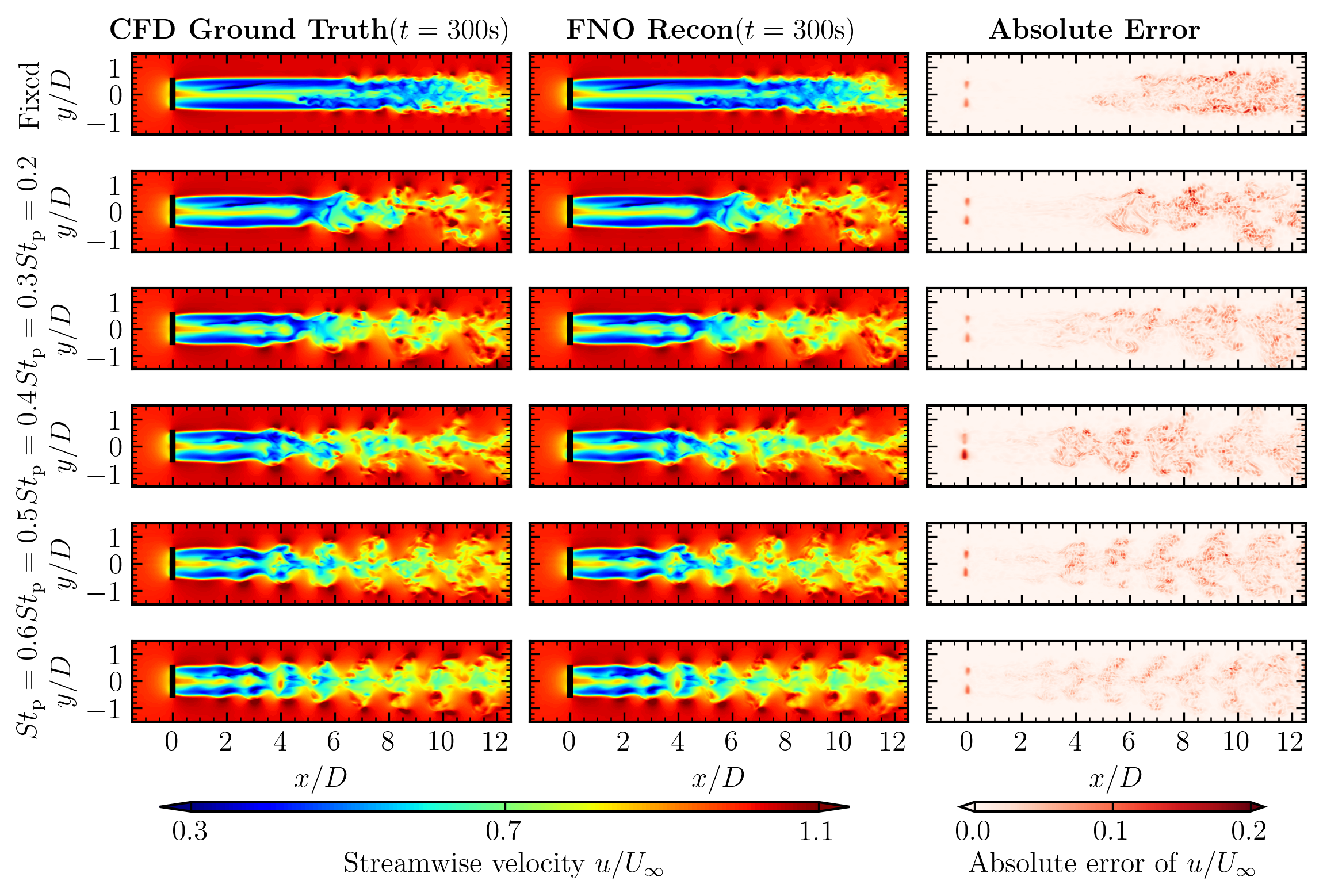}
    \caption{Contours of $u/U_{\infty}$ reconstructed by the FNO at $t=300$~s and the corresponding CFD reference.}
    \label{fig:fnos-300}
\end{figure}

\begin{center}
    \centering
    \includegraphics[width=0.85\linewidth]{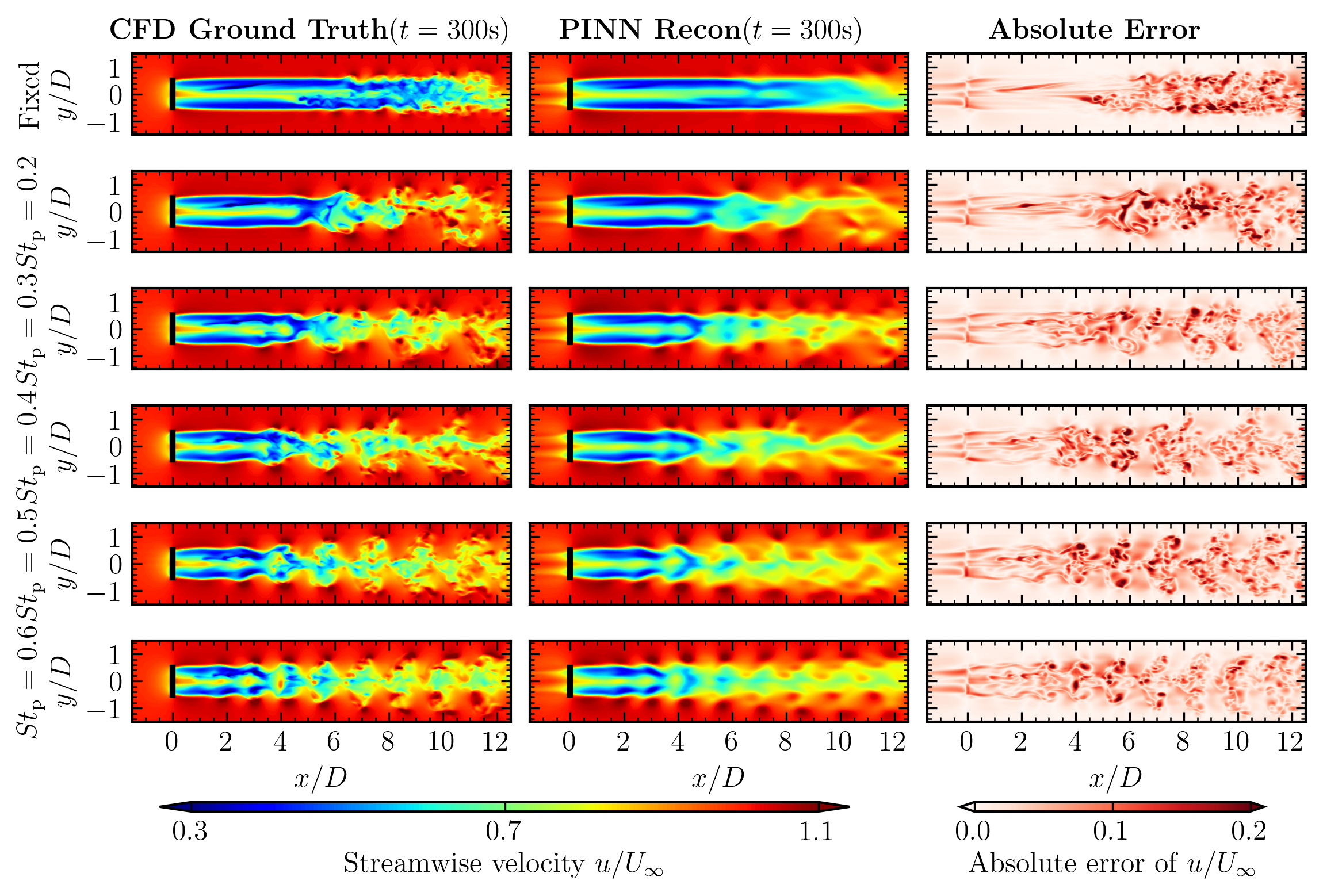}
    \captionof{figure}{Contours of $u/U_{\infty}$ reconstructed by the PINN at $t=300$~s and the corresponding CFD reference.}
    \label{fig:pinns-300}
\end{center}
{\color{black}As summarized in Table~\ref{tab:fair-cost}, the FNO model requires 7.36--10.39~min per case for training, compared with 74.60--123.53~min for the PINN model. The FNO model therefore reduces the training time by approximately one order of magnitude for the implemented configurations while also attaining lower streamwise velocity ($u/U_{\infty}$) prediction errors. The six-case mean MAE is 0.02219 for the FNO model and 0.05288 for the PINN model, and the corresponding final-time values are 0.02784 and 0.07613, respectively. Detailed training-work accounting is provided in Appendix~\ref{app:model-checks}. Both models infer a complete field well within the 1~s offline sampling interval, and their measured GPU and CPU median/P95 latencies are also reported in Appendix~\ref{app:model-checks}.
For the configurations considered here, the FNO provides a more favorable trade-off between predictive accuracy and training time. The PINN, however, has substantially fewer trainable parameters, requires less peak GPU memory during training, and achieves lower full-field inference latency. Nevertheless, the inference times of both models remain well below the 1~s wake-field sampling interval.
}
Figures~\ref{fig:pinns-300} and \ref{fig:fnos-300} compare the reconstructed instantaneous Streamwise velocity fields, $u/U_{\infty}$, at $t=300$~s. Previous work identified wake meandering over {\color{black}a motion Strouhal number range of 0.2--0.6} under prescribed surge or sway motion~\citep{li2022onset}. Within the same frequency range, the present PINN model reconstructs the dominant large scale meandering induced by coupled surge-pitch motion and also captures the principal wake structure in all cases.
\begin{table*}
\centering
\caption{{\color{black}Measured training time and streamwise velocity prediction accuracy for the FNO and PINN models.}}
\label{tab:fair-cost}
\small
\begin{tabular}{l c c c}
\hline\hline
Model & {\color{black}Model selection + final training (min/case)} & Prediction MAE & MAE at the last time step \\
\hline
FNO & 7.36--10.39 & 0.02219 & 0.02784 \\
PINN & 74.60--123.53 & 0.05288 & 0.07613 \\
\hline\hline
\end{tabular}
\end{table*}

As shown in Fig.~\ref{fig:pinns-300}, although the PINN model reproduces the dominant wake motion, it does not resolve all higher-frequency, smaller-scale structures. The largest local discrepancies occur in regions of intense wake evolution, {\color{black}whereas Table~\ref{tab:case-errors} separately quantifies errors over the prediction interval}. The near wake is reproduced more closely than the far wake, where smoothing is more pronounced. The errors do not increase monotonically with $St_{\rm p}$, and therefore do not indicate systematic degradation with increasing motion frequency. 
The FNO reconstructions in Fig.~\ref{fig:fnos-300} agree more closely with the LES reference fields across all cases. {\color{black}The contour maps are used to identify the spatial distribution of the remaining discrepancies, which are concentrated mainly along the boundaries of evolving wake structures. {\color{black}Separately, Table~\ref{tab:case-errors} quantifies prediction-period errors independently of the contour scale and reports} the per-case MAE, RMSE, relative-$L_2$, and maximum absolute error. Over the prediction interval, the six-case mean MAE is 0.02219 for the FNO model and 0.05288 for the PINN model, while the respective per-case RMSE ranges are 0.02467--0.04716 and 0.06529--0.09515.} Relative to the implemented PINNs, the FNOs better preserve the principal coherent structures, sharp velocity gradients, and large-scale wake meandering.

\begin{figure}
    \centering
    \includegraphics[width=0.85\linewidth]{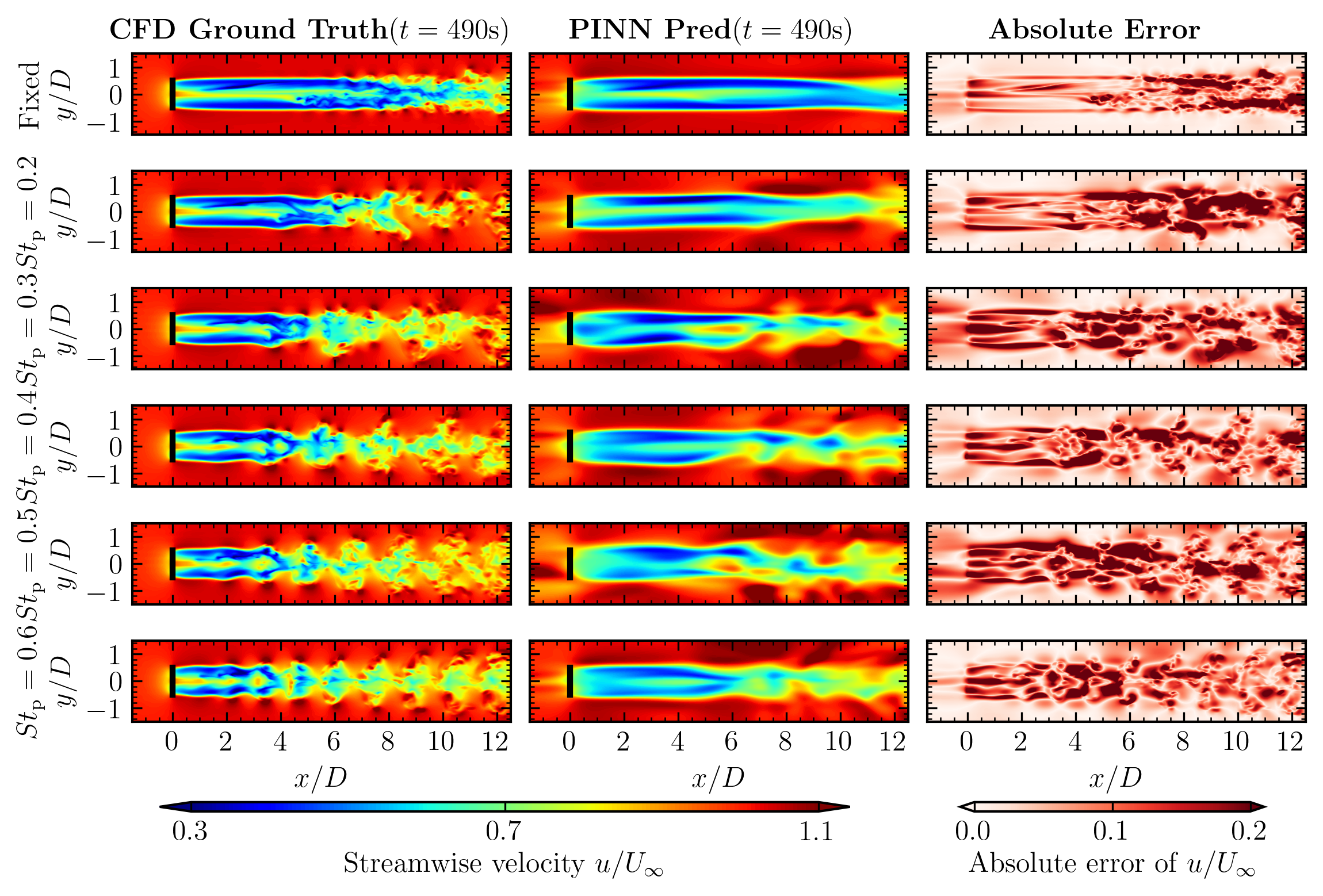}
    \caption{Instantaneous Streamwise velocity contours from the PINN and CFD reference at $t=490$~s.}
    \label{fig:pinns-490}
\end{figure}

\begin{figure}
    \centering
    \includegraphics[width=0.85\linewidth]{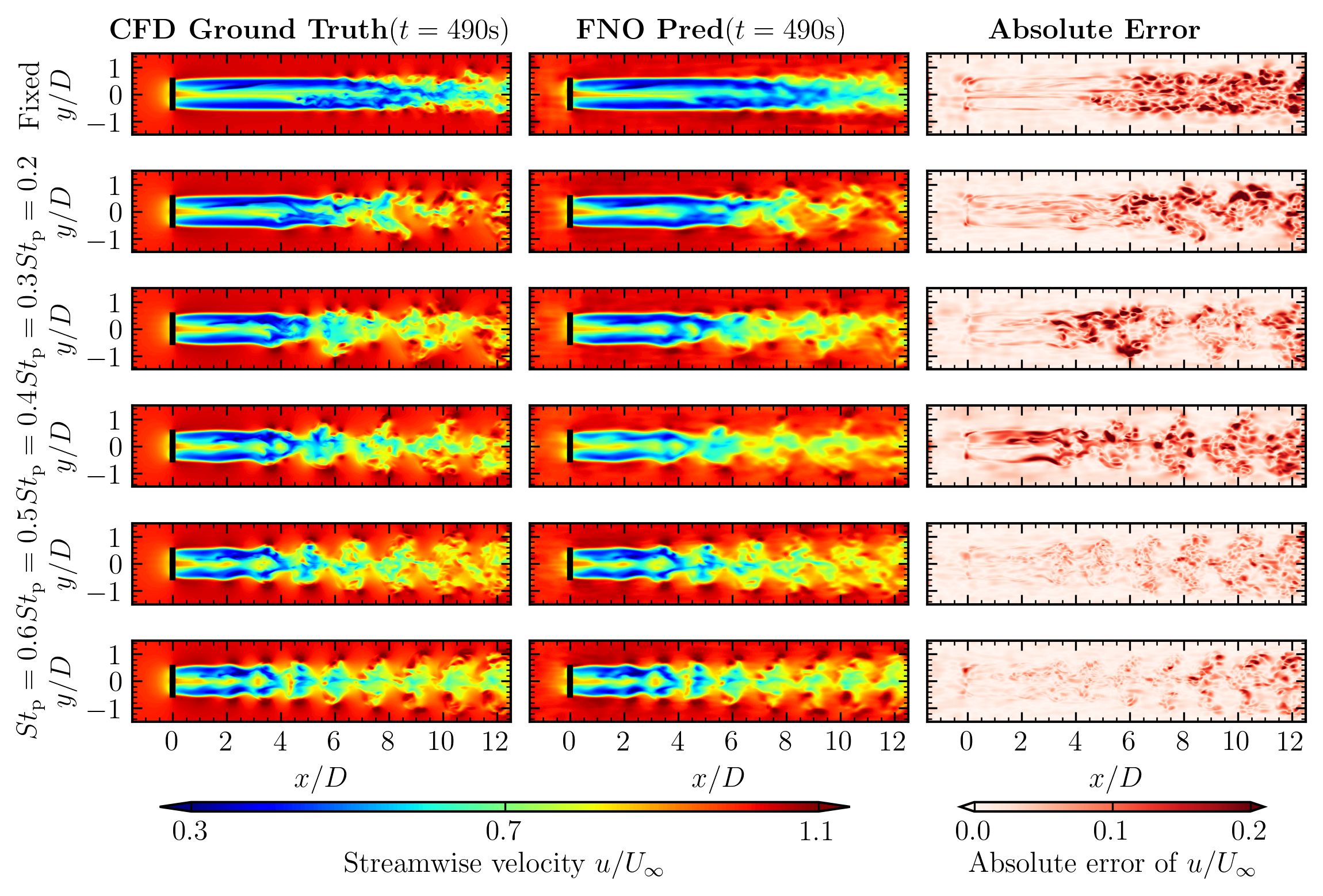}
    \caption{Instantaneous Streamwise velocity contours from the FNO and CFD reference at $t=490$~s.}
    \label{fig:fnos-490}
\end{figure}

The prediction capabilities of the PINN and FNO models
are then evaluated on temporally unseen fields from the same deterministic LES trajectories. No field in this interval is used for model fitting, checkpoint selection, or training-duration selection. Figures~\ref{fig:pinns-490} and \ref{fig:fnos-490} compare the predictions of the trained PINN and FNO models at $t=490$~s. The FNO model preserves the dominant unsteady wake trajectory, such as the large-scale wake meandering, and more of the small-scale coherent structure, whereas the PINN field is visibly smoother. However, the results of PINNs are much smoother and tend to deviate more from the LES reference.

The above discussion shows that during reconstruction, the PINN model captures the dominant wake meandering. During prediction, however, its wake meandering pattern becomes less distinct, and its deviation from the CFD reference increases. By contrast, the FNO model maintains lower errors and a clearer representation of the unsteady wake in both intervals, indicating greater robustness over the prediction interval and supporting its potential for rapid prediction of FOWT wakes.
%
% {\color{black}Shared-model, ensemble, sensitivity, persistence, and leave-one-frequency-out results are reported in Appendices~\ref{app:model-checks} and \ref{app:transfer}; these supplementary analyses are excluded from the primary FNO--PINN comparison.}

\subsection{{\color{black}Time-averaged wake characteristics}}
\begin{figure}
    \centering
    \includegraphics[width=0.8\textwidth]{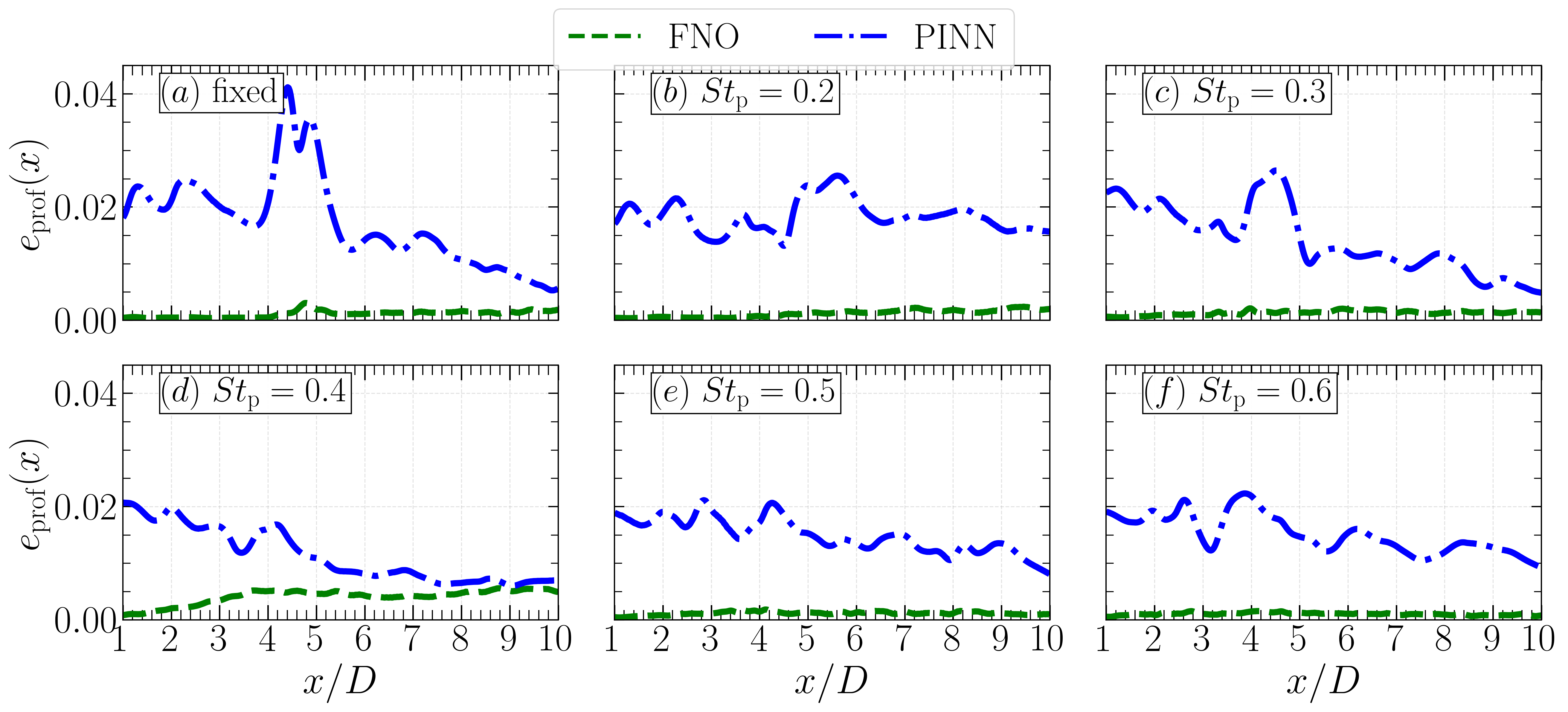}
    \caption{{\color{black}RMSE of the time-averaged transverse velocity deficit along different downstream positions.}}
    \label{fig:mean-profile-error-x}
\end{figure}

{\color{black}Time averages use the common reconstruction interval $t=254,\ldots,449$~s ($N_t=196$).}
{\color{black}Time-averaged wake characteristics describe the persistent velocity deficit and the lateral region that would affect the inflow to a downstream turbine. The mean velocity deficit profiles, defined by $\overline{\Delta u} = U_\infty - \frac{1}{N_t}\sum_{n=1}^{N_t}u(t_n)$, are examined here to assess how accurately the models reproduce the persistent downstream velocity deficit distribution.}
\begin{figure}
    \centering
    \includegraphics[width=0.7\textwidth]{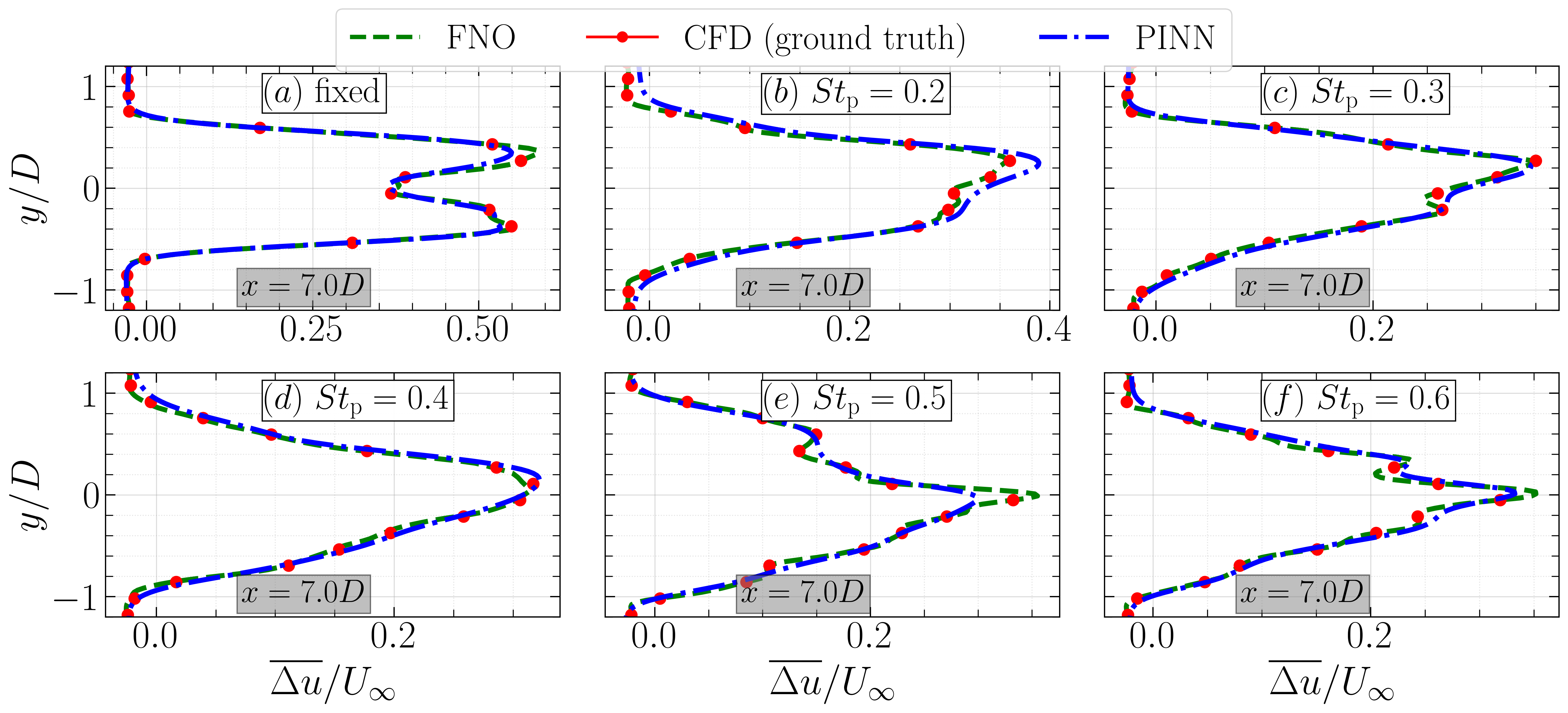}
    \caption{{\color{black}Time-averaged streamwise velocity deficit profiles in the transverse direction at $x/D=7.0$.}}
    \label{fig:mean-profile-x7}
\end{figure}
{\color{black}Figure~\ref{fig:mean-profile-x7} shows the time-averaged transverse wake profiles at $x/D=7$. Both models reproduce the principal mean velocity deficit, showing that the PINN model also provides a useful representation of the mean wake. The FNO model, however, follows the detailed LES profile more closely under all cases, whereas the PINN model produces a broader or smoother profile in several $St_{\rm p}$ cases.

The complete transverse-profile error is therefore evaluated at each streamwise station without filtering or Gaussian fitting. For model $m$ and condition $c$, the local profile root-mean-square error (RMSE) is defined as:
\begin{align}
e_{\rm prof}^{(m,c)}(x_i)
=\left[
\frac{1}{N_y}\sum_{j=1}^{N_y}
\left(
\frac{\overline{\Delta u}^{(m,c)}(x_i,y_j)
-\overline{\Delta u}^{(\mathrm{LES},c)}(x_i,y_j)}
{U_\infty}
\right)^2
\right]^{1/2},
\label{eq:mean-profile-rmse}
\end{align}
where the $N_y$ points satisfy $|y/D|\leq1.2$.
To provide one overall measure of the profile discrepancy for each condition, the local errors over $1\leq x/D\leq10$ are summarized by $E_{\rm prof}^{(m,c)}
=\left[
\frac{1}{N_x}\sum_{i=1}^{N_x}
\left(e_{\rm prof}^{(m,c)}(x_i)\right)^2
\right]^{1/2}$}.
{\color{black}Figure~\ref{fig:mean-profile-error-x} reports the downstream evolution of $e_{\rm prof}$, and Table~\ref{tab:mean-profile-rmse} reports the corresponding aggregate value $E_{\rm prof}$. This figure and table show that the FNO model retains a low profile error from the near wake to the far wake and has a lower aggregate error under all cases considered in the present work. The all-case mean $E_{\rm prof}$ is 0.001746 for the FNO model and 0.016012 for the PINN model, corresponding to an 89.1\% reduction. The largest FNO error occurs at $St_{\rm p}=0.4$ but remains below the PINN error for that condition. The PINN model nevertheless reproduces the broad mean-deficit distribution. Its larger error mainly reflects the smoother or broader profiles and local displacement differences visible in Fig.~\ref{fig:mean-profile-x7}. Thus, the RMSE analysis shows that the FNO model more accurately preserves the mean velocity deficit distribution.}

\begin{table}
\centering
\caption{{\color{black}Aggregate profile RMSE
$E_{\rm prof}$ for the time-averaged transverse velocity deficit.}}
\label{tab:mean-profile-rmse}
\small
\begin{tabular}{c c c}
\hline\hline
Condition & FNO & PINN \\
\hline
Fixed & 0.001259 & 0.019164 \\
$St_{\rm p}=0.2$ & 0.001348 & 0.018565 \\
$St_{\rm p}=0.3$ & 0.001370 & 0.015391 \\
$St_{\rm p}=0.4$ & 0.004274 & 0.012248 \\
$St_{\rm p}=0.5$ & 0.001180 & 0.015186 \\
$St_{\rm p}=0.6$ & 0.001046 & 0.015520 \\
\hline
Six-case mean & 0.001746 & 0.016012 \\
\hline\hline
\end{tabular}
\end{table}

\begin{figure}
    \centering
    \includegraphics[width=0.7\linewidth]{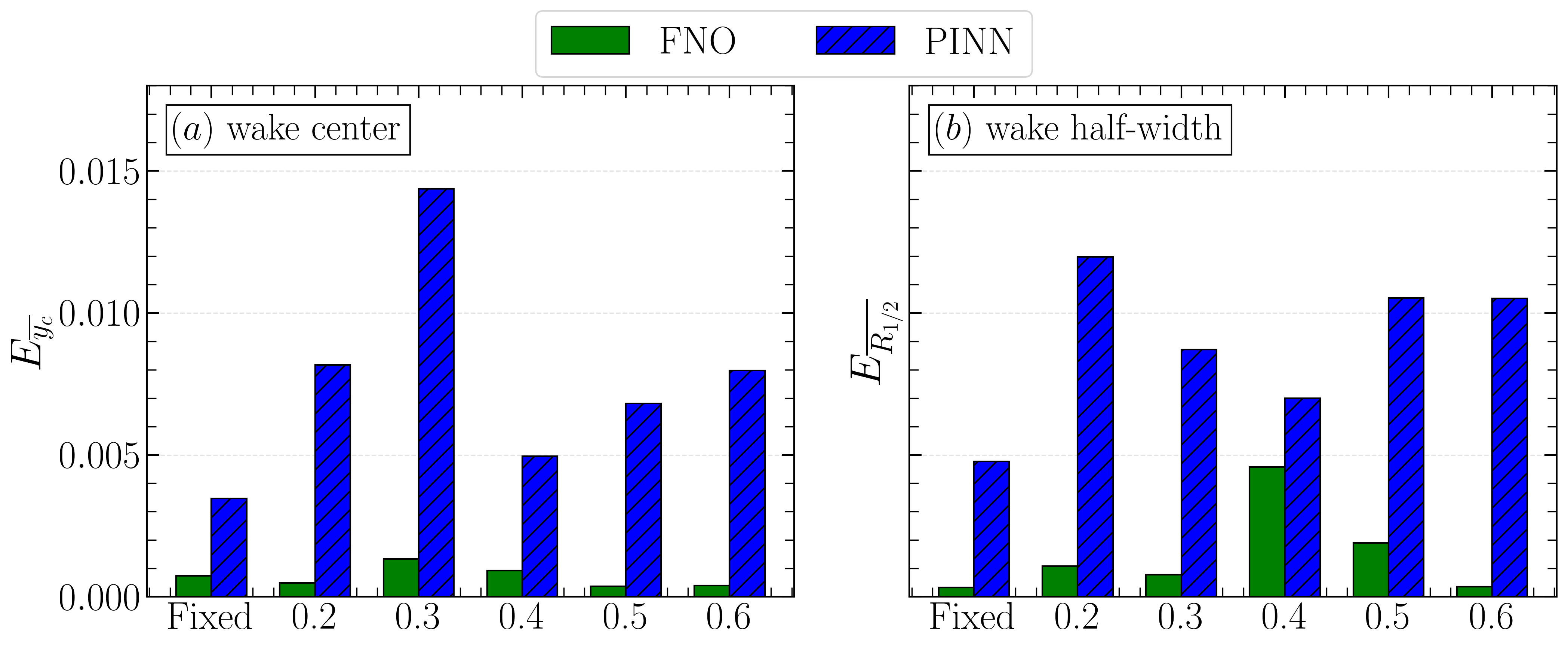}
    \caption{{\color{black}RMSE of mean-wake quantities relative to LES results: (a) wake center and (b) wake half-width.}}
    \label{fig:mean-metric-errors}
\end{figure}

{\color{black}The profile RMSE combines discrepancies in deficit magnitude, lateral displacement, and wake expansion. To identify the physical origin of these discrepancies, the mean wake center and half-width are extracted: the center measures the mean lateral wake displacement, whereas the half-width measures the lateral extent of the velocity deficit region. The filtered profile is fitted using a Gaussian function with a local offset as $\frac{\overline{\Delta u}(x,y)}{U_\infty}
=A(x){\rm e}^{-\frac{1}{2}
\left(\frac{y-\overline{y_c}(x)}{\sigma(x)}\right)^2}+b(x).$
Here, $A(x)$ and $b(x)$ are the fitted deficit amplitude and local offset, $\overline{y_c}(x)$ is the mean wake center, and $\sigma(x)$ is the Gaussian width parameter. The mean wake half-width is $\overline{R_{1/2}}=\sqrt{2\ln2}\,\sigma${\color{black}, measured at half the Gaussian amplitude above the fitted background $b(x)$, i.e., at $\overline{\Delta u}(x,y)/U_\infty=b(x)+A(x)/2$}. }
\begin{center}
    \includegraphics[width=0.85\textwidth]{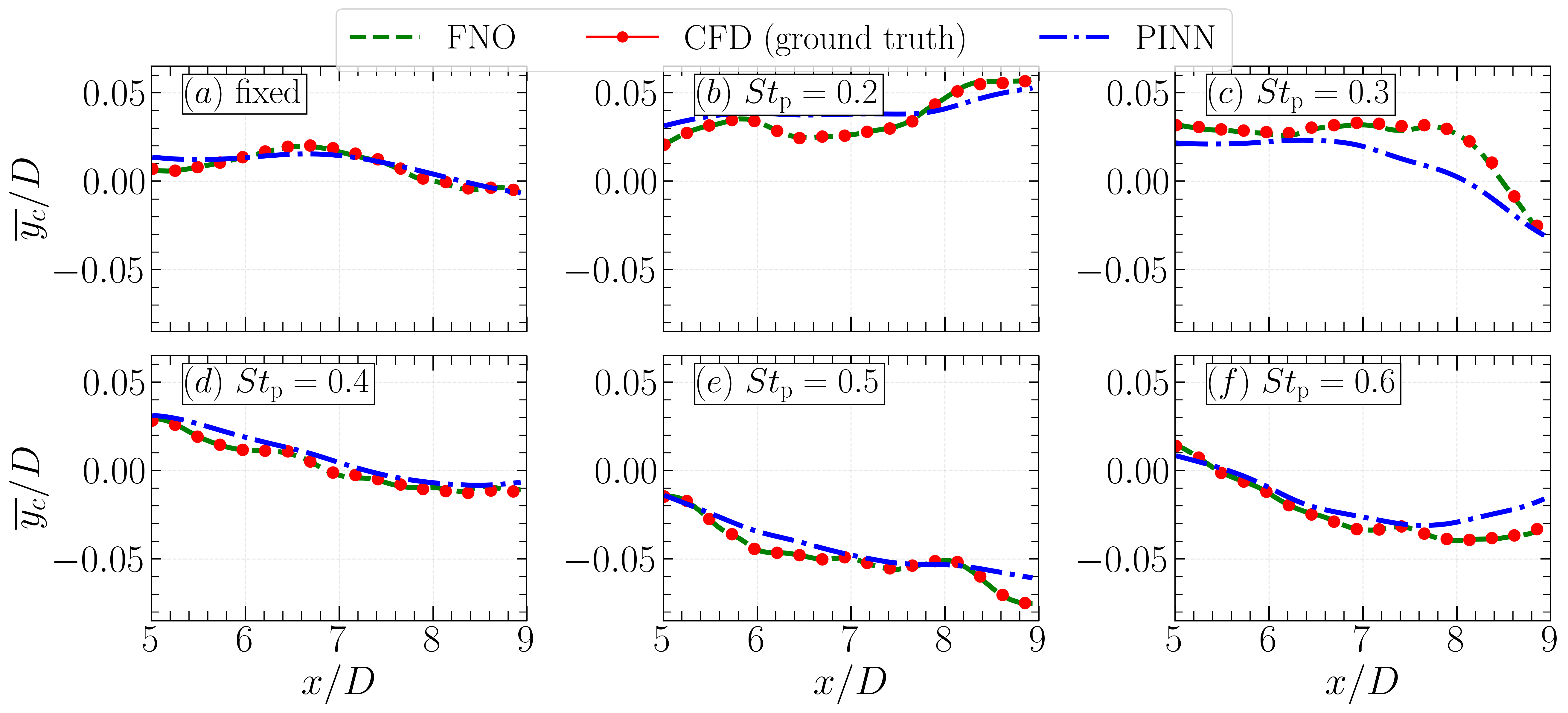}
    \captionof{figure}{{\color{black}Mean wake-center position along different downstream positions.}}
    \label{fig:mean-yc}
\end{center}
\begin{center}
    \includegraphics[width=0.85\textwidth]{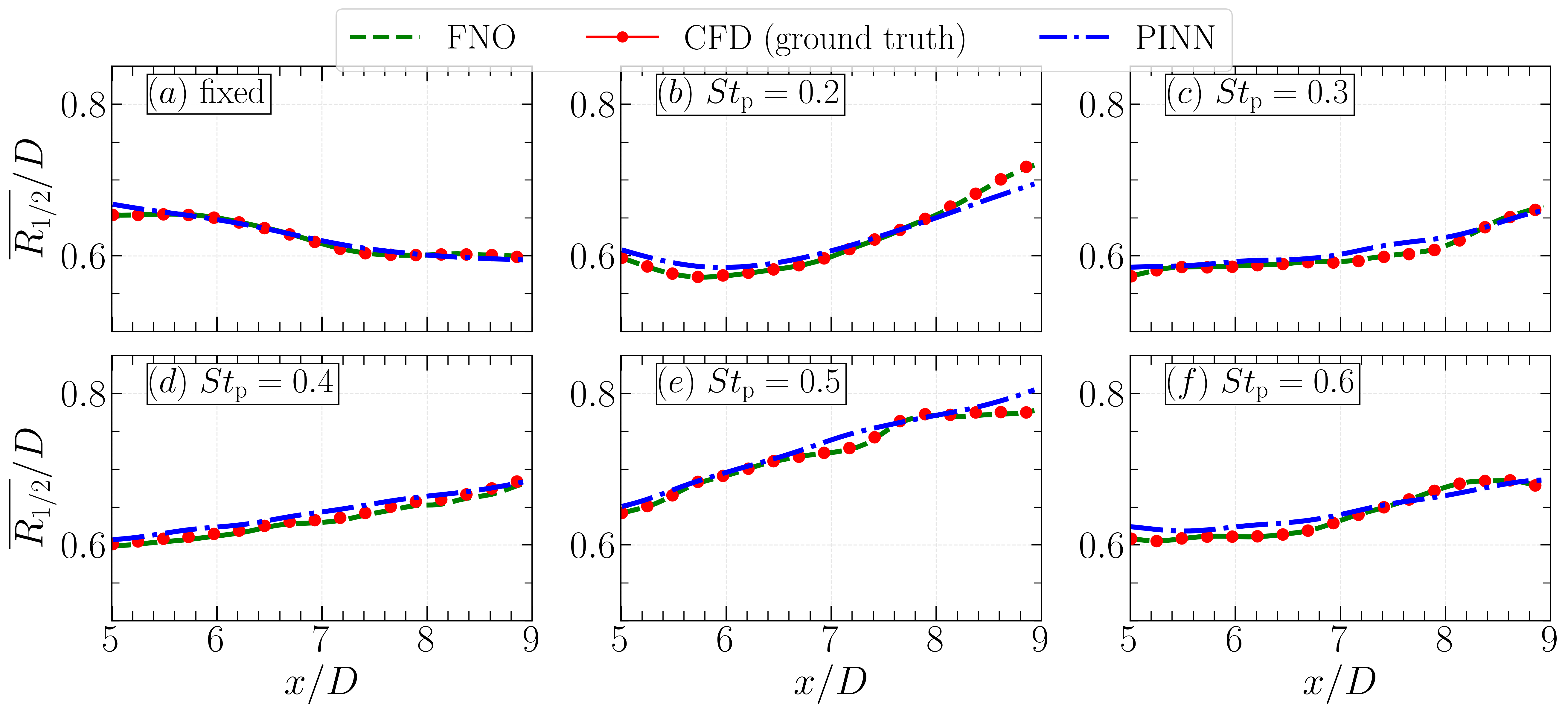}
    \captionof{figure}{{\color{black}Mean wake half-width along different downstream positions.}}
    \label{fig:mean-width}
\end{center}
{\color{black}Figures~\ref{fig:mean-yc} and \ref{fig:mean-width} show that both models reproduce the principal downstream trends in mean wake displacement and expansion. The PINN model captures the direction of the mean lateral displacement and the overall half-width evolution in most cases, whereas the FNO model follows the LES results more closely, particularly for the floating cases. For $q\in\{\overline{y_c}/D,\overline{R_{1/2}}/D\}$, the discrepancy over the fitted Gaussian function is quantified as $E_q^{(m,c)}
 =\left[
 \frac{1}{50}\sum_{i=1}^{50}
 \left(q^{(m,c)}(x_i)-q^{(\mathrm{LES},c)}(x_i)\right)^2
 \right]^{1/2}.$
Figure~\ref{fig:mean-metric-errors} shows the corresponding errors. The all-case mean wake-center errors are 0.000708 for the FNO and 0.007623 for the PINN model, while the corresponding half-width errors are 0.001504 and 0.008917. Relative to the PINN model, the corresponding FNO reductions are 90.7\% and 83.1\%. These results show that both models recover the dominant mean-wake profiles and evolution, while the FNO model more accurately retains the cumulative imprint of motion-induced meandering on the mean wake position and lateral wake expansion.}

\subsection{Instantaneous wake characteristics and prediction errors}
\begin{figure}
    \centering
    \includegraphics[width=0.62\linewidth]{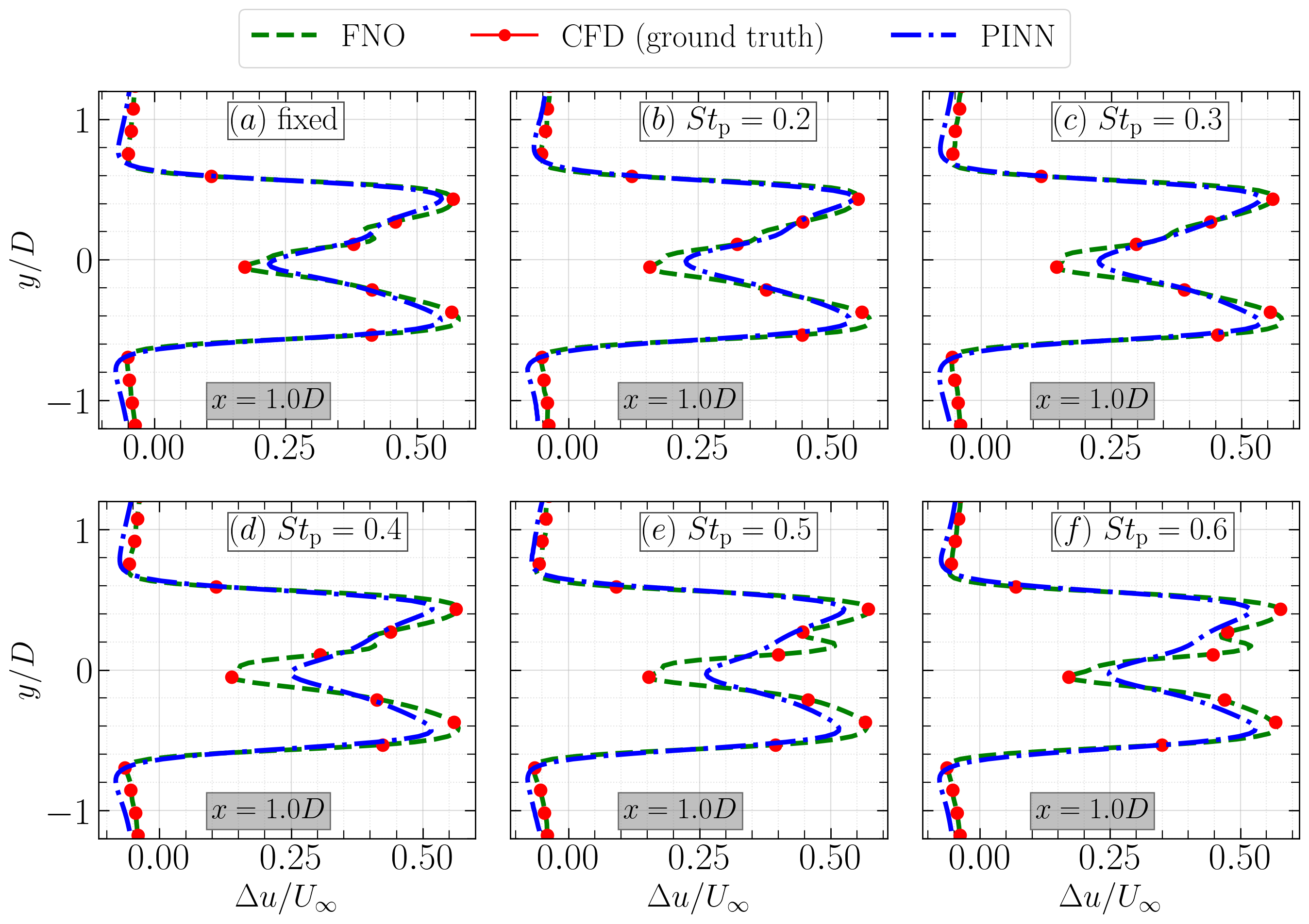}
    \caption{Instantaneous transverse profiles of the normalized streamwise velocity deficit at $x/D=1$ and $t=450$~s.}
    \label{fig:u_1d_profiles}
\end{figure}

\begin{figure}
    \centering
    \includegraphics[width=0.62\linewidth]{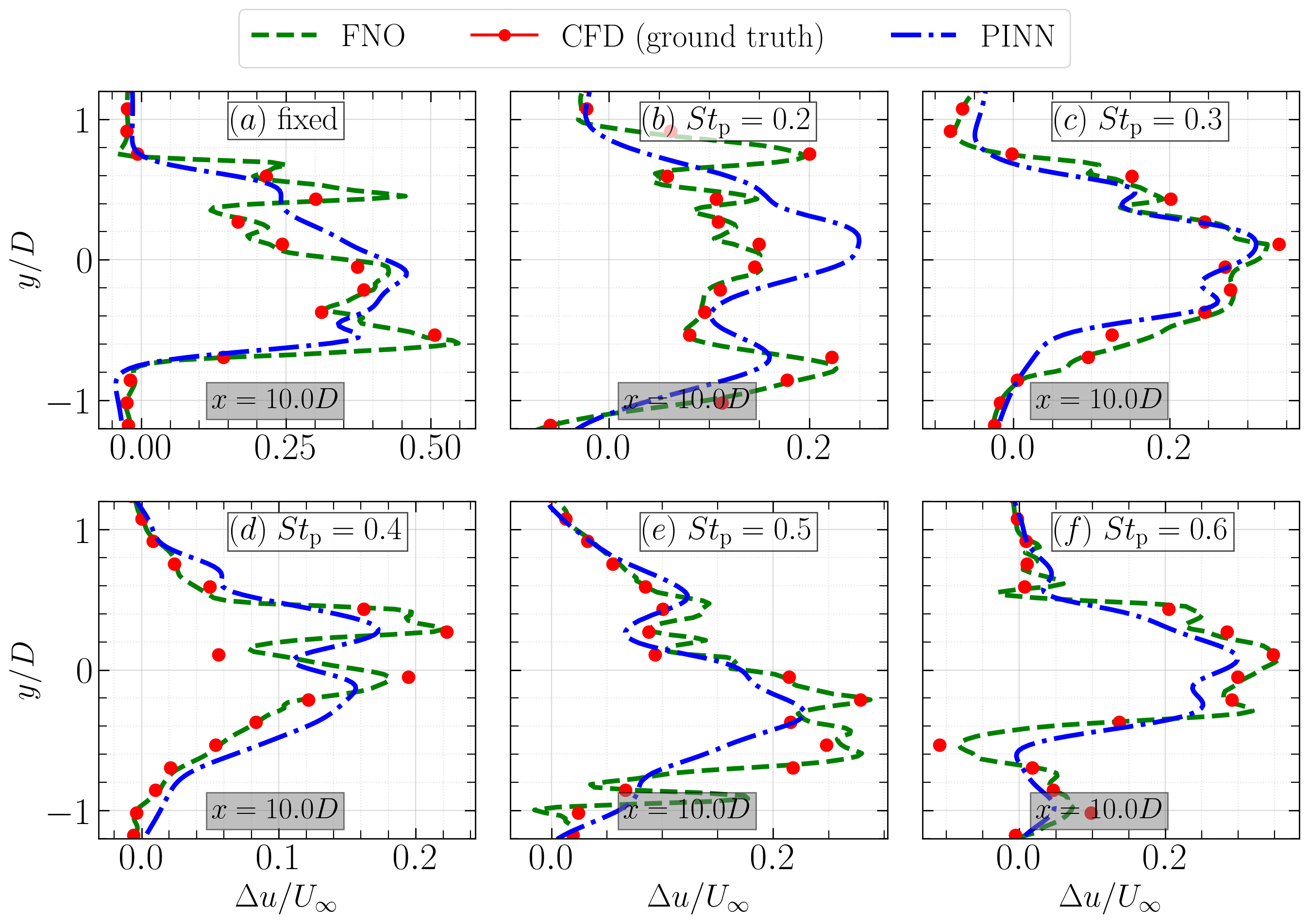}
    \caption{Instantaneous transverse profiles of the normalized streamwise velocity deficit at $x/D=10$ and $t=450$~s.}
    \label{fig:u_10d_profiles}
\end{figure}

The time-averaged results demonstrate that both models recover the persistent wake deficit, with the FNO model providing higher profile accuracy. To examine whether the models also retain the multi-scale dynamics that govern the instantaneous inflow experienced by a downstream turbine, the instantaneous wake profiles, centerline signals, and wake-geometry fluctuations are analyzed below.
Figures~\ref{fig:u_1d_profiles} and \ref{fig:u_10d_profiles} compare the instantaneous normalized streamwise velocity deficit profiles from the PINN, FNO, and CFD reference at $t=450$~s. The stations $x/D=1$ and 10 represent the near- and far-wake regions, respectively. The deficit is defined as $\Delta u = U_{\infty} - u$, where $u$ denotes the instantaneous streamwise velocity.

\begin{figure}
    \centering
    \includegraphics[width=0.8\textwidth]{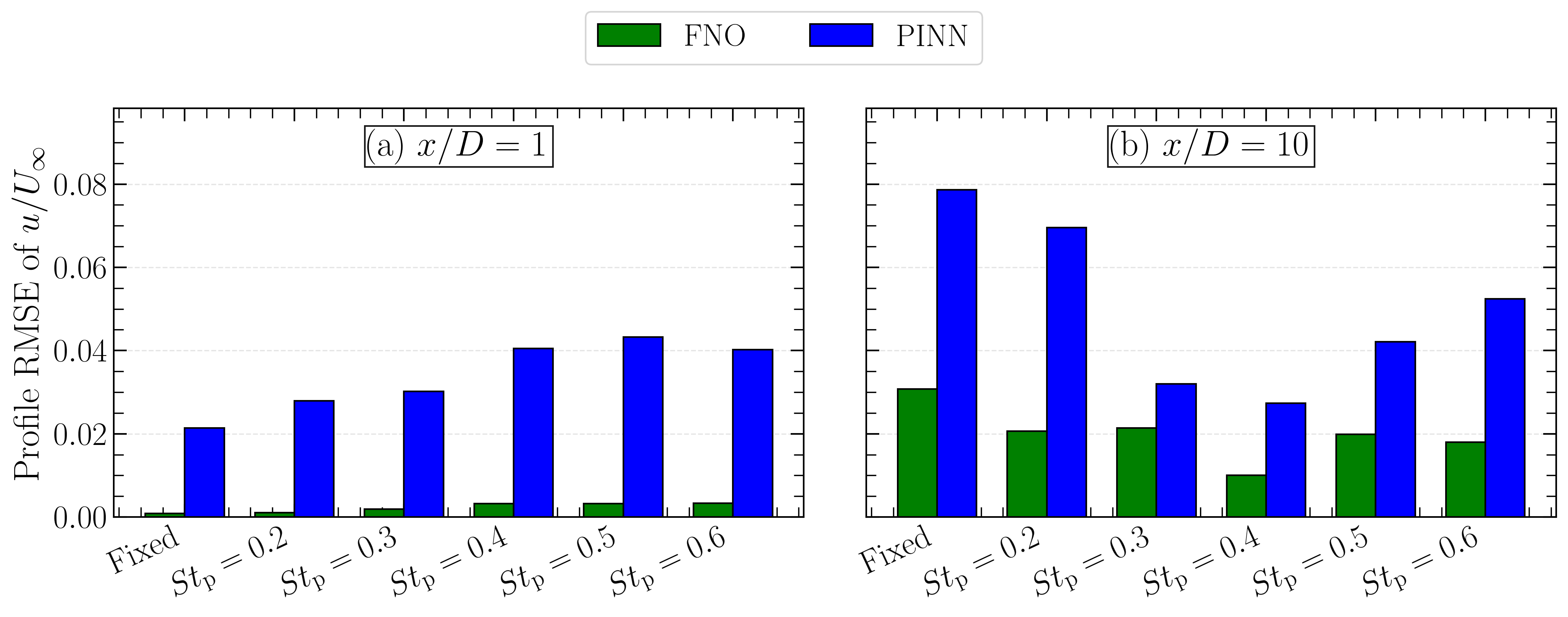}
    \caption{{\color{black}RMSE of the $u/U_\infty$ profiles at $t=450$~s
    for (a) $x/D=1$ and (b) $x/D=10$ for the FNO and PINN.}}
    \label{fig:app-profile-errors}
\end{figure}

At $x/D=1$, both models recover the characteristic double-peak deficit associated with the outer shear layers and blade-tip vortices~\citep{dong2023characteristics}, particularly for the fixed, $St_{\rm p}=0.2$, and $St_{\rm p}=0.3$ cases. The PINN model agrees most closely with LES in the fixed case but overestimates the centerline deficit and smooths the outer-wake variations in some moving cases, including $St_{\rm p}=0.6$. The FNO model follows the LES profiles more closely across the six cases. Because the multi-scale wake dynamics are less developed at this near-wake station, the model differences are smaller than those observed farther downstream.
At $x/D=10$, the difference between the two models is more pronounced because large-scale meandering coexists with smaller-scale turbulent variations. The FNO model tracks the LES deficit magnitude, spatial oscillations, and intermittent variations more closely. The PINN  model captures the overall wake displacement but attenuates the smaller-scale variations, producing smoother profiles. {\color{black}For the present implementation, this behavior indicates scale-selective attenuation of the resolved far-wake fluctuations.} The FNO model therefore reproduces the multi-scale far-wake profiles more accurately.
{\color{black}Figure~\ref{fig:app-profile-errors} quantifies the instantaneous
transverse profile errors at 450~s. At $x/D=1$, the RMSE ranges from 0.00085
to 0.00335 for the FNO model and from 0.02140 to 0.04323 for the PINN model. At $x/D=10$,
the corresponding ranges are 0.01003--0.03075 for the FNO model and
0.02738--0.07860 for the PINN model. The profile-based ranking therefore holds
 across all cases rather than only for the fixed case. The larger model separation at $x/D=10$ is consistent with the coexistence of developed wake meandering and smaller-scale turbulent variations in the far wake.}

\begin{figure}
    \centering
    \includegraphics[width=0.65\linewidth]{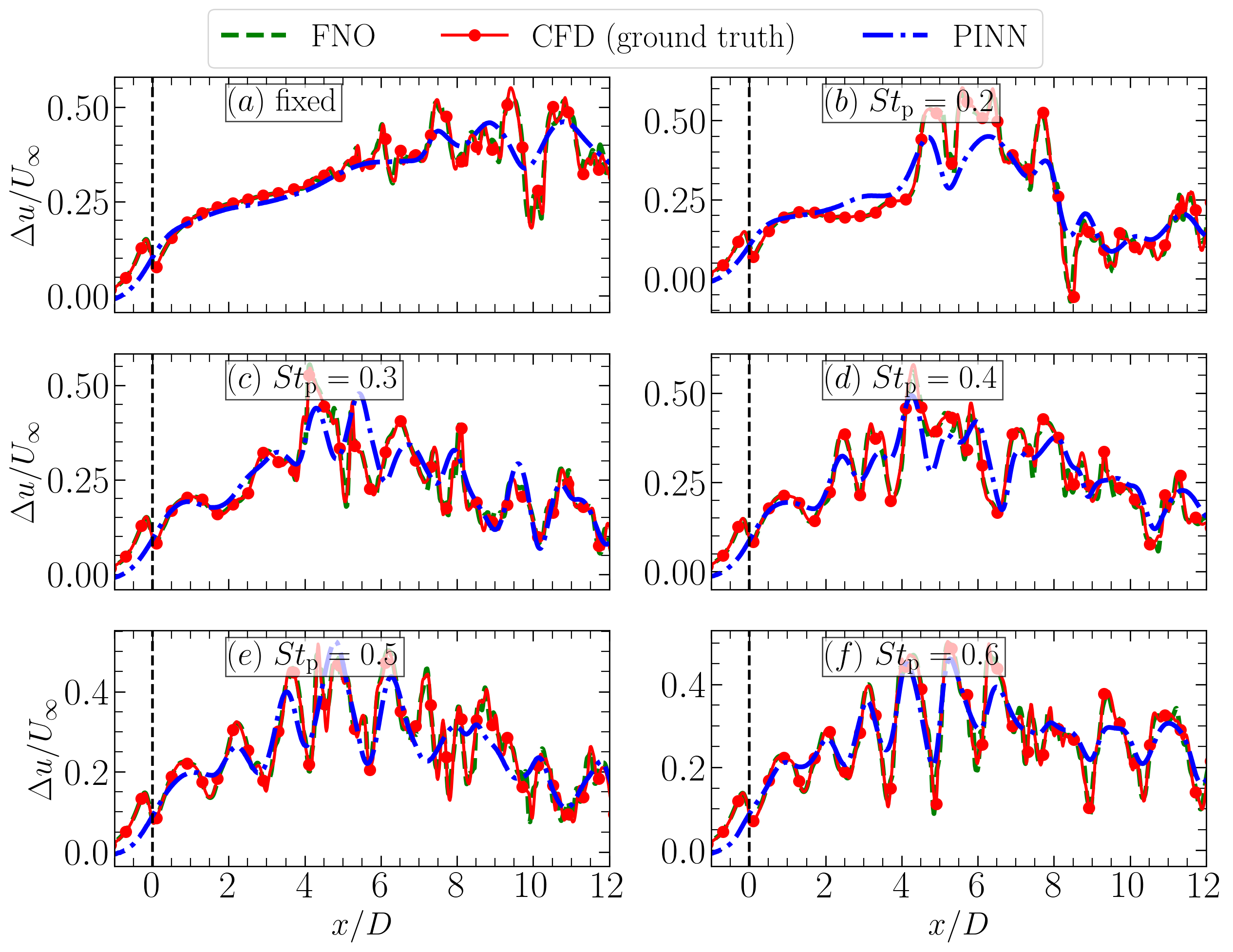}
    \caption{Streamwise evolution of the instantaneous velocity deficit at $t=350$~s along the rotor centerline, $y/D=0$.}
    \label{fig:uc_vs_x}
\end{figure}

\begin{figure}
    \centering
    \includegraphics[width=0.7\linewidth]{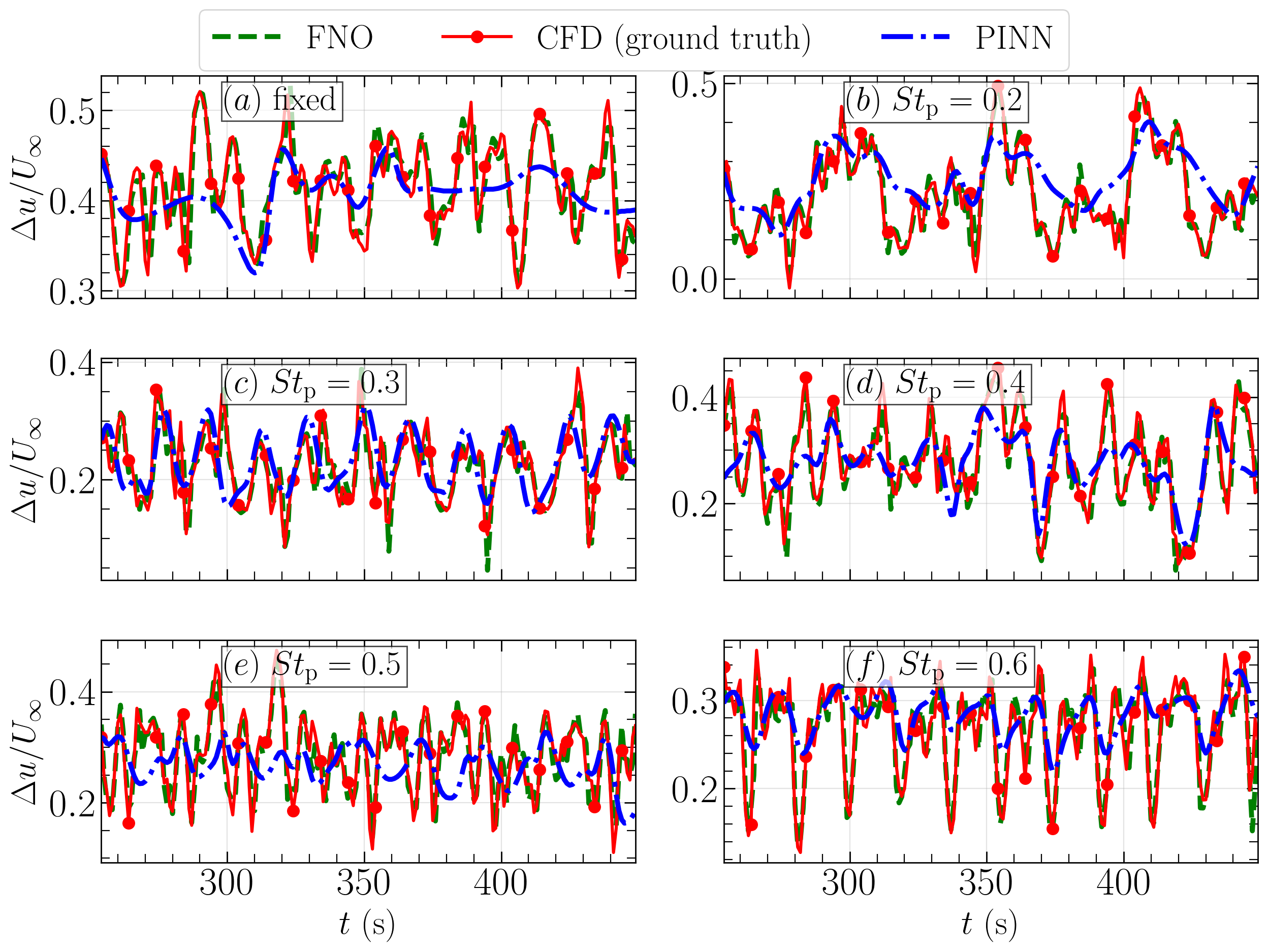}
    \caption{Temporal evolution of the reconstructed streamwise velocity deficit at $(x/D,y/D)=(8,0)$.}
    \label{fig:uc_vs_t}
\end{figure}

Figure~\ref{fig:uc_vs_x} compares the instantaneous normalized streamwise velocity deficit along the rotor centerline, $y/D=0$, at $t=350$~s. As the wake meanders downstream, this centerline cut intersects regions of different deficit intensity and therefore reflects the instantaneous streamwise organization of the wake. Across the six cases, the FNO model follows the LES deficit magnitude and streamwise oscillations closely. The PINN model captures the dominant spatial variation but attenuates the fluctuation amplitude, particularly in regions of strong wake variability.
{\color{black}Figure~\ref{fig:uc_vs_t} compares the temporal evolution of $\Delta u/U_{\infty}$ at $(x/D,y/D)=(8,0)$. For this reconstruction profile, the CFD reference and both models are evaluated over the same time interval. The rapid and intermittent variations reflect the passage of the meandering wake and smaller-scale structures across the probe position. The FNO model reproduces these variations in the LES results, whereas the PINN model captures the dominant phase and temporal trend but underestimates the fluctuation amplitude. Thus, the smoothing observed in the PINN fields also reduces temporal turbulent fluctuations of the far wake.}

To quantify the large-scale unsteady wake dynamics, the wake center, $y_c$, and wake half-width, $R_{1/2}$, are obtained from Gaussian fits to the instantaneous transverse velocity deficit profiles over the common reconstruction interval~\citep{dong2023characteristics}. Before fitting, a transverse box filter of width $1D$ is applied to reduce small-scale fluctuations. The filtered profiles are fitted using:
\begin{align}
\Delta u(x,y) = \Delta u_c(x) \exp \left[ -\frac{1}{2} \left( \frac{y - y_c(x)}{\sigma(x)} \right)^2 \right],
\label{eq:Gaussian}
\end{align}
{\color{black}where $\Delta u_c(x)$ is the fitted center deficit, $y_c(x)$ is the transverse wake-center position, and $\sigma(x)$ is the Gaussian width parameter. {\color{black}Unlike the normalized mean fit, this instantaneous fit uses the dimensional deficit with a zero offset.} The wake half-width is the distance from $y_c$ at which $\Delta u=\Delta u_c/2$, giving $R_{1/2}=\sqrt{2\ln2}\,\sigma(x)$. {\color{black}These fitted quantities describe the large-scale wake structures and complement the full-field errors and complete-profile RMSEs evaluated directly from the velocity fields.}}

\begin{figure}
    \centering
    \includegraphics[width=0.9\linewidth]{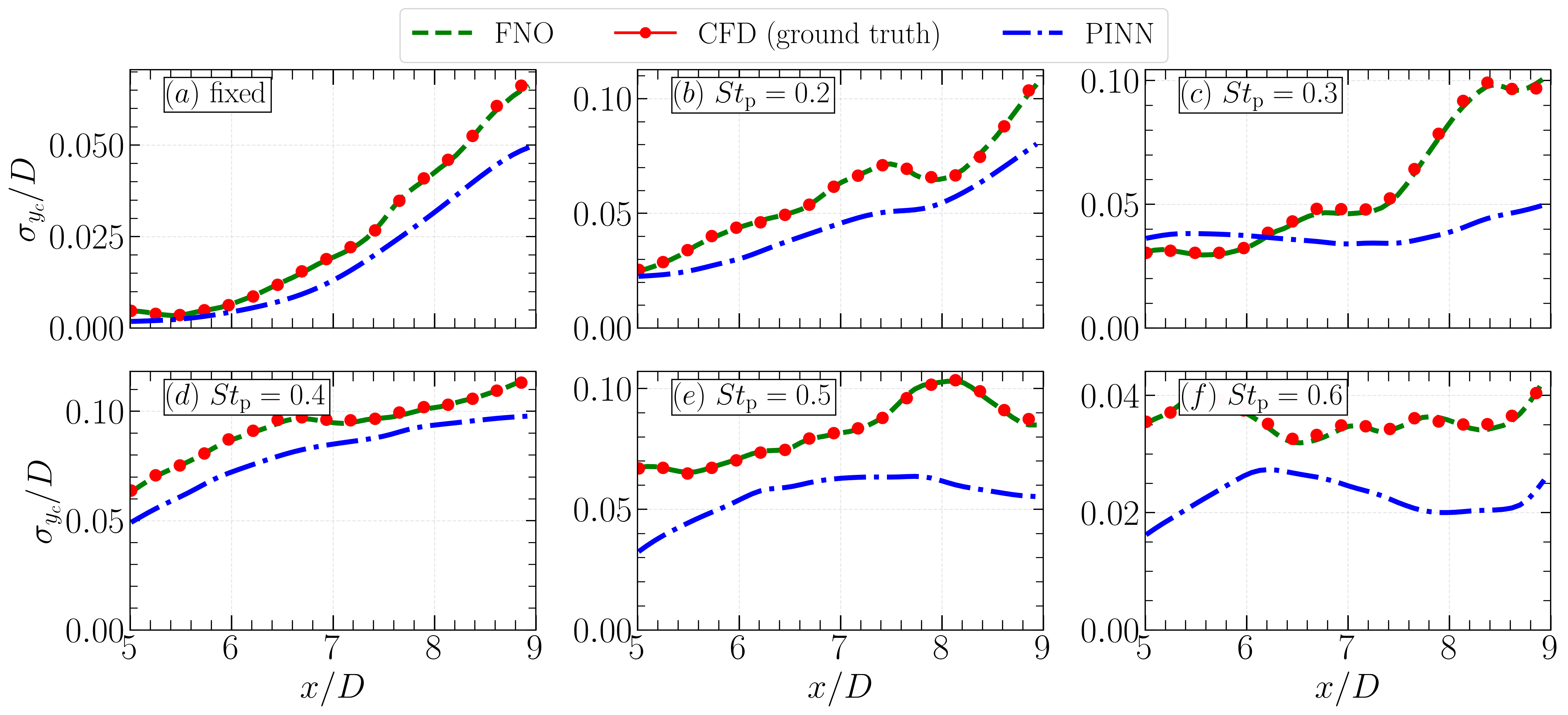}
    \caption{Downstream evolution of the temporal standard deviation of wake-center position, $\sigma_{y_c}/D$, obtained from the CFD reference, FNO, and PINN.}
    \label{fig:std_wake_center}
\end{figure}

Figure~\ref{fig:std_wake_center} shows the downstream evolution of the temporal standard deviation of the fitted wake center, $\sigma_{y_c}/D$. This quantity measures the amplitude of large-scale wake meandering, and hence the lateral variability of the wake encountered downstream. The FNO model follows the LES results across the wake, whereas the PINN model generally underestimates the wake-center fluctuations, particularly in the near wake.
\begin{figure}
    \centering
    \includegraphics[width=0.9\linewidth]{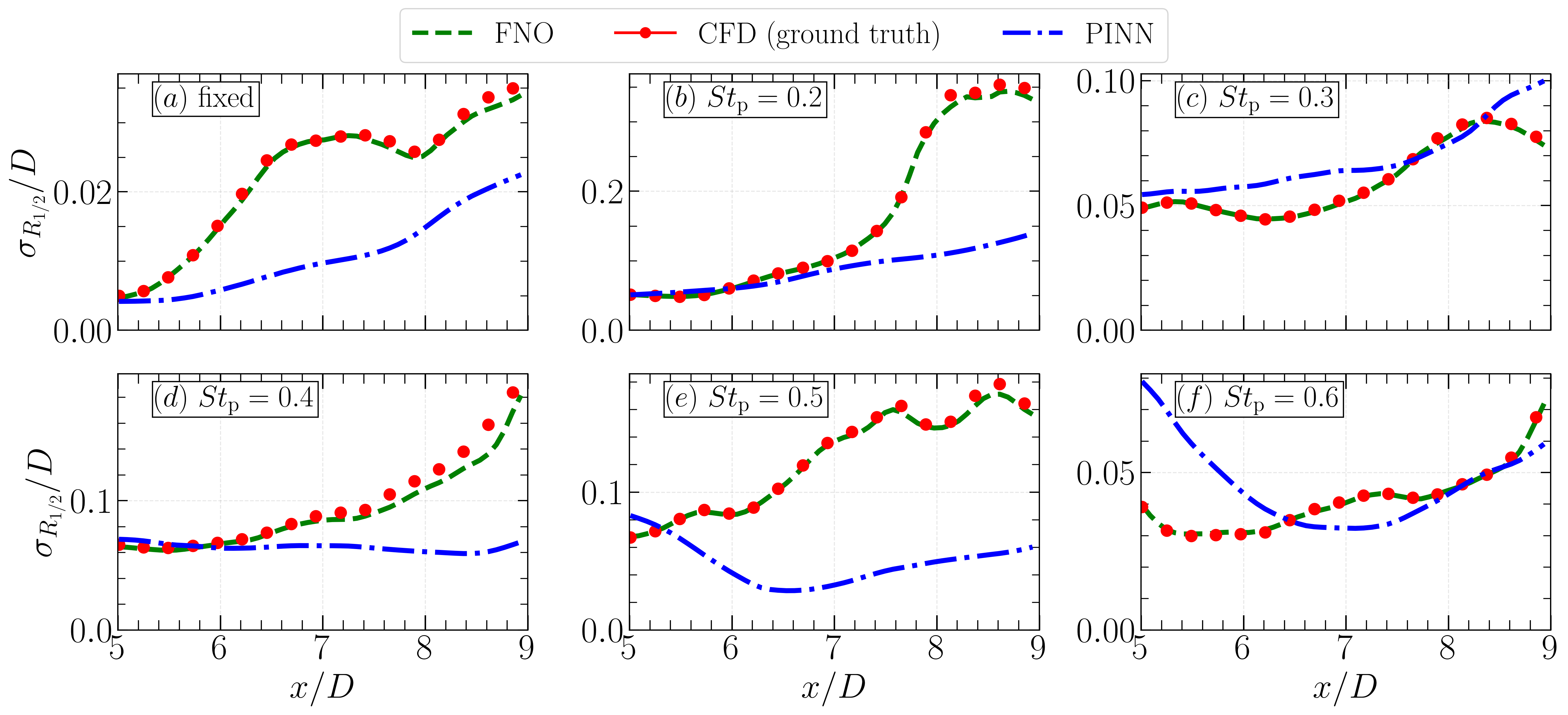}
    \caption{Downstream evolution of the temporal standard deviation of wake half-width, $\sigma_{R_{1/2}}/D$, obtained from the CFD reference, FNO, and PINN.}
    \label{fig:std_wake_width}
\end{figure}
Figure~\ref{fig:std_wake_width} presents the corresponding temporal standard deviation of the fitted wake half-width, $\sigma_{R_{1/2}}/D$, which quantifies the temporal variation in wake expansion and lateral footprint. The FNO model closely follows the LES variation across the studied conditions, whereas the PINN model generally predicts smaller half-width fluctuations. Together, the wake-center and half-width statistics show that both models recover the dominant unsteady wake evolution, but the present PINN model smooths part of the resolved fluctuations and the FNO model retains them more faithfully.
{\color{black}The consequences of these instantaneous wake differences over the prediction interval are quantified using the instantaneous mean absolute error, $\mathrm{MAE}_t$, and the spatiotemporally averaged mean absolute error, $\mathrm{MAE}$:}
\begin{align}
\mathrm{MAE}_t = \frac{1}{N} \sum_{i=1}^{N} |u_{\mathrm{model}, i} - u_i|, \quad
\mathrm{MAE} = \frac{1}{M} \sum_{i=1}^{M} |u_{\mathrm{model}, i} - u_i|,
\label{eq:mae}
\end{align}
{\color{black}where $N=N_x \times N_y$ and
$M=N_x \times N_y \times N_t$ for the complete prediction interval.
Velocities are normalized by $U_{\infty}$, $u_{\mathrm{model},i}$ is the
model output, and $u_i$ is the CFD reference. The analysis also uses:}
\begin{align}
\mathrm{RMSE}&=\left[\frac{1}{M}\sum_{i=1}^{M}
(u_{\mathrm{model},i}-u_i)^2\right]^{1/2}, \quad
\epsilon_{L_2}=\frac{\|\bm u_{\mathrm{model}}-\bm u\|_2}
 {\|\bm u\|_2}.
\end{align}

{\color{black}Table~\ref{tab:case-errors} reports the MAE, RMSE, relative $L_2$ error, and maximum absolute error of the normalized streamwise velocity for every operating condition.}

\begin{center}
\begin{minipage}{0.98\linewidth}
\centering
\captionof{table}{{\color{black}Streamwise-velocity errors over the prediction interval for the FNO and PINN.}}
\label{tab:case-errors}
\small
\resizebox{0.96\textwidth}{!}{
\begin{tabular}{l c c c c c c c c}
\hline\hline
& \multicolumn{4}{c}{FNO} & \multicolumn{4}{c}{PINN} \\
Case & MAE & RMSE & Relative $L_2$ & Max. abs. &
MAE & RMSE & Relative $L_2$ & Max. abs. \\
\hline
Fixed & 0.02598 & 0.04716 & 0.05285 & 0.43092 & 0.03680 & 0.06529 & 0.07317 & 0.56332 \\
$St_{\rm p}=0.2$ & 0.02433 & 0.04437 & 0.04834 & 0.44223 & 0.05647 & 0.09385 & 0.10225 & 0.71550 \\
$St_{\rm p}=0.3$ & 0.02329 & 0.04049 & 0.04405 & 0.45749 & 0.05863 & 0.08911 & 0.09694 & 0.54302 \\
$St_{\rm p}=0.4$ & 0.03031 & 0.04660 & 0.05085 & 0.33615 & 0.05247 & 0.08112 & 0.08851 & 0.63458 \\
$St_{\rm p}=0.5$ & 0.01513 & 0.02474 & 0.02714 & 0.26712 & 0.06375 & 0.09515 & 0.10438 & 0.57783 \\
$St_{\rm p}=0.6$ & 0.01413 & 0.02467 & 0.02688 & 0.29245 & 0.04919 & 0.07520 & 0.08192 & 0.50926 \\
\hline\hline
\end{tabular}}
\end{minipage}
\end{center}

{\color{black}The FNO model has lower MAE, RMSE, relative $L_2$ error, and maximum absolute error than the PINN model under every condition. The PINN errors do not increase monotonically with $St_{\rm p}$, indicating that the model discrepancy cannot be attributed to forcing frequency alone. Maximum absolute error is retained as a secondary diagnostic because it is more sensitive to isolated grid values than the spatially averaged metrics.}

\subsection{Spectral characteristics}
\begin{figure}
    \centering
    \includegraphics[width=0.85\linewidth]{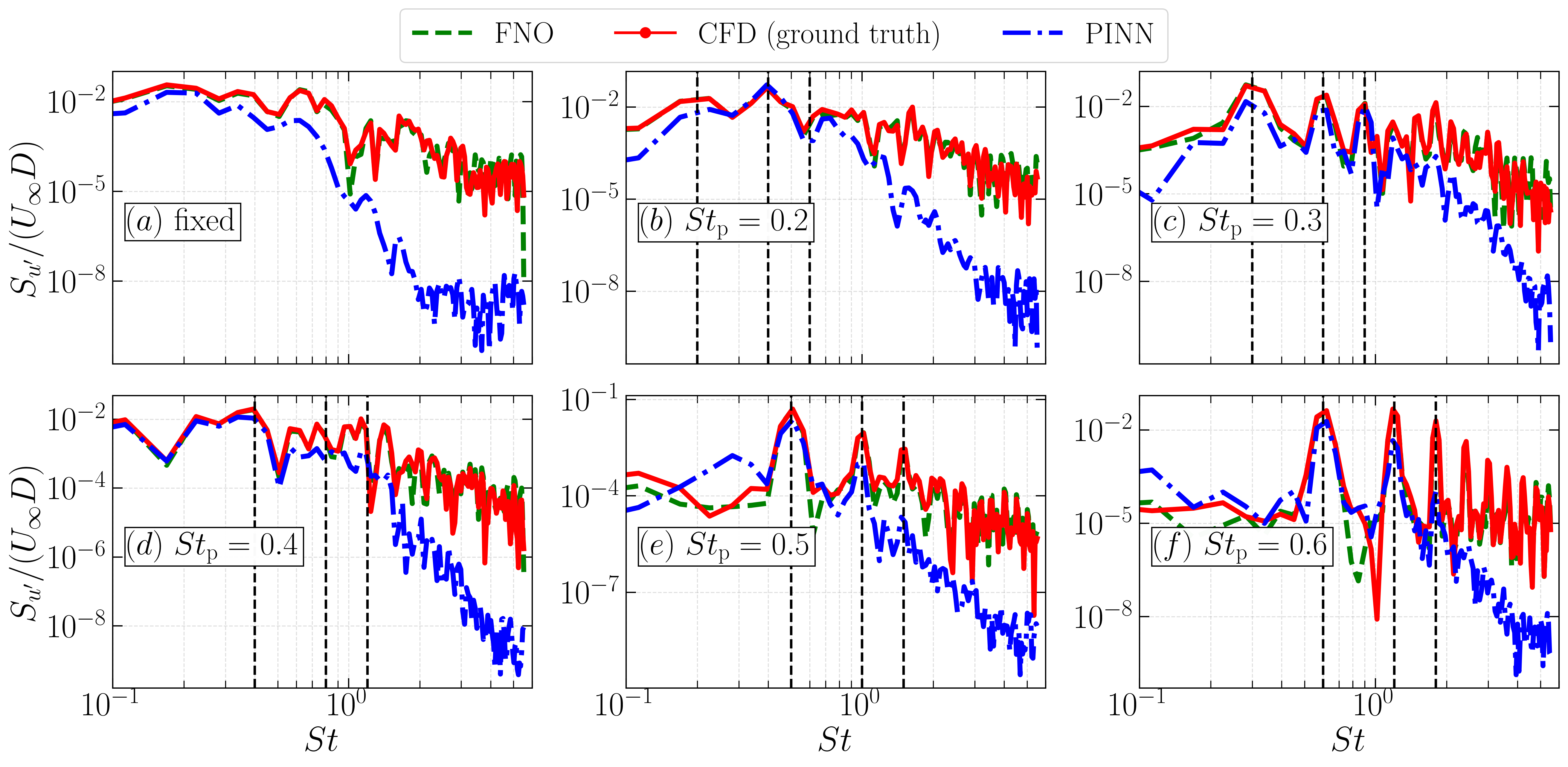}
    \caption{Comparison of $S_{u^\prime}/(U_\infty D)$ among the CFD reference, FNOs, and PINNs
    at $(x/D,y/D)=(7.5,0.5)$.}
    \label{fig:psd}
\end{figure}

\begin{figure}
    \centering
    \includegraphics[width=0.85\linewidth]{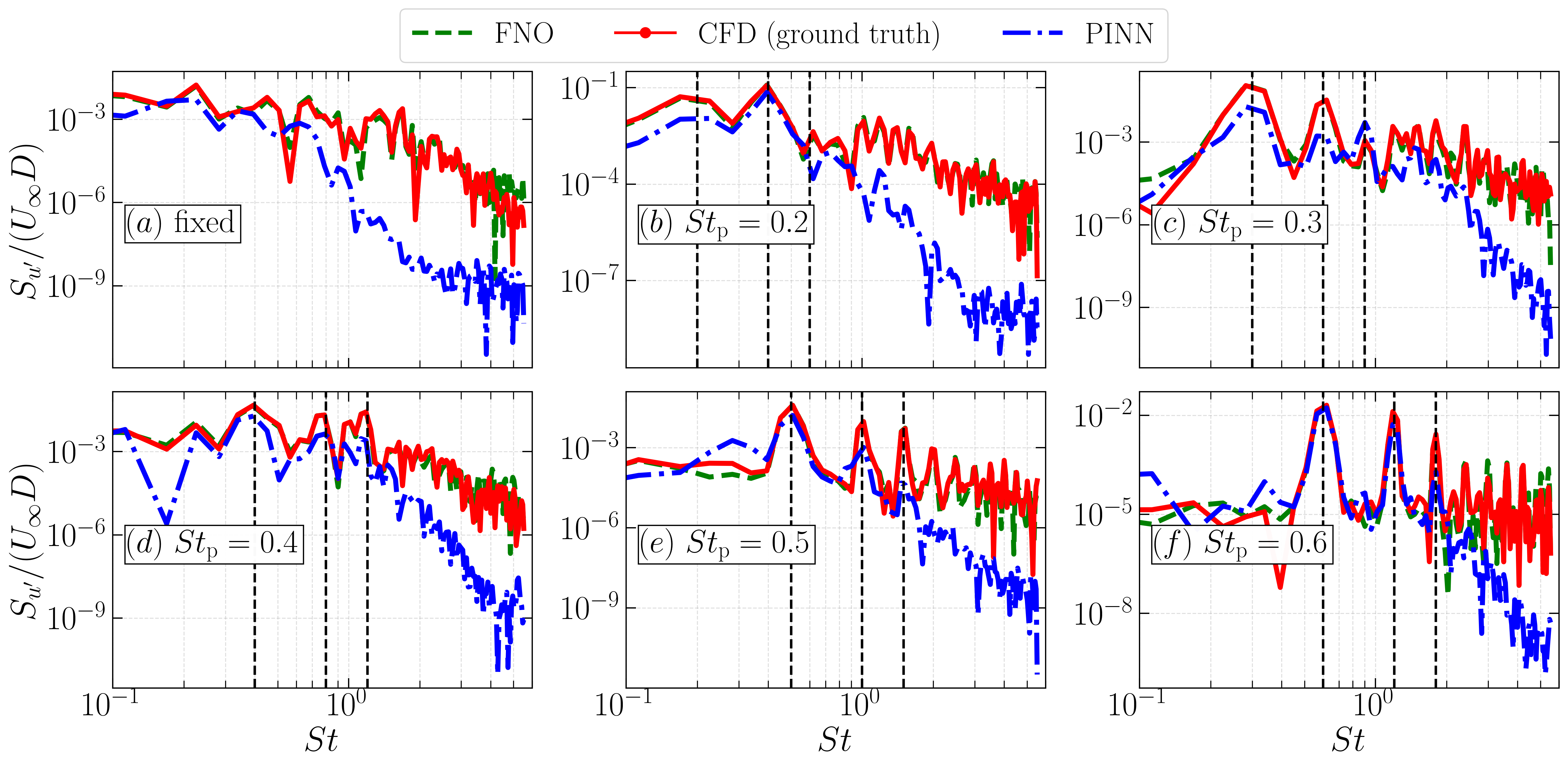}
    \caption{{\color{black}Comparison of $S_{u^\prime}/(U_\infty D)$ among the LES reference, FNOs, and PINNs at $(x/D,y/D)=(6.5,0.5)$.}}
    \label{fig:psd-6p5-edge}
\end{figure}

\begin{figure}
    \centering
    \includegraphics[width=0.85\linewidth]{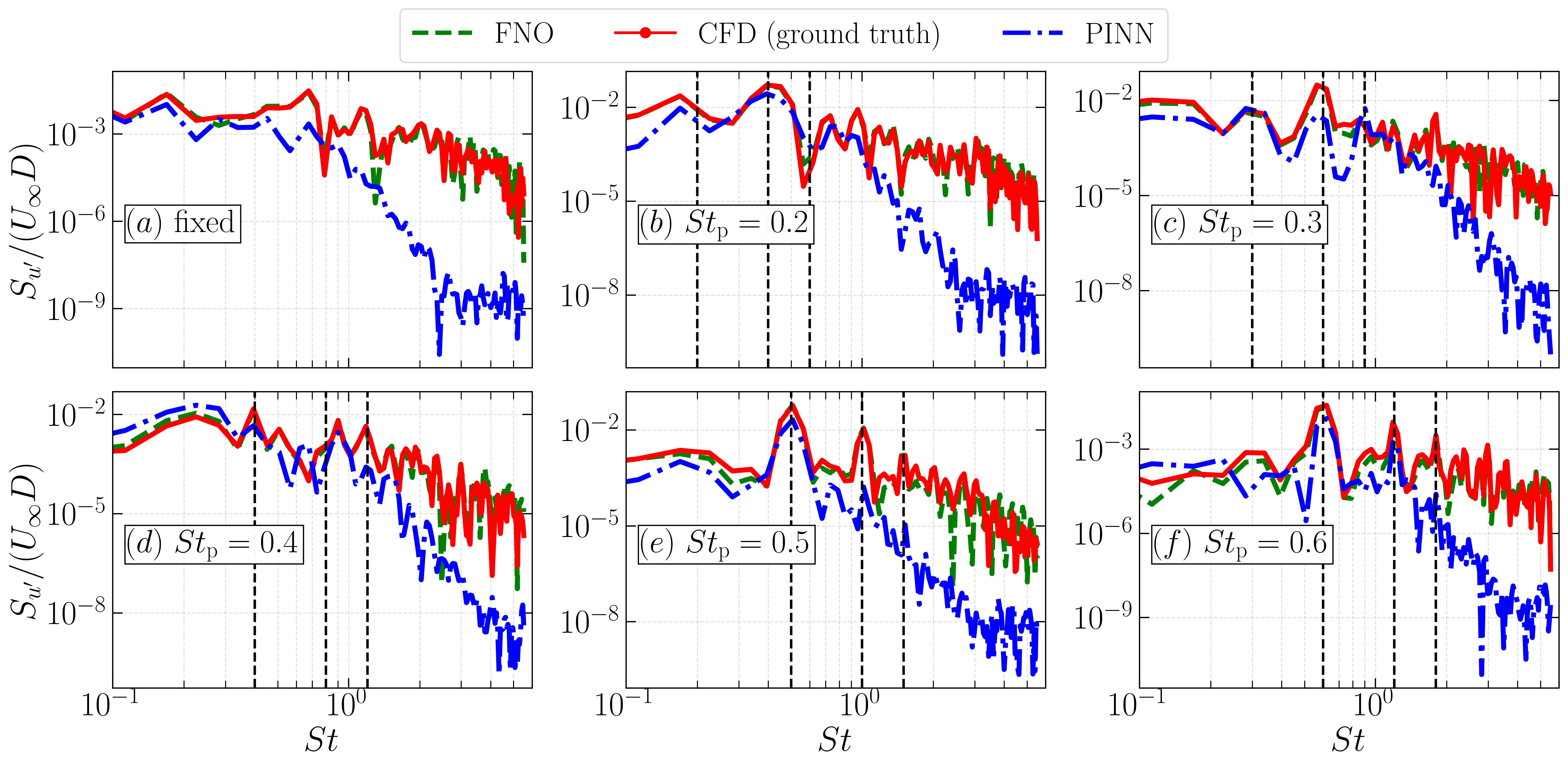}
    \caption{{\color{black}Comparison of $S_{u^\prime}/(U_\infty D)$ among the LES reference, FNOs, and PINNs at $(x/D,y/D)=(10,0.5)$.}}
    \label{fig:psd-10-edge}
\end{figure}

{\color{black}All PSD and FPSD diagnostics use $u'=u-\overline u$ over the common reconstruction interval {\color{black}($t=254,\ldots,449$~s, 196 samples)} at $f_s=1$~Hz. {\color{black}LES fields are stored at 1-s intervals without prior anti-alias filtering.} Welch density scaling is applied with a single full-record Hann segment, zero overlap, temporal-mean removal, and no additional linear detrending. The frequency-bin spacing is $\Delta f\simeq0.0051$~Hz ($\Delta St\simeq0.056$), and the PSD and FPSD are normalized as $S_{u'}/(U_\infty D)$ and $fS_{u'}/U_\infty^2$. The stored-data Nyquist limit is 0.5~Hz ($St\simeq5.53$), irrespective of the 0.025-s LES integration step. {\color{black}Potential aliasing and uncertainty associated with the finite record and limited independent averaging should be considered when interpreting the spectra, particularly at the lowest motion frequencies.} These spectra assess reconstruction over the resolved temporal-frequency range, not prediction-period spectral accuracy or the complete turbulent energy cascade.}
The instantaneous profiles and wake statistics show how accurately each model retains the amplitude and spatial organization of wake fluctuations, but they do not identify {\color{black}{\color{black}the distribution of reconstruction-period fluctuation energy across temporal frequencies}}. {\color{black}The power spectral density (PSD) describes how the fluctuation energy is distributed across the resolved temporal frequencies, while comparison among different downstream positions shows how this spectral content evolves with wake development. The pre-multiplied PSD (FPSD), $fS_{u^\prime}$, represents the contribution to the fluctuation variance per logarithmic frequency interval and therefore more clearly identifies the energy-containing scales, including the wake-meandering frequency and its harmonics.}
{\color{black}Figures~\ref{fig:psd}, \ref{fig:psd-6p5-edge}, and
\ref{fig:psd-10-edge} compare the {\color{black}normalized PSD of $u^\prime$} at three outer-wake downstream positions using Welch's method with Hann window.}
{\color{black}The corresponding PSDs at the three centerline positions, $(x/D,y/D)=(6.5,0)$, $(7.5,0)$, and $(10,0)$, are reported in Appendix~\ref{app:spectra}. Together, these six probes examine whether the spectral differences persist across both downstream and lateral positions. The FPSD is additionally evaluated at the three outer-wake positions, where the wake shear-layer fluctuations are more energetic.}

\begin{figure}
    \centering
    \includegraphics[width=0.85\linewidth]{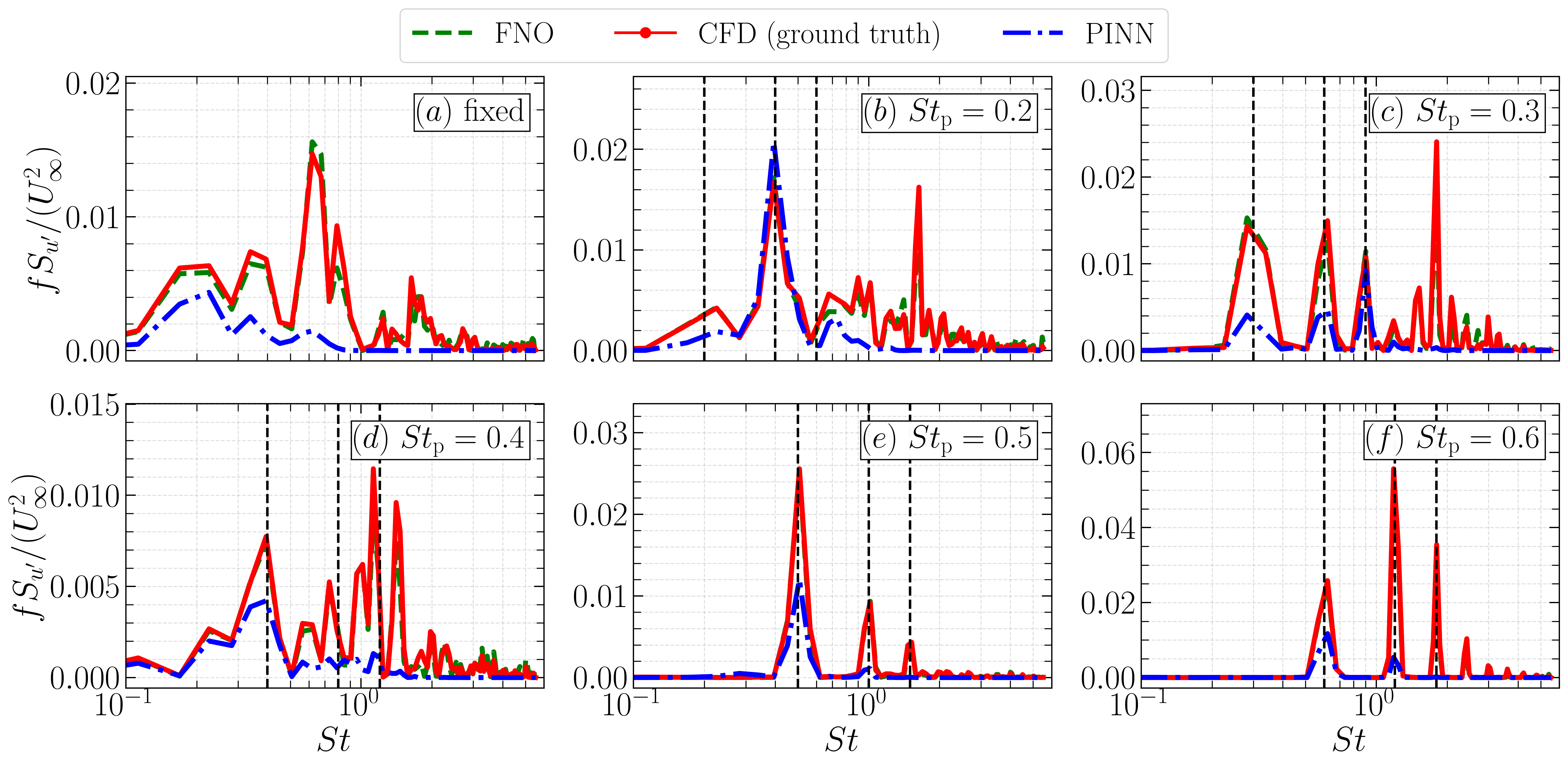}
    \caption{Comparison of the pre-multiplied PSD, $fS_{u^\prime}/U_\infty^2$, for the LES, FNOs, and PINNs at $(x/D,y/D)=(7.5,0.5)$.}
    \label{fig:fpsd}
\end{figure}

\begin{figure}
    \centering
    \includegraphics[width=0.85\linewidth]{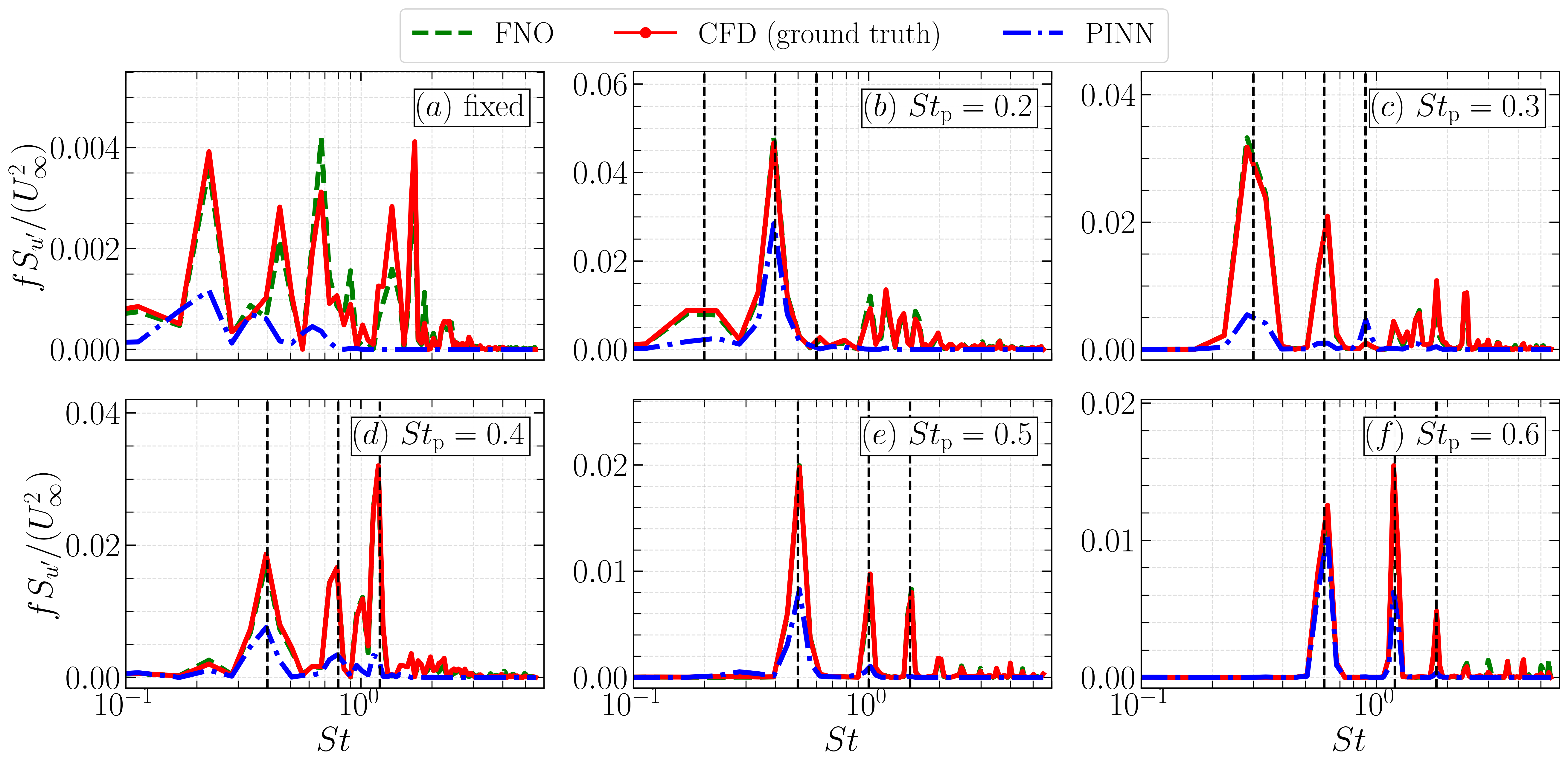}
    \caption{{\color{black}Comparison of the pre-multiplied PSD, $fS_{u^\prime}/U_\infty^2$, for the LES, FNOs, and PINNs at $(x/D,y/D)=(6.5,0.5)$.}}
    \label{fig:fpsd-6p5-edge}
\end{figure}

\begin{figure}
    \centering
    \includegraphics[width=0.85\linewidth]{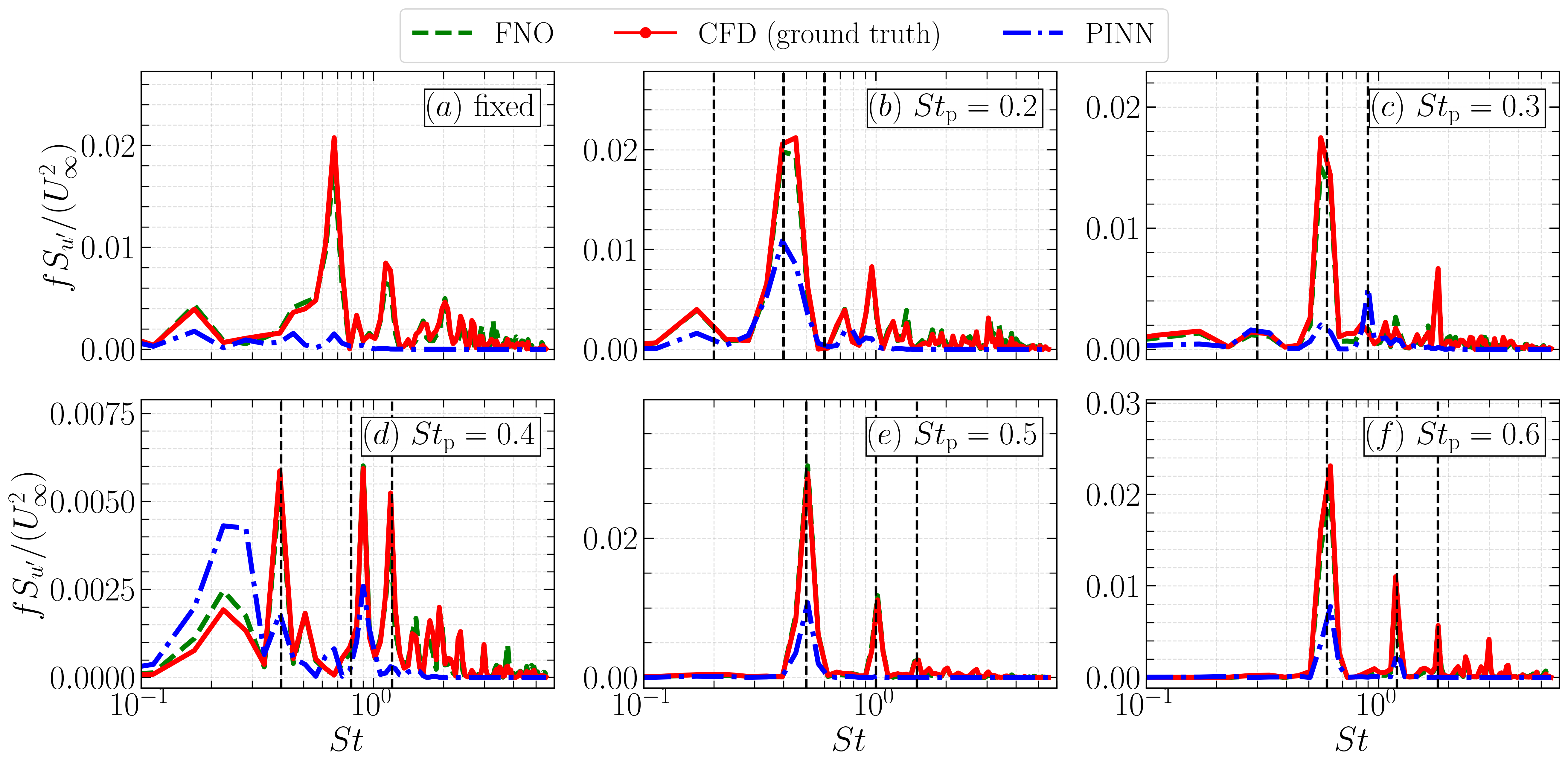}
    \caption{{\color{black}Comparison of the pre-multiplied PSD, $fS_{u^\prime}/U_\infty^2$, for the LES, FNOs, and PINNs at $(x/D,y/D)=(10,0.5)$.}}
    \label{fig:fpsd-10-edge}
\end{figure}

Both models identify the primary coherent structures associated with the wake meandering frequency, showing that they recover the large-scale motion-induced wake dynamics. However, for $St>1$, the implemented PINN model exhibits stronger high-frequency attenuation and a corresponding loss of fluctuations. The FNO model more closely follows the {\color{black}LES} reference at the primary peak, its harmonics, and the higher resolved frequencies. {\color{black}{\color{black}This closer agreement indicates that the FNO more accurately retains the resolved temporal spectral content of the reconstructed LES fields, whereas the stronger high-frequency attenuation of the PINN is consistent with smoother reconstructed fields.}}
{\color{black}The FPSD in Figs.~\ref{fig:fpsd},
\ref{fig:fpsd-6p5-edge}, and \ref{fig:fpsd-10-edge} identifies the
frequency bands that contribute most strongly to the wake fluctuations. Both the FNO and PINN models capture the primary wake meandering frequency ($St_{\rm p}$) and its second and third harmonics ($2St_{\rm p}$ and $3St_{\rm p}$), confirming that both models recover the large-scale motion-induced wake dynamics. However, the PINN model underestimates the energy contained at these scales and at higher frequencies, whereas the FNO model more closely reproduces the LES energy distribution. {\color{black}These scale-dependent energy deficits characterize the smoother PINN reconstruction. }}

\begin{figure}
    \centering
    \includegraphics[width=0.88\textwidth]{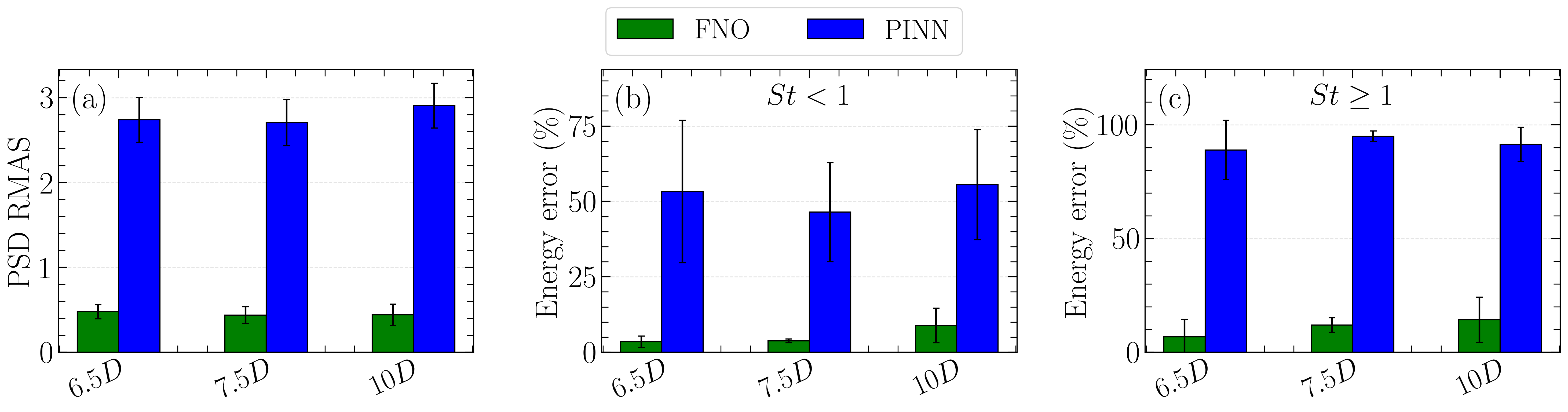}
    \caption{{\color{black}Quantitative PSD errors : (a) PSD RMSE, (b) relative
    integrated $St<1$ energy error, and (c) relative integrated
    $St\geq1$ energy error. Error bars denote one standard deviation across
    the five cases.}}
    \label{fig:spectral-errors}
\end{figure}

{\color{black}For quantitative comparison, let $S^{m}_{u^\prime,j}$ and
$S^{\mathrm{LES}}_{u^\prime,j}$ denote the model and LES PSD values at the
$ j$-th frequency bin, respectively. The PSD error and relative
integrated-energy error over a frequency band $\mathcal{B}$ are defined as}
\begin{align}
\epsilon_{\log S}
&=\left[\frac{1}{N_f}\sum_{j=1}^{N_f}
\left(\log_{10}S^{m}_{u^\prime,j}
-\log_{10}S^{\mathrm{LES}}_{u^\prime,j}\right)^2\right]^{1/2},\quad
\epsilon_E(\mathcal{B})
=\frac{\left|\int_{\mathcal{B}}S^{m}_{u^\prime}(St)\,dSt
-\int_{\mathcal{B}}S^{\mathrm{LES}}_{u^\prime}(St)\,dSt\right|}
{\int_{\mathcal{B}}S^{\mathrm{LES}}_{u^\prime}(St)\,dSt},
\end{align}
{\color{black}where $N_f$ is the number of nonzero frequency bins, and
$\mathcal{B}$ is either $St<1$ or $St\geq1$. Figure~\ref{fig:spectral-errors}
evaluates these quantities at the above three outer-wake
downstream positions and at the three centerline downstream positions reported in
Appendix~\ref{app:spectra}. {\color{black}The bars show the mean over the five moving cases, and the error bars denote one standard deviation across those cases.}}
{\color{black}Averaged over all positions and five moving cases,
$\epsilon_{\log S}$ is 0.455 for the FNO model and 2.742 for the PINN model. For the FNO
and PINN models, respectively, the integrated-energy errors are 5.4\% and 54.2\%
for $St<1$, and 12.2\% and 92.7\% for $St\geq1$. The high-frequency temporal attenuation observed for the PINN model is therefore also evident in the integrated spectral
statistics. Together with the wake evolution statistics, these results show that the PINN smoothing suppresses wake-center ($y_c$) and half-width ($R_{1/2}$) fluctuations, harmonic content, and higher-frequency energy, whereas the FNO model more closely preserves {\color{black}the multi-scale wake dynamics within the resolved temporal-frequency range}.}

{\color{black}Here, the smoothing of the PINN {\color{black}reconstructions} refers to the observed scale-selective attenuation rather than an explicit low-pass filtering operation. The two models represent wake evolution differently: the PINN directly maps space--time coordinates to the flow variables and recovers the dominant large-scale wake evolution, but exhibits stronger attenuation of smaller and more rapidly varying structures. By contrast, the FNO receives four complete wake fields as its temporal history and is trained against dense full-field targets. Its spectral convolutions represent non-local spatial correlations among the retained Fourier modes, while the multi-step training objective constrains their evolution during recursive prediction. These differences are consistent with the broader range of resolved spatial and temporal scales retained by the implemented FNO. Because the models differ in architecture, input representation, and training objective, the present comparison does not isolate a single cause for the observed attenuation, and this interpretation is restricted to the implemented PINN and FNO configurations.}

% \FloatBarrier
\section{Concluding remarks}\label{sec:Conclusions}
In this work, the capabilities of physics-informed neural networks (PINNs) and Fourier neural operators (FNOs) for modeling and predicting the multi-scale dynamic wakes of floating offshore wind turbines (FOWTs) were systematically compared using datasets generated by large-eddy simulation with an actuator-line model (LES--AL). Coupled surge-pitch motions, which are most prevalent for FOWTs, over $St_{\rm p}\in[0,0.6]$ were examined, where $St_{\rm p}=fD/U_{\infty}$, $f$, $D$, and $U_{\infty}$ denote the motion frequency, rotor diameter, and inflow velocity, respectively. The PINN model provides a coordinate-based data-to-data representation regularized by reduced two-dimensional momentum and continuity residuals, whereas the function-to-function FNO model learns the evolution of wake fields from a four-frame state history. 
Their performance was assessed in terms of computational cost, wake-field reconstruction, prediction over unseen periods, time-averaged and instantaneous wake statistical analysis, power spectral density (PSD), and pre-multiplied power spectral density (FPSD). The main findings {\color{black}for the implemented networks and investigated conditions} are summarized as follows:
\begin{enumerate}
    \item The FNO model achieves faster training with lower wake-prediction error than the PINN model. {\color{black}For each operating condition, training requires 7.36-10.39~min for the FNO model and 74.60--123.53~min for the PINN model, reducing the total time by approximately one order of magnitude. Over the temporally unseen 50-s prediction interval, the mean absolute error (MAE) of the normalized streamwise velocity is 0.0222 for the FNO and 0.0529 for the PINN model.}
    \item Both the FNO and PINN models reproduce the dominant {\color{black}reconstruction-period} time-averaged wake evolution, including the mean velocity deficit, wake-center displacement, and wake expansion, but the FNO model is consistently more accurate. For the complete normalized transverse mean velocity deficit profiles, the root-mean-square error (RMSE) relative to LES results, averaged over all cases, is 0.0017 for the FNO and 0.0160 for the PINN model. %The FNO also follows the LES mean wake center and half-width more closely across cases.
    \item The FNO model more accurately reproduces the instantaneous wake structures and their statistical characteristics, such as the evolution of the wake center and wake half-width. At $x/D=10$, the RMSE of the instantaneous normalized streamwise velocity profiles relative to LES results at 450~s ranges from 0.010 to 0.031 for the FNO model, compared with 0.027--0.079 for the PINN model across all $St_p$ cases. {\color{black}Statistics} of the Gaussian-fitted instantaneous profiles further show that the FNO model closely follows the downstream variation in the standard deviations of wake-center displacement ($\sigma_{y_c}$) and wake half-width ($\sigma_{R_{1/2}}$), whereas the PINN model underestimates both fluctuation levels because of field smoothing.
    \item Both models recover the large-scale wake dynamics, including the primary wake meandering frequency ($St_p$) and its second and third harmonics ($2St_{\rm p}$ and $3St_{\rm p}$). {\color{black}The reconstruction-period PSD shows that the FNO model more closely follows the LES within the resolved temporal-frequency range}, whereas the PINN model exhibits pronounced high-frequency attenuation. The FPSD further shows that the FNO model more accurately reproduces the energy contained in the wake meandering and its harmonics. {\color{black}The relative integrated-energy error for $St\geq1$ is 12.2\% for the FNO model and 92.7\% for the PINN model, {\color{black}quantifying the attenuation in the PINN reconstruction over the resolved {\color{black}temporal-frequency} band}.}
\end{enumerate}

{\color{black}These comparisons are conditional on the present LES reference, for which the demonstrated numerical convergence is limited to rotor loading and does not establish convergence of downstream wake statistics or spectra.} In conclusion, the FNO model trains substantially faster than the PINN model. Both models capture the mean wake field and dominant wake meandering, including its second and third harmonics. The FNO model nevertheless provides more accurate mean and instantaneous wake fields, with closer agreement to LES in both the energy distribution across motion-induced scales and {\color{black}the higher resolved temporal frequencies during reconstruction}. Given its advantages in training efficiency and turbulent-scale fidelity, the FNO framework provides a promising approach to high-fidelity turbulent wake modeling {\color{black}with measured compatibility with rapid field inference. In future work, we will assess online sensing, state estimation, communication, noise robustness, and closed-loop controller integration}.

\begin{appendices}
\section{{\color{black} Model sensitivity, robustness, and baseline analyses}}
\label{app:model-checks}
%{\color{black}This appendix first examines whether the retained PINN and FNO configurations reproduce the motion-induced wake dynamics without undue sensitivity to model choices. It then evaluates the robustness of the FNO to independent training realizations, compares the computational requirements of the two frameworks, and tests whether their prediction errors can be explained by short-time persistence of the wake field.}
{\color{black}The PINN sensitivity analysis examines the physics-loss weight, which controls the balance between reproducing the LES velocity fields and satisfying the reduced two-dimensional governing equation residuals. To determine whether the predicted wake dynamics are sensitive to this balance, a representative test varies $\alpha\in\{0.1,1,10\}$ while retaining the network, velocity supervision, residual, coordinate sets, optimizer and checkpoint rule. For $\alpha=0.1$, 1, and 10, respectively, the minimum normalized $u$--$v$ RMSEs over the model-selection evaluation interval are 0.6343, 0.6480, and 0.6888 for
the fixed case, 0.5664, 0.5608, and 0.6363 for the midrange $St_{\rm p}=0.4$ case, and 0.6093, 0.5878, and 0.7406 for the highest-frequency $St_{\rm p}=0.6$ case. The two examined dynamic cases attain their lowest model-selection RMSEs at $\alpha=1$, whereas the fixed case slightly favors $\alpha=0.1$, and all three cases have higher errors at $\alpha=10$. The consistently higher errors at $\alpha=10$ indicate that over-weighting the reduced residual does not improve the representation of the motion-induced wake dynamics sampled from the three-dimensional LES. Therefore, $\alpha=1$ is applied in the present work. }

{\color{black}A supplementary data-only coordinate network ($\alpha=0$, without PDE collocation points) is evaluated for the five moving cases, $St_{\rm p}=0.2$--0.6. It uses five hidden layers of 128 $\tanh$ units with Fourier and motion-phase features, and is fitted through 449~s. Over the full grid at {\color{black}450--500~s}, its mean MAE for $u/U_\infty$ is 0.02473. Its casewise MAEs are lower than the reported PINN values but higher than the FNO values in Table~\ref{tab:case-errors}, preserving the FNO accuracy advantage in these comparisons. The data-only network minimizes velocity mismatch alone, whereas the PINN jointly minimizes velocity mismatch and reduced momentum and continuity residuals. The two-dimensional residuals regularize the representation of the three-dimensional LES wake. The data-only network uses different coordinate encoding, sampling, and training settings from the main PINN{\color{black}. Its comparatively low prediction error illustrates the effectiveness of data fitting for the present LES wake fields. These results characterize the particular model implementations examined here. A predictive benefit specifically attributable to physics regularization remains to be established under matched input features and training settings.}}

{\color{black}Then, the FNO sensitivity analysis is conducted. Table~\ref{tab:app-fno-sensitivity} compares the number of history fields, motion input, spatial modes, hidden channels, training windows, and rollout horizon. %Unless otherwise stated, the reported metric is the mean normalized model-selection RMSE over the fixed, $St_{\rm p}=0.4$, and $St_{\rm p}=0.6$ cases, and the horizon test uses the fixed, $St_{\rm p}=0.3$, and $St_{\rm p}=0.6$ cases.
Therefore, the FNO model uses four history fields, 64 modes in each spatial direction, 32 hidden channels, no explicit phase channels, and a maximum 10-step curriculum. Increasing the number of modes from 64 to 128 reduces the representative model-selection error by only 3.5\% while increasing the parameter count from 17.32 to 68.17 million. Extending the horizon from 10 to 20 steps similarly reduces the model-selection error by only 0.22\% while increasing model-selection time by approximately 58\%. The retained settings therefore balance model-selection accuracy and computational cost.
The FNO model is trained using a multi-step curriculum with the rollout horizon increasing from $K=1$ to 10. The training duration is selected using the mean normalized RMSE over the contiguous model-selection evaluation interval and ranges from 25 to 40 epochs across the six cases. Each final FNO is then initialized from scratch and trained for the selected duration using all available training fields.
}

\begin{center}
\begin{minipage}{0.98\linewidth}
\centering
\captionof{table}{{\color{black}FNO architecture and training sensitivity.}}
\label{tab:app-fno-sensitivity}
\small
\setlength{\tabcolsep}{8pt}
\begin{tabular}{l c c c}
\hline\hline
\makecell[l]{Tested factor} & Settings & {\color{black}\makecell{Mean model-selection\\RMSE}} & \makecell{Retained\\setting} \\
\hline
{\makecell[l]{History fields (64 modes)} }& 4/8 & 0.3815/0.3891 & 4 \\
Explicit phase channels & without/with & 0.3810/0.3815 & without \\
{\makecell[l]{Fourier modes}} & 16/32/64/128 & 0.4577/0.4114/0.3789/0.3655 & 64 \\
Hidden channels & 16/32 & 0.4020/0.3789 & 32 \\
{\makecell[l]{Maximum training windows}} & 50/100/172 & 0.3795/0.3823/0.3789 & 172 \\
{\makecell[l]{Maximum training horizon}} & 10/20/24 & 0.3206/0.3199/0.3201 & 10 \\
\hline\hline
\end{tabular}
\end{minipage}
\end{center}

{\color{black}After finalizing the FNO configuration, three realizations are independently initialized and trained to assess the sensitivity of the model to random initialization. Figure~\ref{fig:app-fno-ensemble} compares the prediction errors of the three individual models and their ensemble. The consistent performance across the independently trained realizations indicates that the FNO predictions are not strongly sensitive to model initialization. Averaging the predictions further reduces the six-case aggregate and final-time MAEs from the mean values of 0.02219 and 0.02765 for the individual models to 0.01896 and 0.02252, respectively. However, this improvement comes at approximately three times the total training cost. Therefore, to balance predictive accuracy and computational efficiency, a single FNO realization is used in the present work.}
{\color{black}{\color{black}The implemented FNO and PINN contain 17,315,491 and 66,949 trainable parameters, respectively.} For reproducibility, the primary FNO model uses 5,750-7,160 updates at 0.0768-0.0871~s per update, whereas the PINN uses 204,800-320,000 updates at 0.0218-0.0251~s per update. The corresponding FNO training exposes 20.62-29.45 billion supervised scalar values, while the PINN exposes 26.84-41.94 billion supervised and 40.27-62.91 billion PDE scalar values. These counts document the native implementations but are not equivalent measures of floating-point work because the PINN physics loss requires automatic differentiation through first and second derivatives. The measured peak training GPU memory is 7.39~GiB for the FNO and 2.53~GiB for the PINN.}
{\color{black}The field inference latency was also measured for both models. For the FNO and PINN models, respectively, the median/P95 values are 2.31/2.98 and 0.62/0.79~ms on the GPU, and 143.46/183.98 and 46.66/57.78~ms on the CPU. Both models therefore infer a complete wake field well within the 1~s offline sampling interval, {\color{black}demonstrating computational compatibility with rapid inference. This offline benchmark does not establish online or closed-loop control performance}.}
{\color{black}To determine whether the low FNO error arises from learned wake evolution rather than short-time temporal continuity, a training free persistence baseline is constructed by holding the last available LES field at $t=449$~s fixed throughout the prediction interval $\widehat{u}(x,y,449+\tau)=u_{\mathrm{LES}}(x,y,449)$, and $\tau=1,\ldots,{\color{black}51}~\mathrm{s}$. This baseline retains the instantaneous spatial organization of the wake at initialization but contains no subsequent evolution of wake meandering or multiscale velocity fluctuations.}
\begin{center}
    \centering
    \includegraphics[width=0.65\linewidth]{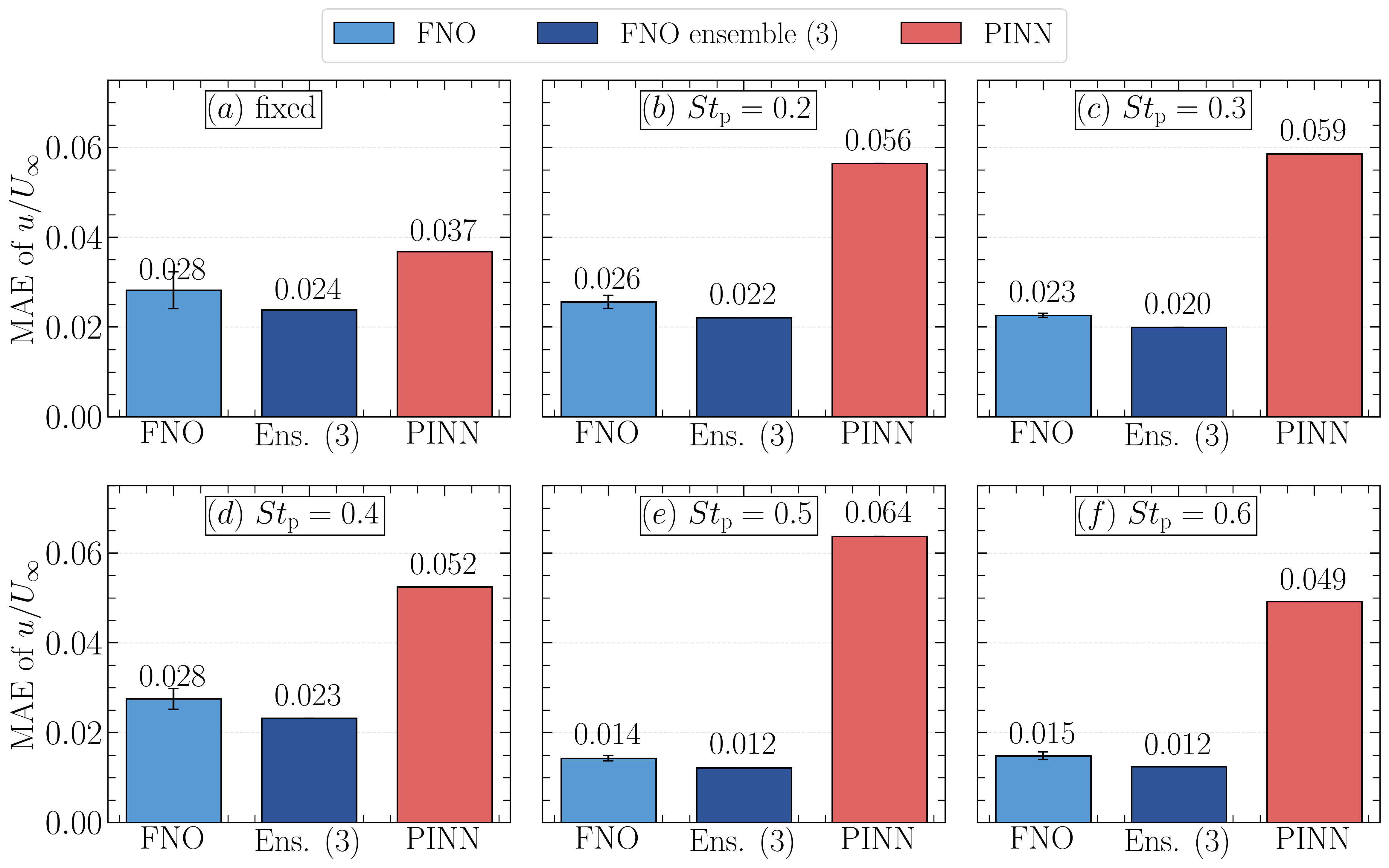}
    \captionof{figure}{{\color{black}MAE of the streamwise velocity over the prediction interval for the FNO, FNO ensemble, and PINN models.}}
    \label{fig:app-fno-ensemble}
\end{center}

\begin{center}
\begin{minipage}{\linewidth}\centering
\captionof{table}{{\color{black}Streamwise velocity MAE at selected prediction lead times for the FNO model, PINN model, and persistence baseline.}}
\label{tab:app-baselines}
\small
\begin{tabular}{c c c c}
\hline\hline
$\tau$ (s) & FNO & PINN & Persistence \\
\hline
1  & 0.0078 & 0.0306 & 0.0135 \\
5  & 0.0151 & 0.0334 & 0.0423 \\
10 & 0.0182 & 0.0374 & 0.0533 \\
20 & 0.0219 & 0.0465 & 0.0398 \\
50 & 0.0277 & 0.0761 & 0.0455 \\
\hline\hline
\end{tabular}
\end{minipage}
\end{center}

{\color{black}Table~\ref{tab:app-baselines} shows that the FNO model consistently outperforms the persistence baseline at all reported prediction times. Its interval-averaged and final time MAEs are 0.02219 and 0.0277, respectively, compared with 0.04391 and 0.0455 for persistence. This improvement indicates that the FNO model captures the temporal evolution of the wakes of FOWTs rather than simply carrying the initial velocity field forward. By contrast, the corresponding PINN errors of 0.05288 and 0.0761 exceed those of the persistence baseline, consistent with its smoother predictions and reduced ability to represent the instantaneous wake dynamics.}

\section{{\color{black} Centerline spectra of wake dynamics}}
\label{app:spectra}
{\color{black}The outer wake probes at $y/D=0.5$ in the main text sample the energetic shear layer region, whereas centerline probes test how the models retain temporal fluctuations as the meandering velocity deficit region moves across the nominal wake axis. To determine whether the spectral conclusions depend on lateral position, Figs.~\ref{fig:app-psd-6p5-center}-\ref{fig:app-psd-10-center} report the PSD at $(x/D,y/D)=(6.5,0)$, $(7.5,0)$, and $(10,0)$. All three figures use the same periods, 1~Hz sampling, mean removal, full record Hann window, and Welch density scaling as the main analysis.
For the moving cases, both models retain the dominant low-frequency peaks associated with wake meandering. The FNO model more closely follows the LES peak structure, harmonics, and higher-frequency fluctuations, whereas the PINN model exhibits stronger attenuation beyond the dominant low-frequency content, especially when $St>1$. The agreement between the centerline and outer wake results shows that this spectral difference persists across both sampled lateral regions and along the downstream development of the wake. }
\begin{center}\begin{minipage}{\linewidth}\centering
    \includegraphics[width=0.74\linewidth]{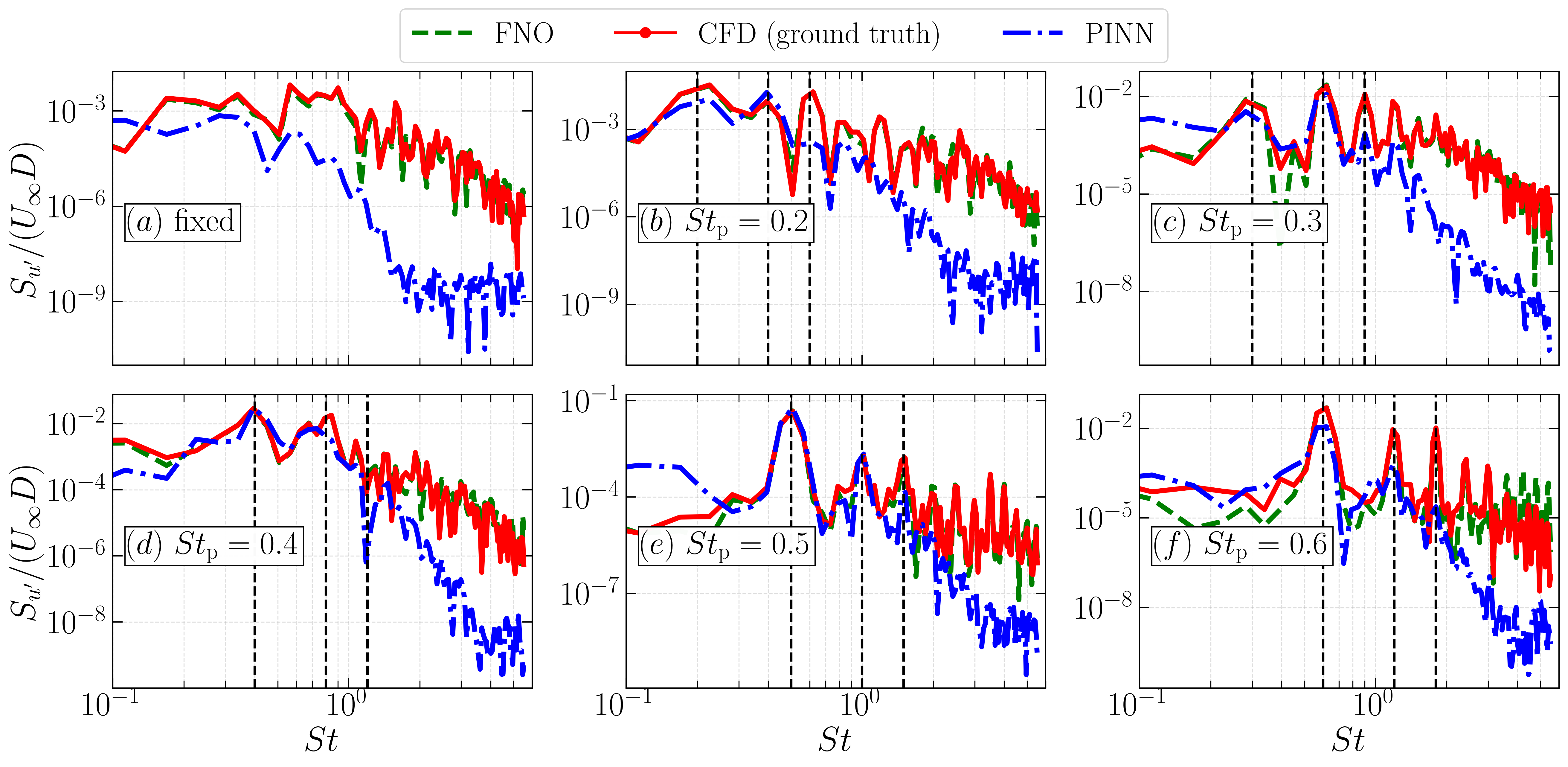}
    \captionof{figure}{{\color{black}PSD of the $u^\prime$, $S_{u^\prime}/(U_\infty D)$, for the LES reference, FNO, and PINN at $(x/D,y/D)=(6.5,0)$.}}
    \label{fig:app-psd-6p5-center}
\end{minipage}\end{center}
\begin{center}\begin{minipage}{\linewidth}\centering
    \includegraphics[width=0.74\linewidth]{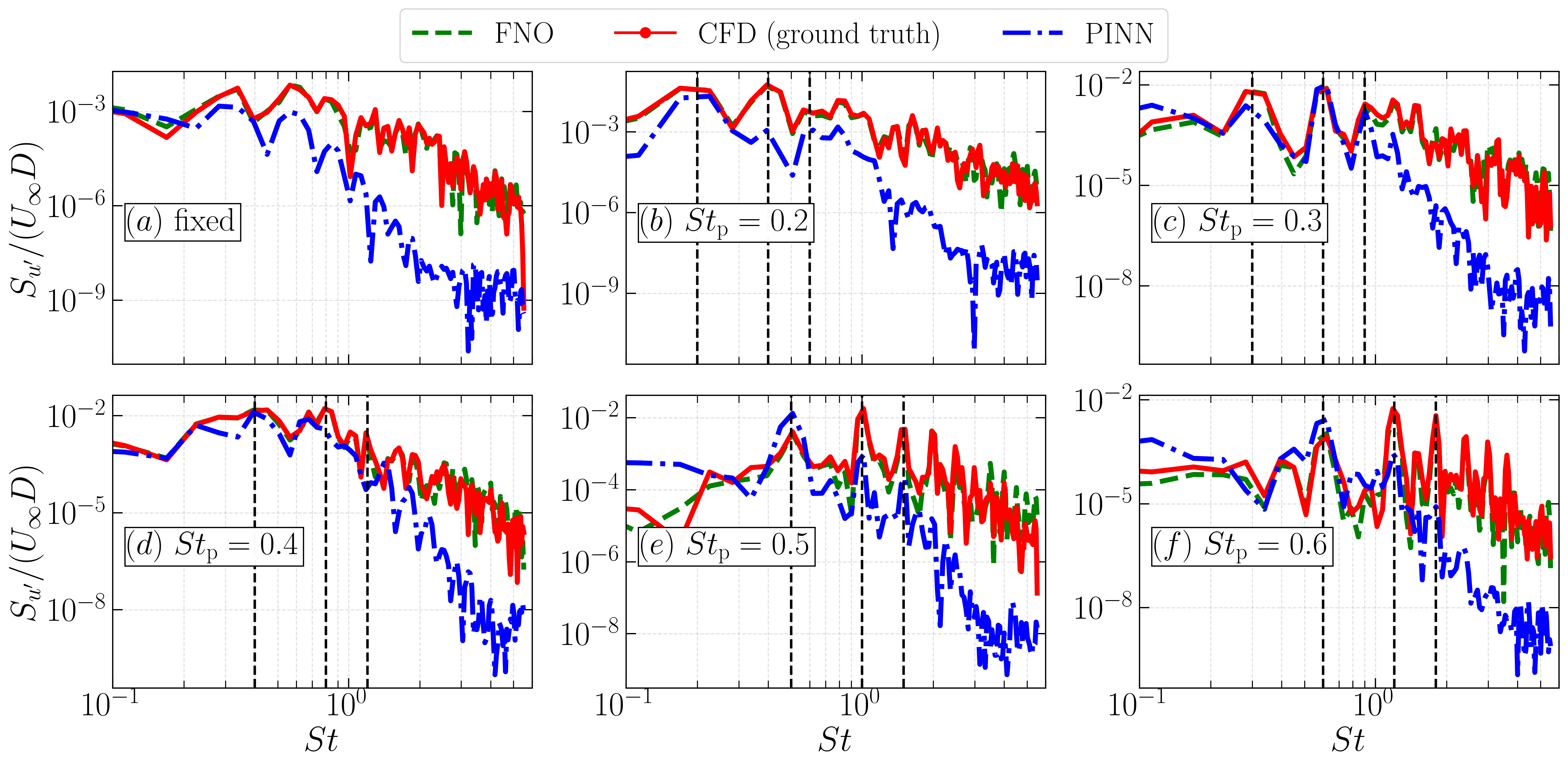}
    \captionof{figure}{{\color{black}PSD of the $u^\prime$, $S_{u^\prime}/(U_\infty D)$, for the LES reference, FNO, and PINN at $(x/D,y/D)=(7.5,0)$.}}
    \label{fig:app-psd-7p5-center}
\end{minipage}\end{center}
\begin{center}\begin{minipage}{\linewidth}\centering
    \includegraphics[width=0.74\linewidth]{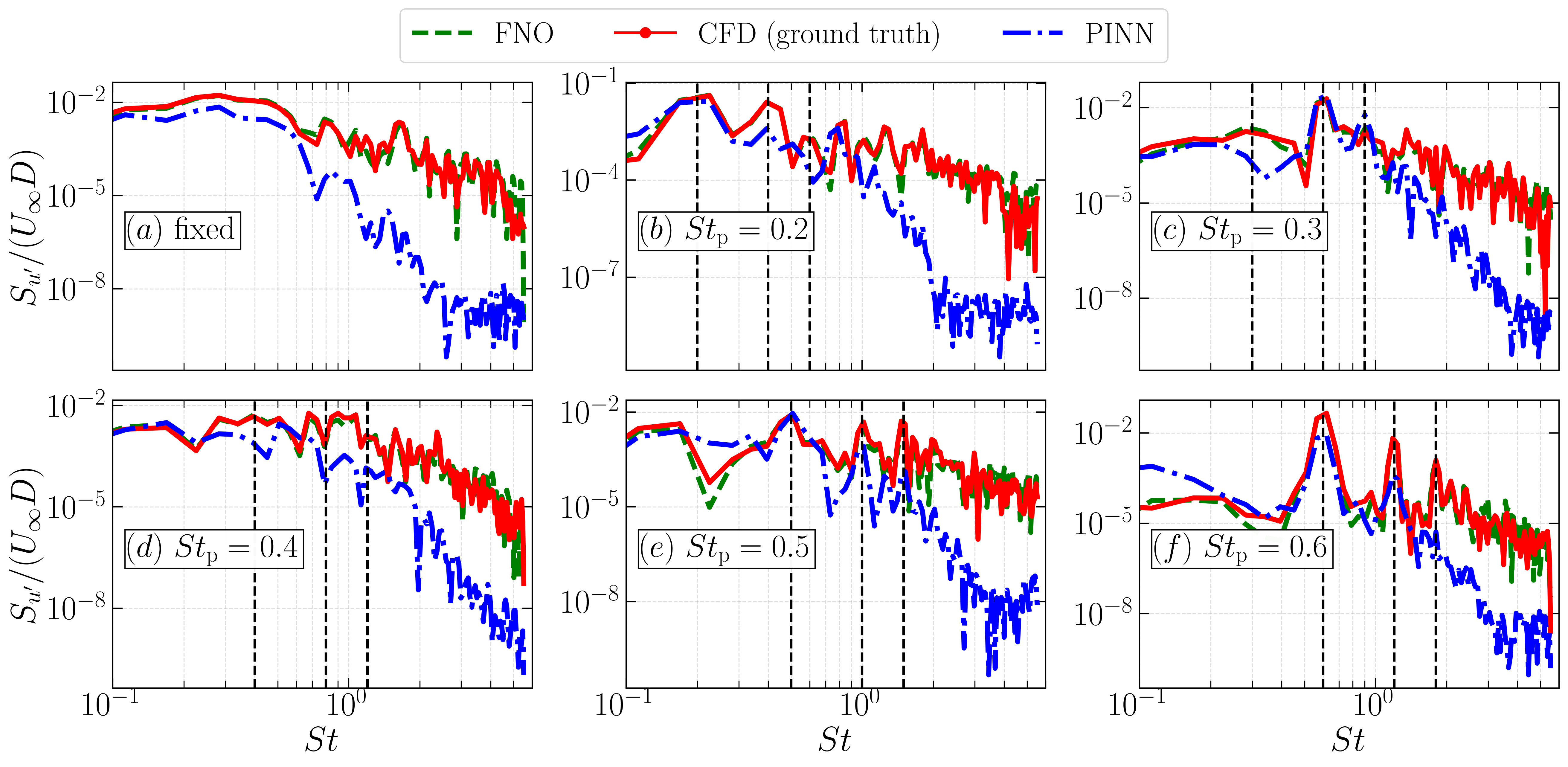}
    \captionof{figure}{{\color{black}PSD of the $u^\prime$, $S_{u^\prime}/(U_\infty D)$, for the LES reference, FNO, and PINN at $(x/D,y/D)=(10,0)$.}}
    \label{fig:app-psd-10-center}
\end{minipage}\end{center}

% \refstepcounter{section}
\section{{\color{black} Jointly trained FNO for wake dynamics modeling and cross-condition generalization
}}
\label{app:transfer}

{\color{black}
This appendix first evaluates whether a single jointly trained FNO can represent wake evolution across all six prescribed-motion conditions and then examines its generalization to unseen conditions through held-out-condition tests. The four-field input history is augmented with $[St_{\rm p}/0.6,\sin\phi_t,\cos\phi_t]$, where $\phi_t=2\pi St_{\rm p}U_{\infty}t/D$. The normalized frequency and phase encode the imposed motion time scale and instantaneous phase, respectively, while the velocity-field history describes the evolving wake state. These descriptors are broadcast over the spatial grid. The jointly trained model otherwise adopts the same four-layer, 32-channel architecture, 64 retained modes in each spatial direction, and training procedure as the case-specific FNOs used in the main text.
The six-case averaged and final-time streamwise-velocity MAEs of the jointly trained FNO are 0.02518 and 0.03509, respectively, compared with 0.02219 and 0.02784 for the case-specific FNOs. This modest loss of accuracy reveals a trade-off between cross-condition model sharing and case-specific prediction accuracy: the jointly trained FNO captures common motion-induced wake dynamics using a single parameterized model, whereas the case-specific FNOs more accurately resolve the downstream response of each individual condition. The case-specific FNOs are therefore retained and applied in the main analysis to ensure higher predictive accuracy and to provide a controlled comparison with the corresponding PINNs.
}

\begin{center}
    \includegraphics[width=0.7\linewidth]{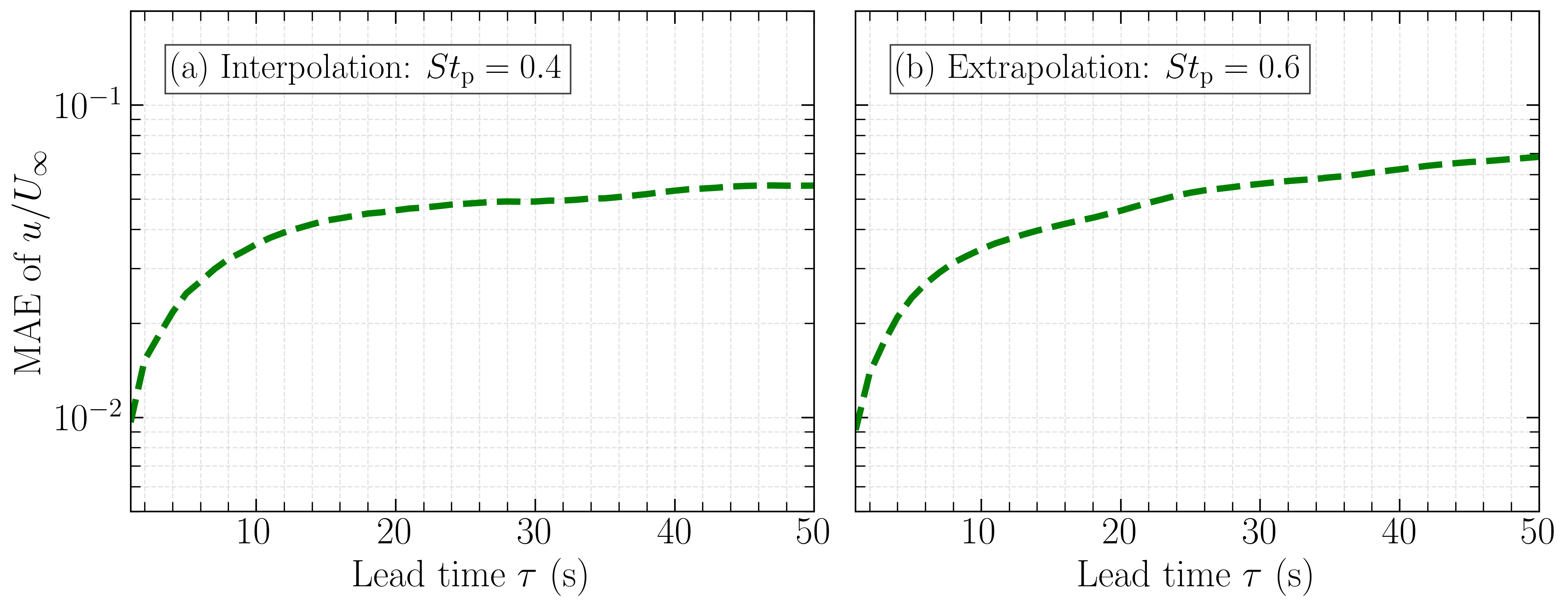}
    \captionof{figure}{{\color{black}Evolution of the streamwise velocity MAE in the held-out-condition tests.}}
    \label{fig:app-ood-growth}
\end{center}

{\color{black}To assess generalization to unseen motion frequencies, two FNOs are trained separately, with one complete motion condition withheld in each test. For the interpolation test, the entire $St_{\rm p}=0.4$ condition is excluded from training, whereas the highest-frequency condition, $St_{\rm p}=0.6$, is excluded for the extrapolation test. At inference, each rollout is initialized using four reference fields from the held-out condition. These fields are used only for initialization, and no data from the held-out condition are used for training or fine-tuning. 
The tests therefore evaluate whether the wake dynamics learned from the remaining frequencies can predict the subsequent evolution of an observed wake state at an unseen motion frequency. 
Figure~\ref{fig:app-ood-growth} shows that the prediction errors increase with lead time for both held-out frequencies. For the interpolation test at $St_{\rm p}=0.4$, the averaged and final-time MAEs are 0.04428 and 0.05525, respectively, compared with 0.03031 and 0.04059 for the corresponding case-specific FNO. For the extrapolation test at $St_{\rm p}=0.6$, the respective errors are 0.04848 and 0.06830, compared with 0.01413 and 0.01773 for the case-specific FNO. This generalization gap indicates that the motion descriptors and initial wake history encode common features of the wake dynamics but do not fully determine the frequency-dependent downstream response. The larger degradation at $St_{\rm p}=0.6$ further suggests that extrapolation beyond the training-frequency range is more challenging than interpolation within it. Together with the jointly trained-model results, these tests demonstrate that the FNO learns wake-evolution features that are partially transferable across the prescribed motion frequencies.}

\section{{\color{black}Instantaneous wake profiles at a later prediction time}}
\label{app:late-profiles}
{\color{black}Figures~\ref{fig:app-profile-490-near}--\ref{fig:app-profile-490-rmse} extend the comparisons in Figs.~\ref{fig:u_1d_profiles}--\ref{fig:app-profile-errors} to $t=490$~s. The FNO rollout is initialized with LES fields at 446--449~s, so this is its 41st autoregressive prediction step, 41~s after the last reference field. The profiles compare the same case-specific FNO realization and the implemented $\alpha=1$ PINN with LES. The RMSE uses all 150 transverse grid points at each station, without filtering or Gaussian fitting. At $x/D=1$, the six-case RMSE ranges are 0.01129--0.04032 for FNO and 0.05346--0.12670 for PINN. At $x/D=10$, the corresponding ranges are 0.02194--0.07883 and 0.05193--0.16690. FNO retains a lower profile RMSE in all twelve case--station comparisons. The larger errors than at 450~s show the loss of accuracy with prediction time, while the ranking at 490~s supports the later-time validity of the FNO--PINN comparison.}

\begin{center}\begin{minipage}{\linewidth}\centering
\includegraphics[width=0.67\linewidth]{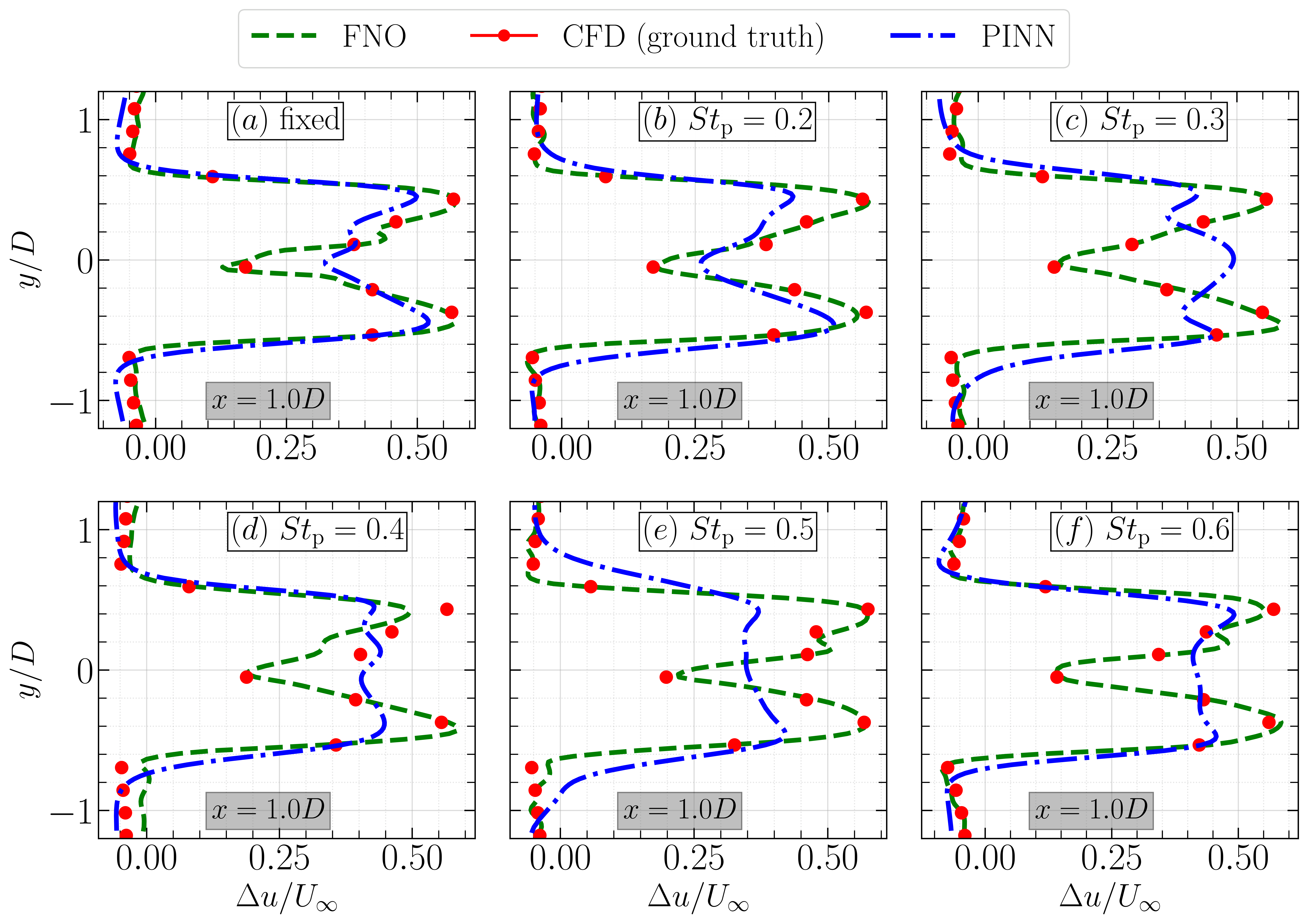}
\captionof{figure}{{\color{black}Instantaneous transverse profiles of the normalized streamwise velocity deficit at $x/D=1$ and $t=490$~s.}}
\label{fig:app-profile-490-near}
\end{minipage}\end{center}
\begin{center}\begin{minipage}{\linewidth}\centering
\includegraphics[width=0.67\linewidth]{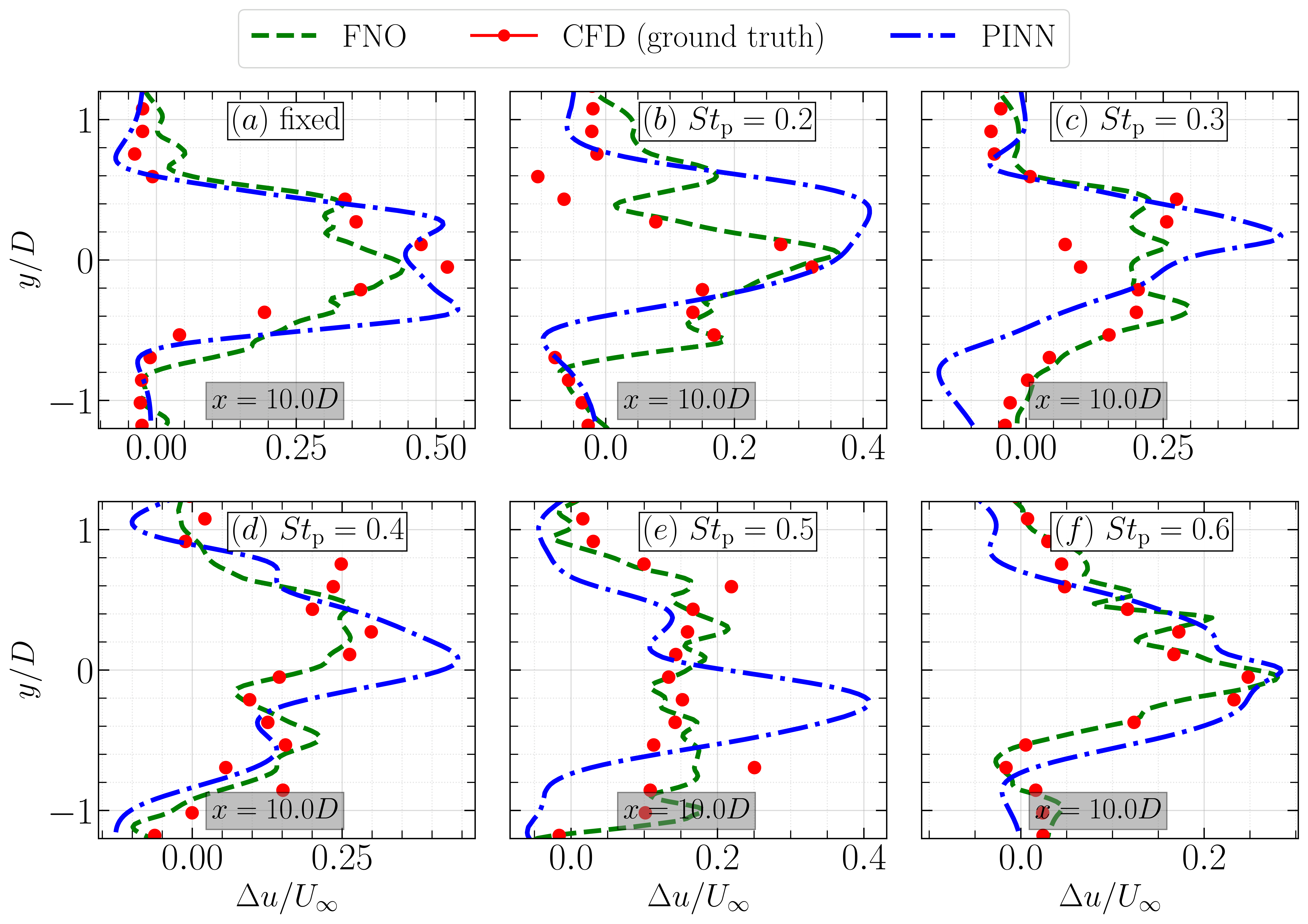}
\captionof{figure}{{\color{black}Instantaneous transverse profiles of the normalized streamwise velocity deficit at $x/D=10$ and $t=490$~s.}}
\label{fig:app-profile-490-far}
\end{minipage}\end{center}
\begin{center}\begin{minipage}{\linewidth}\centering
\includegraphics[width=0.8\linewidth]{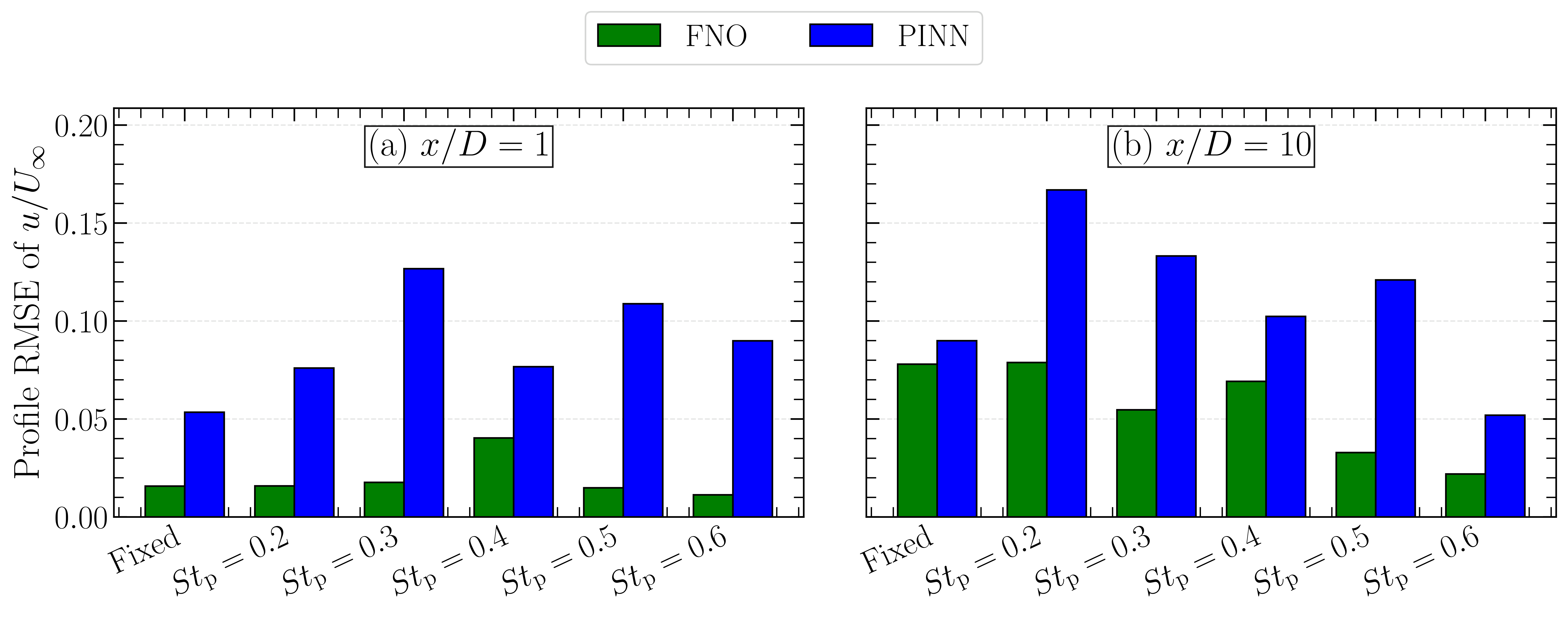}
\captionof{figure}{{\color{black}RMSE of the $u/U_\infty$ profiles at $t=490$~s for (a) $x/D=1$ and (b) $x/D=10$ for the FNO and PINN.}}
\label{fig:app-profile-490-rmse}
\end{minipage}\end{center}

\end{appendices}

\section*{Acknowledgement}
This work was supported by the {\color{black}National Key Research and Development Program of China for ``Key Technologies and Software Development for Multi-Scenario Wind Farm Planning and Design'' (No. 2024YFB4205700), the National Natural Science Foundation of China (Nos. 52606283, 12202456)}, the Natural Science Foundation of Jiangsu Province, China (No. BK20250336), China Postdoctoral Science Foundation (No. 2025M773143), the Fundamental Research Funds for the Central Universities, College of Renewable Energy, HoHai University (Nos. B250201091, B250201221).

\section*{Declaration of competing interest}\label{sec:Declaration of Interests}
The authors declare that they have no known competing financial interests or personal relationships that could have appeared to influence the work reported in this paper.

% \section{Author contributions}
% \textbf{Guodan Dong:} Conceptualization, Data curation, Investigation, Methodology,
% Software, Visualization, Writing - original draft.
% \textbf{Jianhua Qin:} Investigation, Methodology, Writing - review $\&$ editing.
% \textbf{Chang Xu:} Investigation, Methodology, Writing - review $\&$ editing.

\section*{Data availability}
{\color{black}{\color{black}A representative subset of LES snapshots with temporal indices, representative trained weights, and sample post-processing scripts for wake diagnostics and spectral analysis} are available from the first and corresponding authors upon reasonable request.}

\printcredits
% \bibliographystyle{cas-model2-names}
\bibliographystyle{elsarticle-num}

\begin{thebibliography}{10}
\expandafter\ifx\csname url\endcsname\relax
  \def\url#1{\texttt{#1}}\fi
\expandafter\ifx\csname urlprefix\endcsname\relax\def\urlprefix{URL }\fi
\expandafter\ifx\csname href\endcsname\relax
  \def\href#1#2{#2} \def\path#1{#1}\fi

\bibitem{GWEC2025Report}
{Global Wind Energy Council}, Global wind report 2025, Tech. rep., Global Wind
  Energy Council (GWEC), Bonn, Germany (2025).

\bibitem{barthelmie2007modelling}
R.~J. Barthelmie, O.~Rathmann, S.~T. Frandsen, K.~Hansen, E.~Politis,
  J.~Prospathopoulos, K.~Rados, D.~Cabez{\'o}n, W.~Schlez, J.~Phillips, et~al.,
  Modelling and measurements of wakes in large wind farms, Journal of Physics:
  Conference Series 75~(1) (2007) 012049.
\newblock \href {https://doi.org/10.1088/1742-6596/75/1/012049}
  {\path{doi:10.1088/1742-6596/75/1/012049}}.

\bibitem{dong2026reinforcement}
G.~Dong, J.~Qin, C.~Wu, C.~Xu, X.~Yang, Reinforcement learning-enhanced genetic
  algorithm for wind farm layout optimization, Renewable Energy 259 (2026)
  125093.
\newblock \href {https://doi.org/10.1016/j.renene.2025.125093}
  {\path{doi:10.1016/j.renene.2025.125093}}.

\bibitem{wang2023evolution}
T.~Wang, C.~Cai, X.~Wang, Z.~Wang, Y.~Chen, C.~Hou, S.~Zhou, J.~Xu, Y.~Zhang,
  Q.~Li, Evolution mechanism of wind turbine wake structure in yawed condition
  by actuator line method and theoretical analysis, Energy Conversion and
  Management 281 (2023) 116852.

\bibitem{yang2026wind}
X.~Yang, F.~Sotiropoulos, J.~N. S{\o}rensen, Wind farm fluid mechanics for
  high-penetration wind energy, Renewable and Sustainable Energy Reviews 226
  (2026) 116260.
\newblock \href {https://doi.org/10.1016/j.rser.2025.116260}
  {\path{doi:10.1016/j.rser.2025.116260}}.

\bibitem{sant2026progress}
T.~Sant, Progress in the modelling and management of offshore wind farm wakes:
  A literature review, Ocean Engineering 352 (2026) 124544.

\bibitem{stevens2017flow}
R.~J. Stevens, C.~Meneveau, Flow structure and turbulence in wind farms, Annual
  Review of Fluid Mechanics 49 (2017) 311--339.

\bibitem{huanqiang2024investigation}
H.~Zhang, X.~Gao, H.~Lu, Q.~Zhao, X.~Zhu, Y.~Wang, F.~Zhao, Investigation of a
  new 3d wake model of offshore floating wind turbines subjected to the
  coupling effects of wind and wave, Applied Energy 365 (2024) 123189.

\bibitem{wenfeng2025investigation}
W.~Li, Z.~Zhao, Y.~Liu, H.~Liu, S.~Wei, A.~Kashif, G.~Dong, Y.~Liu, Y.~Ma,
  Investigation of dynamic wake model of a floating offshore wind turbine under
  heave, surge and pitch motion, Renewable Energy 254 (2025) 123665.
\newblock \href {https://doi.org/10.1016/j.renene.2025.123665}
  {\path{doi:10.1016/j.renene.2025.123665}}.

\bibitem{he2025novel}
G.~He, H.~Sun, R.~He, A novel analytical wake model for floating offshore wind
  turbines with pitch motion effects, Renewable Energy 256 (2026) 124090.
\newblock \href {https://doi.org/10.1016/j.renene.2025.124090}
  {\path{doi:10.1016/j.renene.2025.124090}}.

\bibitem{luo2025innovative}
Z.~Luo, L.~Wang, Y.~Fu, J.~Yuan, J.~Xu, A.~C. Tan, Innovative sparse data
  reconstruction approaches for yawed wind turbine wake flow via data-driven
  and physics-informed machine learning, Physics of Fluids 37~(3) (2025)
  035158.
\newblock \href {https://doi.org/10.1063/5.0256953}
  {\path{doi:10.1063/5.0256953}}.

\bibitem{liu2026phywakenet}
X.~Liu, Z.~Li, X.~Yang, Phywakenet: a dynamic wake model accounting for
  aerodynamic force oscillations, Wind Energy Science 11~(3) (2026) 771--793.
\newblock \href {https://doi.org/10.5194/wes-11-771-2026}
  {\path{doi:10.5194/wes-11-771-2026}}.

\bibitem{zhang2025advanced}
B.~Zhang, L.~Wang, J.~Ge, Z.~Luo, J.~Yuan, Z.~Wang, J.~Xu, Advanced wake
  modeling in wind farm: A physics-informed framework with virtual lidar
  measurements, Physics of Fluids 37~(6) (2025) 065133.
\newblock \href {https://doi.org/10.1063/5.0267955}
  {\path{doi:10.1063/5.0267955}}.

\bibitem{dong2023characteristics}
G.~Dong, J.~Qin, Z.~Li, X.~Yang, Characteristics of wind turbine wakes for
  different blade designs, Journal of Fluid Mechanics 965 (2023) A15.
\newblock \href {https://doi.org/10.1017/jfm.2023.385}
  {\path{doi:10.1017/jfm.2023.385}}.

\bibitem{dong2026impacts}
G.~Dong, X.~Yang, Impacts of blade designs on wake dynamics in tandem wind
  turbine configurations, Journal of Fluid Mechanics 1040 (2026) A42.
\newblock \href {https://doi.org/10.1017/jfm.2026.11891}
  {\path{doi:10.1017/jfm.2026.11891}}.

\bibitem{martinez2015large}
L.~A. Mart{\'\i}nez-Tossas, M.~J. Churchfield, S.~Leonardi, Large eddy
  simulations of the flow past wind turbines: actuator line and disk modeling,
  Wind Energy 18~(6) (2015) 1047--1060.

\bibitem{dong2022predictive}
G.~Dong, Z.~Li, J.~Qin, X.~Yang, Predictive capability of actuator disk models
  for wakes of different wind turbine designs, Renewable Energy 188 (2022)
  269--281.

\bibitem{li2025effects}
W.~Li, Z.~Zhao, G.~Dong, Y.~Liu, H.~Liu, S.~Wei, K.~Ali, Y.~Liu, Y.~Ma, Effects
  of heave frequency and amplitude on wake evolution of floating offshore wind
  turbine in smooth flow conditions, Ocean Engineering 340 (2025) 122401.
\newblock \href {https://doi.org/10.1016/j.oceaneng.2025.122401}
  {\path{doi:10.1016/j.oceaneng.2025.122401}}.

\bibitem{goccmen2025data}
T.~G{\"o}{\c{c}}men, J.~Liew, E.~Kadoche, N.~Dimitrov, R.~Riva, S.~J. Andersen,
  A.~W. Lio, J.~Quick, P.-E. R{\'e}thor{\'e}, K.~Dykes, Data-driven wind farm
  flow control and challenges towards field implementation: A review, Renewable
  and Sustainable Energy Reviews 216 (2025) 115605.
\newblock \href {https://doi.org/10.1016/j.rser.2025.115605}
  {\path{doi:10.1016/j.rser.2025.115605}}.

\bibitem{taira2017modal}
K.~Taira, S.~L. Brunton, S.~T. Dawson, C.~W. Rowley, T.~Colonius, B.~J. McKeon,
  O.~T. Schmidt, S.~Gordeyev, V.~Theofilis, L.~S. Ukeiley, Modal analysis of
  fluid flows: An overview, AIAA journal 55~(12) (2017) 4013--4041.

\bibitem{li2022onset}
Z.~Li, G.~Dong, X.~Yang, Onset of wake meandering for a floating offshore wind
  turbine under side-to-side motion, Journal of Fluid Mechanics 934 (2022).

\bibitem{zhou2023high}
L.~Zhou, J.~Wen, Z.~Wang, P.~Deng, H.~Zhang, High-fidelity wind turbine wake
  velocity prediction by surrogate model based on d-pod and lstm, Energy 275
  (2023) 127525.
\newblock \href {https://doi.org/10.1016/j.energy.2023.127525}
  {\path{doi:10.1016/j.energy.2023.127525}}.

\bibitem{parekh2023uncertainty}
J.~Parekh, R.~W. C.~P. Verstappen, Uncertainty quantification analysis for
  simulation of wakes in wind-farms using a stochastic {RANS} solver, compared
  with a deep learning approach, Computers \& Fluids 257 (2023) 105867.
\newblock \href {https://doi.org/10.1016/j.compfluid.2023.105867}
  {\path{doi:10.1016/j.compfluid.2023.105867}}.

\bibitem{li2024transformerwake}
H.~Li, Q.~Yang, T.~Li, Wind turbine wake prediction modelling based on
  transformer-mixed conditional generative adversarial network, Energy 291
  (2024) 130403.
\newblock \href {https://doi.org/10.1016/j.energy.2024.130403}
  {\path{doi:10.1016/j.energy.2024.130403}}.

\bibitem{hasegawa2020cnn}
K.~Hasegawa, K.~Fukami, T.~Murata, K.~Fukagata, Cnn-lstm based reduced order
  modeling of two-dimensional unsteady flows around a circular cylinder at
  different reynolds numbers, Fluid Dynamics Research 52~(6) (2020) 065501.

\bibitem{raissi2019physics}
M.~Raissi, P.~Perdikaris, G.~E. Karniadakis, Physics-informed neural networks:
  A deep learning framework for solving forward and inverse problems involving
  nonlinear partial differential equations, Journal of Computational physics
  378 (2019) 686--707.

\bibitem{zhang2021three}
J.~Zhang, X.~Zhao, Three-dimensional spatiotemporal wind field reconstruction
  based on physics-informed deep learning, Applied Energy 300 (2021) 117390.

\bibitem{cobelli2023physics}
P.~Cobelli, K.~Shukla, S.~Nesmachnow, M.~Draper, Physics informed neural
  networks for wind field modeling in wind farms, in: Journal of Physics:
  Conference Series, Vol. 2505, IOP Publishing, 2023, p. 012051.

\bibitem{zhang2023digital}
J.~Zhang, X.~Zhao, Digital twin of wind farms via physics-informed deep
  learning, Energy Conversion and Management 293 (2023) 117507.
\newblock \href {https://doi.org/10.1016/j.enconman.2023.117507}
  {\path{doi:10.1016/j.enconman.2023.117507}}.

\bibitem{hasanpoor2025physics}
S.~Hasanpoor, D.~A. Romero, J.~E. Moran, C.~H. Amon, Physics-informed deep
  convolutional hierarchical encoder-decoder neural network for flow field
  prediction in wind farms, Energy and AI 21 (2025) 100553.

\bibitem{wang2021fourierfeatures}
S.~Wang, H.~Wang, P.~Perdikaris, On the eigenvector bias of {Fourier} feature
  networks: From regression to solving multi-scale {PDEs} with physics-informed
  neural networks, Computer Methods in Applied Mechanics and Engineering 384
  (2021) 113938.
\newblock \href {https://doi.org/10.1016/j.cma.2021.113938}
  {\path{doi:10.1016/j.cma.2021.113938}}.

\bibitem{sitzmann2020siren}
V.~Sitzmann, J.~N.~P. Martel, A.~W. Bergman, D.~B. Lindell, G.~Wetzstein,
  Implicit neural representations with periodic activation functions, in:
  Advances in Neural Information Processing Systems, Vol.~33, 2020, pp.
  7462--7473.

\bibitem{wang2021gradient}
S.~Wang, Y.~Teng, P.~Perdikaris, Understanding and mitigating gradient flow
  pathologies in physics-informed neural networks, SIAM Journal on Scientific
  Computing 43~(5) (2021) A3055--A3081.
\newblock \href {https://doi.org/10.1137/20M1318043}
  {\path{doi:10.1137/20M1318043}}.

\bibitem{wang2024piratenets}
S.~Wang, B.~Li, Y.~Chen, P.~Perdikaris, {PirateNets}: Physics-informed deep
  learning with residual adaptive networks, Journal of Machine Learning
  Research 25~(402) (2024) 1--51.

\bibitem{jagtap2020xpinn}
A.~D. Jagtap, G.~E. Karniadakis, Extended physics-informed neural networks
  ({XPINNs}): A generalized space-time domain decomposition based deep learning
  framework for nonlinear partial differential equations, Communications in
  Computational Physics 28~(5) (2020) 2002--2041.
\newblock \href {https://doi.org/10.4208/cicp.OA-2020-0164}
  {\path{doi:10.4208/cicp.OA-2020-0164}}.

\bibitem{wang2024dynamic}
L.~Wang, M.~Chen, Z.~Luo, B.~Zhang, J.~Xu, Z.~Wang, A.~C. Tan, Dynamic wake
  field reconstruction of wind turbine through physics-informed neural network
  and sparse lidar data, Energy 291 (2024) 130401.

\bibitem{zhang2024reconstruction}
J.~Zhang, X.~Zhao, Reconstruction of dynamic wind turbine wake flow fields from
  virtual {L}idar measurements via physics-informed neural networks, Journal of
  Physics: Conference Series 2767~(9) (2024) 092017.
\newblock \href {https://doi.org/10.1088/1742-6596/2767/9/092017}
  {\path{doi:10.1088/1742-6596/2767/9/092017}}.

\bibitem{gafoor2025physics}
A.~Gafoor~CTP, S.~Kumar~Boya, R.~Jinka, A.~Gupta, A.~Tyagi, S.~Sarkar, D.~N.
  Subramani, A physics-informed neural network for turbulent wake simulations
  behind wind turbines, Physics of Fluids 37~(1) (2025) 015110.
\newblock \href {https://doi.org/10.1063/5.0245113}
  {\path{doi:10.1063/5.0245113}}.

\bibitem{wang2026multi}
L.~Wang, M.~Dong, L.~Wang, C.~Huang, D.~Song, X.~Fan, J.~Yang, T.~Wang,
  S.~Chen, Q.~Li, Multi-scale wake modeling based on physics-informed neural
  networks and transfer learning, Applied Energy 406 (2026) 127318.
\newblock \href {https://doi.org/10.1016/j.apenergy.2025.127318}
  {\path{doi:10.1016/j.apenergy.2025.127318}}.

\bibitem{jin2026sparse}
Y.~Jin, S.~Ye, J.~Cui, Sparse data wind field wake reconstruction based on
  parametric adaptive physics-informed neural networks, Physics of Fluids
  38~(1) (2026) 015110.
\newblock \href {https://doi.org/10.1063/5.0308204}
  {\path{doi:10.1063/5.0308204}}.

\bibitem{zhang2026novel}
X.~Zhang, T.~Li, Q.~Yang, X.~Zhou, E.~Benini, A novel non-dimensional multiple
  operating conditions physics-informed neural networks for wind turbine wake
  and power prediction in complex terrains, Energy (2026) 141301.

\bibitem{zhang2025novel}
X.~Zhang, C.~Zhang, X.~Cai, Y.~Zhu, C.~Luo, A novel spatiotemporal fourier
  neural operator for dynamic wake prediction, Energy 341 (2025) 139233.
\newblock \href {https://doi.org/10.1016/j.energy.2025.139233}
  {\path{doi:10.1016/j.energy.2025.139233}}.

\bibitem{kovachki2023neural}
N.~Kovachki, Z.~Li, B.~Liu, K.~Azizzadenesheli, K.~Bhattacharya, A.~Stuart,
  A.~Anandkumar, Neural operator: Learning maps between function spaces with
  applications to pdes, Journal of Machine Learning Research 24~(89) (2023)
  1--97.

\bibitem{azizzadenesheli2024neural}
K.~Azizzadenesheli, N.~Kovachki, Z.~Li, M.~Liu-Schiaffini, J.~Kossaifi,
  A.~Anandkumar, Neural operators for accelerating scientific simulations and
  design, Nature Reviews Physics 6~(5) (2024) 320--328.

\bibitem{lu2021learning}
L.~Lu, P.~Jin, G.~Pang, Z.~Zhang, G.~E. Karniadakis, Learning nonlinear
  operators via deeponet based on the universal approximation theorem of
  operators, Nature machine intelligence 3~(3) (2021) 218--229.

\bibitem{li2020fourier}
Z.~Li, N.~Kovachki, K.~Azizzadenesheli, B.~Liu, K.~Bhattacharya, A.~Stuart,
  A.~Anandkumar, Fourier neural operator for parametric partial differential
  equations, arXiv preprint arXiv:2010.08895 (2020).

\bibitem{pathak2022fourcastnet}
J.~Pathak, S.~Subramanian, P.~Harrington, S.~Raja, A.~Chattopadhyay,
  M.~Mardani, T.~Kurth, D.~Hall, Z.~Li, K.~Azizzadenesheli, et~al.,
  Fourcastnet: A global data-driven high-resolution weather model using
  adaptive fourier neural operators, arXiv preprint arXiv:2202.11214 (2022).

\bibitem{meng2023fast}
D.~Meng, Y.~Zhu, J.~Wang, Y.~Shi, Fast flow prediction of airfoil dynamic stall
  based on fourier neural operator, Physics of Fluids 35~(11) (2023).

\bibitem{li2025attention}
L.~Li, W.~Zhang, Y.~Li, C.~Jiang, Y.~Wang, An attention-enhanced fourier neural
  operator model for predicting flow fields in turbomachinery cascades, Physics
  of Fluids 37~(3) (2025) 036121.
\newblock \href {https://doi.org/10.1063/5.0254681}
  {\path{doi:10.1063/5.0254681}}.

\bibitem{li2023iufno}
Z.~Li, W.~Peng, Z.~Yuan, J.~Wang, Long-term predictions of turbulence by
  implicit {U-Net} enhanced fourier neural operator, Physics of Fluids 35~(7)
  (2023) 075145.
\newblock \href {https://doi.org/10.1063/5.0158830}
  {\path{doi:10.1063/5.0158830}}.

\bibitem{wang2024channelFNO}
Y.~Wang, Z.~Li, Z.~Yuan, W.~Peng, T.~Liu, J.~Wang, Prediction of turbulent
  channel flow using fourier neural operator-based machine-learning strategy,
  Physical Review Fluids 9~(8) (2024) 084604.
\newblock \href {https://doi.org/10.1103/PhysRevFluids.9.084604}
  {\path{doi:10.1103/PhysRevFluids.9.084604}}.

\bibitem{renn2023forecasting}
P.~I. Renn, C.~Wang, S.~Lale, Z.~Li, A.~Anandkumar, M.~Gharib, Forecasting
  subcritical cylinder wakes with fourier neural operators, arXiv preprint
  arXiv:2301.08290 (2023).

\bibitem{lee2025channelFNO}
E.~Lee, S.~Lee, Channel-assisted fourier neural operator for high-fidelity
  far-wake prediction of a wind turbine under yawed conditions, Physics of
  Fluids 37~(9) (2025) 097143.
\newblock \href {https://doi.org/10.1063/5.0287914}
  {\path{doi:10.1063/5.0287914}}.

\bibitem{germano1991dynamic}
M.~Germano, U.~Piomelli, P.~Moin, W.~H. Cabot, A dynamic subgrid-scale eddy
  viscosity model, Physics of Fluids A: Fluid Dynamics 3~(7) (1991) 1760--1765.

\bibitem{sorensen2002numerical}
J.~N. Sorensen, W.~Z. Shen, Numerical modeling of wind turbine wakes, J. Fluids
  Eng. 124~(2) (2002) 393--399.

\bibitem{frontera2022cfd}
P.~Frontera, Cfd simulation of a floating wind turbine with openfoam: An fsi
  approach based on the actuator line and relaxation zone methods, Msc thesis,
  Delft University of Technology, Delft, The Netherlands (2022).

\bibitem{jonkman2009definition}
J.~Jonkman, S.~Butterfield, W.~Musial, G.~Scott, Definition of a 5-mw reference
  wind turbine for offshore system development, Tech. Rep. NREL/TP-500-38060,
  National Renewable Energy Laboratory, Golden, CO (2009).
\newblock \href {https://doi.org/10.2172/947422} {\path{doi:10.2172/947422}}.

\bibitem{glauert1935airplane}
H.~Glauert, Airplane propellers, in: Aerodynamic Theory, Springer, Berlin,
  Heidelberg, 1935, pp. 169--360.
\newblock \href {https://doi.org/10.1007/978-3-642-91487-4_3}
  {\path{doi:10.1007/978-3-642-91487-4_3}}.

\bibitem{martinez2017optimal}
L.~A. Mart{\'\i}nez-Tossas, M.~J. Churchfield, C.~Meneveau, Optimal smoothing
  length scale for actuator line models of wind turbine blades based on
  gaussian body force distribution, Wind Energy 20~(6) (2017) 1083--1096.

\bibitem{marten2013qblade}
D.~Marten, J.~Wendler, G.~Pechlivanoglou, C.~N. Nayeri, C.~O. Paschereit,
  Qblade: an open source tool for design and simulation of horizontal and
  vertical axis wind turbines, International Journal of Emerging Technology and
  Advanced Engineering 3~(3) (2013) 264--269.

\bibitem{make2015analyzing}
M.~Make, G.~Vaz, Analyzing scaling effects on offshore wind turbines using cfd,
  Renewable Energy 83 (2015) 1326--1340.
\newblock \href {https://doi.org/10.1016/j.renene.2015.05.048}
  {\path{doi:10.1016/j.renene.2015.05.048}}.

\bibitem{onel2021investigation}
H.~C. Onel, I.~H. Tuncer, Investigation of wind turbine wakes and wake recovery
  in a tandem configuration using actuator line model with les, Computers \&
  Fluids 220 (2021) 104872.
\newblock \href {https://doi.org/10.1016/j.compfluid.2021.104872}
  {\path{doi:10.1016/j.compfluid.2021.104872}}.

\bibitem{tu2023aerodynamic}
Y.~Tu, K.~Zhang, Z.~Han, D.~Zhou, O.~Bilgen, Aerodynamic characterization of
  two tandem wind turbines under yaw misalignment control using actuator line
  model, Ocean Engineering 281 (2023) 114992.
\newblock \href {https://doi.org/10.1016/j.oceaneng.2023.114992}
  {\path{doi:10.1016/j.oceaneng.2023.114992}}.

\end{thebibliography}

% Mark bibliography entries added or corrected during revision.
\makeatletter
\let\DGDoriginalbibitem\bibitem
\newcommand{\DGDbluebibkeys}{glauert1935airplane,bouckaert2021net,GWEC2025Report,barthelmie2007modelling,li2025effects,he2025novel,luo2025innovative,liu2026phywakenet,yang2026wind,zhang2025advanced,wang2025effects,li2025effects,goccmen2025data,gafoor2025physics,wang2026multi,li2025attention,zhang2025novel,wang2021fourierfeatures,sitzmann2020siren,wang2021gradient,wang2024piratenets,jagtap2020xpinn,parekh2023uncertainty,li2024transformerwake,pathak2022fourcastnet,li2023iufno,wang2024channelFNO,lee2025channelFNO,marten2013qblade,make2015analyzing,onel2021investigation,tu2023aerodynamic,dong2026impacts,dong2026reinforcement}
\newcommand{\DGDsetbibcolor}[1]{%
    \IfStrEq{#1}{jonkman2009definition}{\color{black}}{\color{black}}}
\newcommand{\DGDbluebibitem}[1]{%
    \DGDsetbibcolor{#1}\DGDoriginalbibitem{#1}}
\def\DGDbluebibitemopt[#1]#2{%
    \DGDsetbibcolor{#2}\DGDoriginalbibitem[#1]{#2}}
\renewcommand{\bibitem}{%
    \@ifnextchar[\DGDbluebibitemopt\DGDbluebibitem}
\makeatother


\color{black}

\end{document}